\documentclass[12pt]{report}
\usepackage{graphics}
\usepackage{setspace}
\usepackage{geometry}
\def\be{\begin{equation}}
\def\ee{\end{equation}}
\def\bc{\begin{center}}
\def\ec{\end{center}}

\begin{document}
\begin{titlepage}
\begin{center}
{\Large\bf NOVEL DYNAMICAL PHENOMENA \\
\vskip .5cm
IN MAGNETIC SYSTEMS}\\ [5.5cm]
{\large\bf SOHAM BISWAS}\\
\vskip .5cm
 {\large\bf DEPARTMENT OF PHYSICS\\
 UNIVERSITY OF CALCUTTA\\
INDIA\\
 2011} \\
\end{center}

\end{titlepage}

\thispagestyle{empty}


\newpage
\thispagestyle{empty}
\begin{center}
{\large\bf Acknowledgement}
\end{center}
\vspace{-0.2cm}
I express my sincere gratitude to Prof. Parongama Sen for her invaluable guidance, advice and continuous
encouragement for the past five years. In these years her words became true that ``the relationship between a student and a 
supervisor is much more than simply a student-teacher relationship''. I have enjoyed not only academic but also non-academic 
discussions with her. I thank Prof Deepak Dhar, Prof. Bikas Chakrabarti, Prof. Indrani Bose, Prof Subinay Dasgupta, Prof 
Purusattam Ray and Prof. P.K.Mohanti for some valuable discussions regarding my work.

I also thank my collaborators Prof Purusattam Ray and Dr. Anjan Kumar Chandra for all the help and support that I received from them. 

I thank the faculty members of the Department of Physics, University of Calcutta for their support in various forms.

I thank all my friends of the Department of Physics, University of Calcutta.
I shared my room in the Department with Roshni, Sanchari, Bipasha, Biplob, Chirashree, Pankaj and Anindya. 
I have thoroughly enjoyed the primarily non-academic and some academic discussions with them.

I also thank the University of Calcutta, the University Grants Commission (UGC/520/Jr. Fellow (RFSMS) and the Council of 
Scientific and Industrial Research (SRF-DIRECT: sanction no. 09/028 (0818)/2010 EMR-I) for providing me the financial support. 
DST project SR/S2/CMP-56/2007 is also acknowledged for the financial support and also for the providing computational facilities 
for this work.

Finally, I thank my parents for their support.

\newpage
\pagenumbering{roman}
\tableofcontents
\newpage

\date{\today}
\pagenumbering{arabic}

\chapter{Introduction}
\label{CIntro}
\section{Magnetic system : The Models}
A magnet may be regarded as consisting of a set of magnetic dipoles residing on the vertices of a 
crystal lattice. We often refer to the magnetic dipoles as spins. The spins are able to exchange 
energy through interactions between themselves and with other degrees of freedom of the crystal lattice (e.g., via
spin orbit coupling).
There are many models which have been proposed to represent magnetic systems.
Ising, XY and Heisenberg models are three spin models which may be considered to be the most basic models among them \cite{golden}.

The simplest model describing interactions between spins, is the {\em Ising Model}, proposed by E. Ising  
in 1925 to represent magnetic systems 
 and alloys and magnetic phase transitions from ferromagnetic/antiferromagnetic to paramagnetic states \cite{Ising}. 
The model consists of a system of
magnetic dipoles placed on a hypercubic lattice, that can be either up or down and
interact among themselves only by nearest neighbour interactions. The Hamiltonian of the Ising model 
with no external magnetic field can be written as 
\begin{equation}
 H = -  \sum_{<ij>} J_{ij}S_i S_j ,
\label{hamiltonian}
 \end{equation}
where the interaction between the $i$th and $j$th spins is denoted by $J_{ij}$.
This model with nearest neighbour uniform interaction exhibits a temperature-induced continuous phase transition at a non-zero 
temperature in two and higher dimensions, but not 
in one dimension. Though in this model a spin can have only two states (either up or down),  
still it is possible to obtain phase transition and critical behaviour in a realistic manner from this simple model which has made
it one of the most studied models in history of condensed matter physics.

In the Heisenberg model, spins $S_{i}$ obey a continuous symmetry instead of discrete symmetry.
That means the $S_{i}$ are $vectors$. 
The Hamiltonian of equation (\ref{hamiltonian1}) can be generalized as the Hamiltonian of Heisenberg and XY model.
\begin{equation}
 H = -  \sum_{<ij>} J_{ij}\vec{S_i}. \vec{S_j} .
\label{hamiltonian1}
 \end{equation}
In the  Heisenberg model $S_{i}$ are allowed to point in all directions ($4\pi$ steradians), rather than 
 having only up or down state of the spins ($S_{i}=\pm1$). 
The Hamiltonian of the XY Model is also given by equation (\ref{hamiltonian1}) but the spins are unit vectors confined 
to rotate in a plane. Since the Heisenberg and XY models involve more than one component of spin, these models 
are essentially quantum in nature as the different components of the spin do not commute. The Ising model on the 
other hand, is purely classical in nature. 

Ising model can be generalized to the $q$-state classical spin model popularly known as Potts model \cite{potts} 
where lattice spins can take $q$ different discrete values.
In this model, a system of spins are considered to be confined in a plane, with each spin pointing to one of the 
equally spaced directions specified by the angles
\[
 \Theta_n = \frac{2\pi n}{q}, ~~~~ n=0,1,2,....,q-1
\]
The interaction Hamiltonian of the Potts model is given by 
\begin{equation}
 H= - \sum_{<ij>} J_{ij} \delta_{S_i,S_j},
\end{equation}
where the interaction between spins $S_i$ and $S_j$, is denoted by $J_{ij}$ and $\delta_{kr}$ is the Kronecker delta given by
\[
 \delta_{\alpha, \beta} = \frac{1}{q} [1+(q-1)e^\alpha e^\beta]
\]
and $\alpha=0,1,...,q-1$ are $q$ unit vectors pointing in the $q$ symmetric directions of a hypertetrahedron in $q-1$ dimensions.
For $q=2$, Potts model is equivalent to the 
Ising model. Though Potts model is a simple extension of the Ising model, it has a much richer phase structure, 
which makes it an important testing ground for new theories and algorithms in the study of critical phenomena \cite{Wu}.

Ising model does not have any intrinsic dynamics as it is a classical model. The dynamics of Ising model can only 
be induced by the influence of some external agents (change of temperature or field etc.). 
In this thesis to study the dynamics of the magnetic systems we will restrict ourselves in the study of the dynamics 
 of Ising like classical spin systems only.

\section{Dynamical phenomena}
Dynamics of spin models  is a much studied
phenomenon and has emerged as a rich field of
present-day research. Models having identical static critical behavior may display different
behavior when dynamic critical phenomena are considered \cite{Ho_Ha}.
Our primary focus is on a prototypical system that is initially in a homogeneous high-temperature disordered phase 
and the temperature is quenched (suddenly dropped) to below the critical temperature.
The quenching phenomenon below the critical temperature is an important dynamical feature.
Because of the complexity of the domain coarsening process at the level of discrete spins, considerable effort has been devoted in
constructing a complementary approach that is based on a phenomenological description at the continuum
level. But here in this thesis we have studied the quenching dynamics of Ising spin like system which exhibit discrete 
symmetry, not the continuous one. The details of the dynamics and the methodology is discussed in the next chapter 
(Chapter \ref{Cdyn} ) of the thesis.

When a homogeneous system is quenched to below the critical temperature, a coarsening mosaic of
ordered-phase domains forms, as the distinct broken-symmetry phases compete with each other in their
quest to select the low-temperature thermodynamic equilibrium state \cite{gunton,bray}. As a result of this competition,
equilibrium is never reached for an infinite system. Instead, self-similar behavior typically arises, where the
domain mosaic looks the same at different times but only its overall length scale changes. This self-similarity
is an important simplifying feature that is characteristic of coarsening. So one of the very interesting phenomena 
widely studied in the quenching process is the domain growth \cite{gunton,bray}.

For the dynamics of zero temperature quenching the scenario is a little bit different.
In  one dimension, a zero temperature quench  of the  Ising model
ultimately leads to the equilibrium
configuration, i.e., all spins point up (or down) for the finite system size. 
 The average domain size $D$ increases in time $t$ as $D(t)\sim t^{1/z}$,
where $z$ is the dynamical exponent associated with the growth.
In two or higher dimensions, however, the system does not
always reach equilibrium \cite{Krap_Redner} even for the finite system size, although the scaling relations
still hold good. But even in one dimension, if the interactions are not restricted to nearest neighbours only, 
the dynamical behaviour may change considerably, often leading to absence of scaling altogether. 
We have therefore studied the zero temperature quenching dynamics of Ising spin like models in several systems where 
the interactions are more complicated than simple nearest neighbour type.

Apart from the domain growth  phenomenon, another important dynamical
 behavior commonly studied is persistence \cite{satya1,derrida}. In Ising model, in a zero temperature quench, persistence 
is simply the probability that a spin has not flipped till time $t$ and
is given by $P(t)\sim t^{-\theta}$. $\theta$ is called the persistence exponent and
is unrelated to any other previously known static or dynamic exponents. A general discussion of persistence and  
its scaling etc., has been included in the next chapter (Chapter \ref{Cdyn}).

With the understanding developed in connection with the dynamics of nonlinearly coupled many body 
systems in Physics for the last four decades, people started to study the macroscopic dynamics of 
various social systems or networks. One of the first models in sociophysics was 
proposed by   Schelling \cite{schell} in 1971 to simulate social segregation,  
 which  was  similar (in purpose) to  phase separation models studied by physicists.  
In recent years, building on the development of the kinetic theory of 
gases  and statistical mechanics, physicists   have begun to incorporate a statistical thermodynamic 
perspective in models of social physics 
in which individuals are viewed as some effective atoms/molecule-like units (having spatial and
dynamical properties)  and the law of large numbers yields social behaviours. 
Microscopic human behaviour is assumed to be represented in such models by real numbers.
When these numbers are discrete and have binary choices, the social system can be modeled 
 as a magnetic model where the Ising spin variables can represent the states of the individuals and 
the interactions among them by spin-spin interactions \cite{staufer,sznajdweron}.
Existence of a phase transition from a heterogeneous society to a homogeneous society  \cite{baron,galam} in many  opinion dynamics   models also
adds to the interest of Sociophysics.

The dynamics of Ising model can also be mapped to a random walk problem as domain coarsening is identical to 
a reaction diffusion system \cite{derrda_reac_diff}. The motions of the domain walls in one dimension (with nearest neighbour 
interactions only), can be viewed as the 
motions of the particles $A$ with the reaction $A + A \rightarrow \emptyset$.  This means the 
particles are walkers and when two particles come on top of each other they are annihilated.
The annihilation reaction ensures domain coalescence and coarsening.

\section{The outline of the thesis}
The work reported in this thesis includes studies on the zero temperature quenching dynamics of
 Ising spin like models as well as opinion dynamics, which involves Ising spin like variables. 
A random walk problem is also included in this thesis, as domain coarsening in one dimension can be mapped to 
a reaction diffusion system. 

In chapter  \ref{Cdyn} we have presented a general discussion on the features 
associated with the quenching dynamics, coarsening phenomena, persistence etc. and the numerical methods we have used to study them.

In chapter \ref{Cannni} we have presented our investigation on the dynamics of a two dimensional axial next nearest neighbour
Ising  (ANNNI) model following
a quench to zero temperature. The Hamiltonian is given by
\begin{equation}
H = -J_0\sum_{i,j=1}^L S_{i,j}S_{i+1,j} - J_1\sum_{i,j=1} [S_{i,j} S_{i,j+1} -\kappa S_{i,j} S_{i,j+2}].
\end{equation}
For $\kappa <1$, the system does not reach the 
equilibrium ground state but slowly evolves to a metastable state.
For $\kappa > 1$, the system shows a behaviour similar to the 
two dimensional ferromagnetic Ising model in the sense that it freezes to 
a striped state with a finite probability. The persistence probability shows algebraic 
 decay here with an exponent $\theta = 0.235 \pm 0.001$ while the
dynamical exponent of growth $z=2.08\pm 0.01$.
For $\kappa =1$, the system belongs to a completely different
dynamical class; it always evolves to the true ground state with  the
persistence and dynamical exponent having unique values.
Much of the dynamical phenomena can be
understood by  studying the dynamics and distribution of the number of domains walls. 
We have also compared the dynamical behaviour to that of a Ising   model in which both the nearest and 
next nearest neighbour interactions are ferromagnetic \cite{annni}.

Randomness is known to affect the dynamical behaviour of complex systems to a large extent. 
In the next chapter (chapter \ref{Crand})  we have presented our investigation on how the nature of randomness affects the
dynamics in a zero temperature quench of Ising model on two types of random networks. 
In both the networks, which are embedded in a one dimensional space, the first neighbour connections exist
and the average degree is four per node.  
In the random model A,
the second neighbour connections are rewired with a probability $p$ while in the random model B,
additional connections between neighbours at Euclidean distance $l ~ (l >1)$  are introduced with 
a probability $P(l) \propto l^{-\alpha}$. We find that for both models, 
the dynamics leads to freezing such that the system gets locked in a disordered state. The point at which
the disorder of the nonequilibrium steady state is maximum is located. 
Behaviour of  dynamical quantities like residual energy, 
order parameter and persistence are
discussed and compared. 
Overall, the behaviour of physical quantities are similar although subtle 
differences are observed due to the difference in the nature of randomness.

In chapter \ref{Cmodel}, we have proposed a new model of binary opinion for opinion dynamics  
in which the opinion of the individuals change according to 
 the state of their neighbouring domains.
If the neighbouring domains have opposite opinions, then the opinion of the 
domain with the larger size is followed (Model I).
Starting from a random configuration, the system evolves to a homogeneous 
state. The dynamical evolution  show novel scaling behaviour with the 
 persistence exponent $\theta \simeq 0.235$
and dynamic exponent $z \simeq1.02 \pm 0.02$. Here we have obtained a new dynamical class. 
Introducing disorder in Model I through a parameter called rigidity parameter $\rho$ 
(probability that people are completely rigid and never change 
their opinion), the transition to a heterogeneous society  
at $\rho = 0^{+}$ is obtained. Close to $\rho =0$, the   equilibrium values of the 
dynamic variables show power law scaling behaviour with $\rho$.
 We have also discussed the effect of having both quenched and annealed disorder in the system \cite{model}.
Further, by mapping Model I to a system of random walkers in one dimension 
with a tendency to walk towards their nearest neighbour with probability  $\epsilon$, we 
find that for any $\epsilon > 0.5$, the Model I dynamical behaviour is prevalent at long times \cite{rwalk}.

 In chapter \ref{Ccrossover}, a parameter $p$ is defined to modify the dynamics introduced in chapter \ref{Cmodel} such that
a spin can sense domain sizes up to $R = pL/2$ in a one dimensional system 
of size $L$. For the cutoff factor $p \to 0$, the dynamics is Ising like and the   domains grow with time $t$ diffusively  as 
$ t^{1/z}$ with $z=2$, while  for $p=1$, the original model I showed
ballistic dynamics with $z \simeq 1$. For intermediate values of $p$, 
the domain growth, magnetization and persistence show model I like  
behaviour up to a macroscopic crossover time $ t_1 \sim pL/2$.  Beyond $t_1$,
characteristic power law variations of the dynamic quantities are 
no longer observed. The total time
to reach equilibrium is found to be  $t = apL + b(1-p)^3L^2$, from which we 
conclude that the later time behaviour is diffusive. 
We have also considered the case when  a random but quenched 
 value of  $p$ is used  for each spin for which  ballistic  behaviour is once again obtained \cite{crossover}.

\chapter{Dynamical Phenomena in Ising systems}
\label{Cdyn}

\section{Introduction}
Dynamics of spin models, as mentioned in the previous chapter, has emerged as a rich field of present day research.
When a system is at the critical point or close to the critical point, anomalies occur in a large variety of dynamical properties 
and models having identical static critical behavior may display different dynamical behavior when the system is close to the 
critical point.
The dynamical properties of a system are quantities which depend on the equations of motion and are not determined simply by the 
equilibrium properties.

Over last few decades, a number of theoretical ideas like (i) the conventional theory of critical slowing down, 
(ii) the `mode-coupling theory' of transport phenomena, (iii) the hypothesis of dynamical scaling and universality and 
(iv) the renormalization group approach to critical dynamics etc. have been proposed and discussed for understanding of dynamic 
critical phenomena. Brief reviews of these concepts may be found in \cite{2Ho_Ha}.

\section{Equilibrium dynamics }
A system is in equilibrium when its bulk properties remain constant or at least
fluctuate closely around a constant mean value over a time period long enough in the context of the study.
The equilibrium statistical mechanics can be explored totally once the partition function $Z$ of the system is known.
But even when the system is in thermal equilibrium, to calculate any thermodynamic quantity 
we need the knowledge of the variation of $Z$ with temperature and other 
parameters affecting the system (like external magnetic field). If the partition
function can be calculated exactly, the problem is said to have an exact solution \cite{2Ising, 2baxter}.
But whenever we are studying the dynamics of the system, most of the times a probabilistic description is required for the 
formation of of equation of motion. Often it is not possible to compute the probability
distribution functions analytically in explicit form, because of the complexity of
the problem and we need to go for a number of approximate techniques which include series expansions, field
theoretical methods and computational methods. The focus of this chapter will be mainly the computational methods, explicitly 
the method of Monte Carlo Simulation \cite{MC, new_brk}

 \subsection{Monte Carlo Study}

Numerical simulation can be regarded as an experiment made on computer.
For stochastic systems in which the number of degrees of freedom is large and analytical methods
are not very efficient, computer simulation becomes a very useful method. Monte-Carlo methods aim at a
numerical estimation of probability distributions as well as of averages, that can be calculated from them, 
making use of (pseudo) random numbers \cite{MC}. Whenever we are considering any classical stem (say for example Ising spin system) 
to calculate the average of any observable 
quantity $O$, not all configurations are equally likely, rather their probabilities are proportional to the Boltzmann factor 
$\exp (-\beta E)$. Thus the ensemble average of the quantity of interest over all states $\mu$ of the system (weighting each other 
with its own Boltzmann probability) is given by 
\begin{equation}
 \langle O \rangle = \frac{\sum_{\mu} O_\mu \exp(-\beta E_\mu)}{\sum_{\mu} \exp(-\beta E_\mu)}
\label{av}
\end{equation}
where $\beta = 1/kT$, $k$ being the Boltzmann’s constant and $E_\mu$ is the energy of the state $\mu$.

One can choose $N$ such states $\{ \mu_1,\mu_2,..,\mu_N \}$ randomly and take the above average.
However, it should be remembered that for the canonical ensemble the fluctuations in the energy vanishes 
as $1/\sqrt{L}$ ($L$ be the system size) which means that only a few states with energy very close to the average energy 
will occur with high probability and contribute to the average. Thus it will be meaningful to device a method by 
which one can generate the states which are more probable.

These states can be dynamically evolved from arbitrary initial states. 
The mechanism of generating a new state $\delta$ from the initial state $\mu$ of the system, in a random fashion 
using a $`$transition probability' $w_{\mu \delta}$ is called a \textbf{Markov process} \cite{markov}.
For a Markov process all the transition probabilities should satisfy the following two conditions : 

1. They should not vary over time.

2. They should depend only on the properties of the current states $\mu$ and $\delta$, and
not on any other states the system has passed through (history independent).
The transition probabilities $P(\mu \rightarrow \delta)$ must satisfy also the constraint
\[
\sum_\delta P(\mu \rightarrow \delta) =1
\]
If $P_A (t)$ is the probability of a state $A$ at time $t$, 
then the master equation can be written as
\begin{equation}
\frac{dP_A}{dt} = \sum_B w_{BA} P_B (t) - \sum_B w_{AB} P_A (t),
\label{mseq}
\end{equation}
where the first term is a gain term and the second one is the loss term. $w'$ s denote the transition probabilities.
The above equation (Equ. \ref{mseq}) can be written, for discrete times,
\begin{equation}
 P_A(t+1) -P_A(t)= \sum_B w_{BA} P_B (t) - \sum_B w_{AB} P_A (t).
\end{equation}
At steady state, which is expected to occur for large times, the RHS is zero which gives the condition
\[
 w_{BA} P_B (t \rightarrow \infty) = w_{AB} P_A (t \rightarrow \infty).
\]
Since at equilibrium, the probabilities are given by the Boltzmann’s expression, we have,
\begin{equation}
  w_{BA} \exp(-\beta E_B) = w_{AB} \exp(-\beta E_A)
\label{detbal}
\end{equation}
The above condition (Equ \ref{detbal}) is known as the \textbf{principle of detailed balance}.

That means the system is not sampling all the states with equal probability, but sampling them according 
to the Boltzmann probability distribution.
This process of choosing states which are more probable than just choosing a set of random states is known as 
\textbf{importance sampling}. Thus in an importance sampling, average values are calculated using the formula
\begin{equation}
 \langle O \rangle = \frac{1}{N}\sum_{\mu=1}^N O_\mu
\label{impav}
\end{equation}
instead of equation \ref{av}.

There can be several choices for the transition probabilities. Metropolis and Glauber are important among them.
In this thesis we have basically studied the nonequilibrium quenching dynamics of Ising spin systems using Glauber dynamics.
The detail of these are discussed in next section of this chapter.

\section{Quenching dynamics}
The phenomena in which the temperature of a system is suddenly dropped from very high ($T \rightarrow \infty$) to a very low 
value ($T \sim 0$), is called quenching. For an Ising spin system very high temperature means the system is in 
completely random disordered phase. The behaviour of Ising spin system following a deep quench below the critical temperature 
comprises a central topic in the study of the nonequilibrium dynamics of the system nowadays. Systems quenched from a disordered 
phase into an ordered phase do not order instantaneously. Instead, the length scale of ordered regions grows with time as
the different broken symmetry phases compete each other to select the equilibrium state \cite{2bray}.

The nonequilibrium process is very complex and critical to understand \cite{priv}. Here the probability distributions are not the simply 
Boltzmann distributions (as in equilibrium process) and changing at each and every time step.
This process may be roughly devided in two different categories. First, the dynamical system which evolve according to some given 
dynamical rules exist and there is no so called Hamiltonian describing the system. Second the systems, where the equilibrium state is 
known and one starts far from  equilibrium. Then we evolve the system according to a rule which has been determined from 
the equilibrium dynamics of the system (these rules lead the system to the equilibrium many times) and observe what happens.
 Thus in the second case we use the transition probabilities determined from the equilibrium dynamics of the system and we also use  
the formula given by equation \ref{impav} to calculate the average of any observable quantity $O$. One of the example pf the 
second case is the quenching dynamics of Ising spin system, in which we are interested.

Many rigorous and nonrigorous results have been obtained on different questions arising in this context: 
the formation of domains, their subsequent evolution (discussed in Sec. \ref{domain}), spatial and temporal
scaling properties, the persistence properties at zero and positive temperature (Sec. \ref{per}); the observed aging phenomena in
both disordered and ordered systems and many others \cite{2bray,new}.

It may be noted that Glauber dynamics can be used when the order parameter is not conserved. For system with conserved order 
parameter, the dynamical evolution can be studied using e.g, the Kawasaki exchange dynamics \cite{kawa}. 
In this thesis we shall present our investigation on those systems, where the order parameter is not conserved. So in this thesis 
we shall discuss the quenching phenomena considering the evolution of the Ising spin system using 
Glauber Dynamics \cite{glauber}.

\subsection{Glauber dynamics}
Let us take a Ising spin system with Hamiltonian 
\begin{equation}
 H = -  \sum_{<ij>} J_{ij}S_i S_j ,
\label{2hamiltonian}
 \end{equation}
where $S_i = \pm 1$ and the interaction between the $i$th and $j$th spins is denoted by $J_{ij}$.
 We start from a completely random disordered phase (that means spins are completely uncorrelated and $ S_i = \pm 1$ equiprobably)
 which evolve by very low finite temperature Glauber dynamics \cite{glauber} corresponding to a quench from very high temperature 
($T \rightarrow \infty$) to very low one ($T \sim 0$). For each initial spin configuration, one realization of the dynamics was 
performed until the final state has been reached. As Glauber dynamics is essentially a single spin flip dynamics, one may consider 
single spin flips to generate the configuration B from the configuration A. Thus only one spin in configuration A is flipped 
to get the configuration B. The choice of transition probability $w_{AB}$  in Glauber dynamics is simply
\begin{equation}
 w_{AB} = 1/2 (1 - \tanh(\beta \bigtriangleup E))
\label{gl}
\end{equation}
where $\bigtriangleup E = E_B - E_A$ and $\beta=1/kT$

So the precise steps for performing Monte Carlo simulation following Glauber dynamics will be as follows :

1. Pick up a spin at random.

2. Calculate the change of energy $\triangle E$ which is essentially $E_{flipped} - E_{present}$.

3. Flip the spin according to the probability given by equation (\ref{gl}).

For zero temperature Glauber dynamics (that means the Ising spin system is quenched to $T=0$ temperature) 
the rule of spin flip will be as follows (obtained by putting $T=0$ in equation \ref{gl}) :

1. If $\bigtriangleup E <0$ : The spin will flip ($w_{AB} =1$)

2. If $\bigtriangleup E >0$ : The spin will not flip ($w_{AB} =0$)

3. If $\bigtriangleup E = 0$ : The spin will flip with probability 1/2 ($w_{AB} =1/2$)

For a system of $N$ spins, one Monte Carlo time step is said to be completed after $N$ such flippings. Here the process of update 
we have considered (pick up a spin at random and update it according to the rule) is known as \textit{random update}.
In the process of random update, in one MC time step a single spin can be picked up more than once and there may exist few spins 
which could not be picked up at all. There is another process of update, known as sequential update where each and every spin of 
the spin system are used to be picked up sequentially one after another for update.

Random update process can be of two types named (a) random sequential update and (b) random parallel update. In the random update process 
either all spins are randomly selected and updated at each time step, or only one spin is randomly selected and updated in each time 
step. We refer to the first update rule as random parallel update, and
the second as random sequential update. The time interval between updates is taken $\bigtriangleup t = 1$ in parallel updates, and 
$\bigtriangleup t = 1/N$ in sequential updates, such that in both cases $O(N)$ spins are updated per unit time. Here $N$ is the 
number of spins or the system size. Glauber dynamics was 
originally introduced as a sequential updating process \cite{glauber} and in one dimension evolution under this dynamics with 
random sequential updating is already well known and can be derived analytically \cite{su}. The process of updating of all the spins of 
the spin system simultaneously, at one go following the rule of the dynamics, is known as parallel update \cite{parallel}.

Here in this thesis we have used the process of random sequential update everywhere.

\subsection{Domain coarsening}
\label{domain}

\begin{figure}[hbpt]
\centerline{
{\resizebox*{6.3cm}{!}{\includegraphics{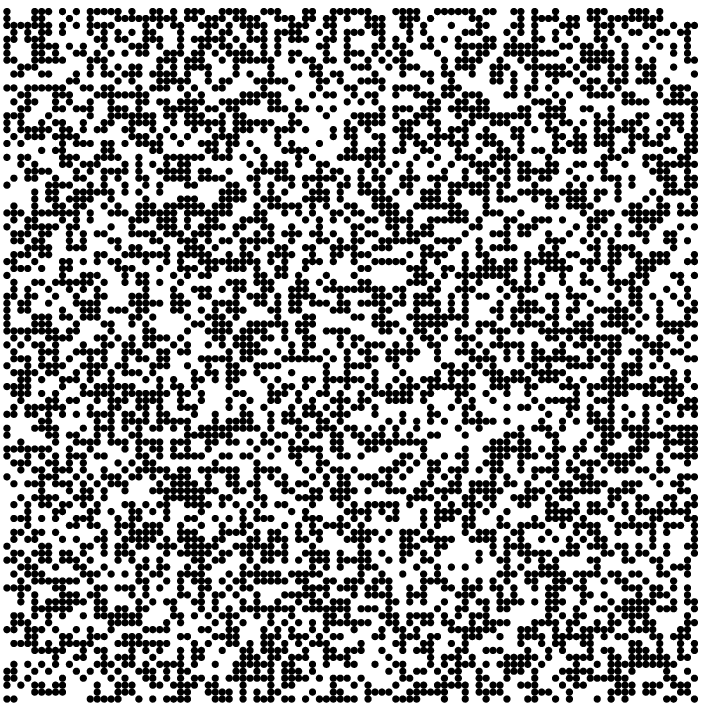}}}
\hspace{2.4cm}
{\resizebox*{6.3cm}{!}{\includegraphics{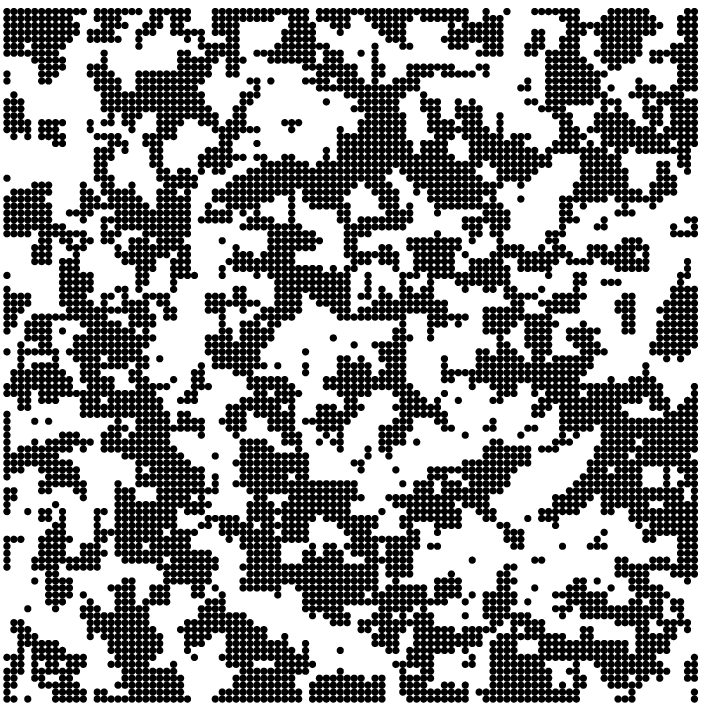}}}
}
\hspace{3cm} $t_1$ \hspace{8cm} $t_2$
\vskip 0.7cm
\centerline{
{\resizebox*{6.3cm}{!}{\includegraphics{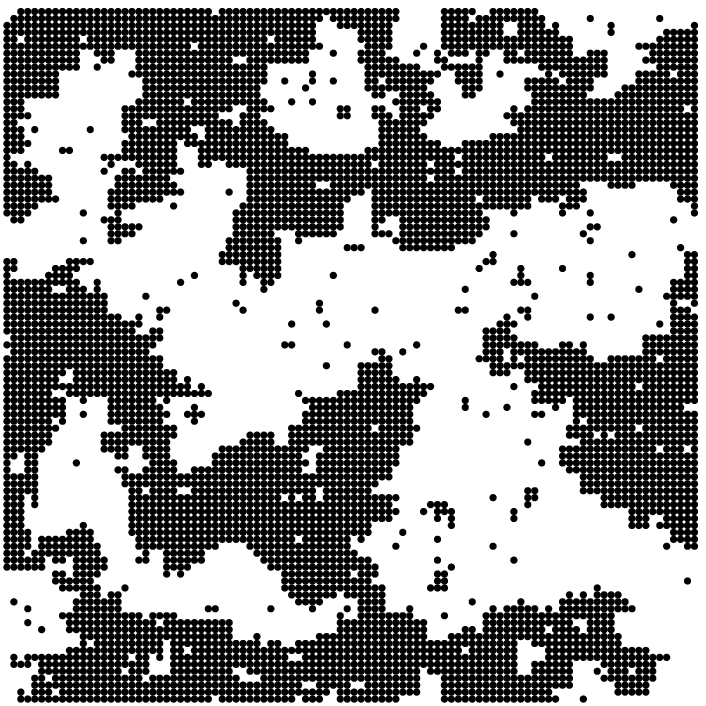}}}
\hspace{2.4cm}
{\resizebox*{6.3cm}{!}{\includegraphics{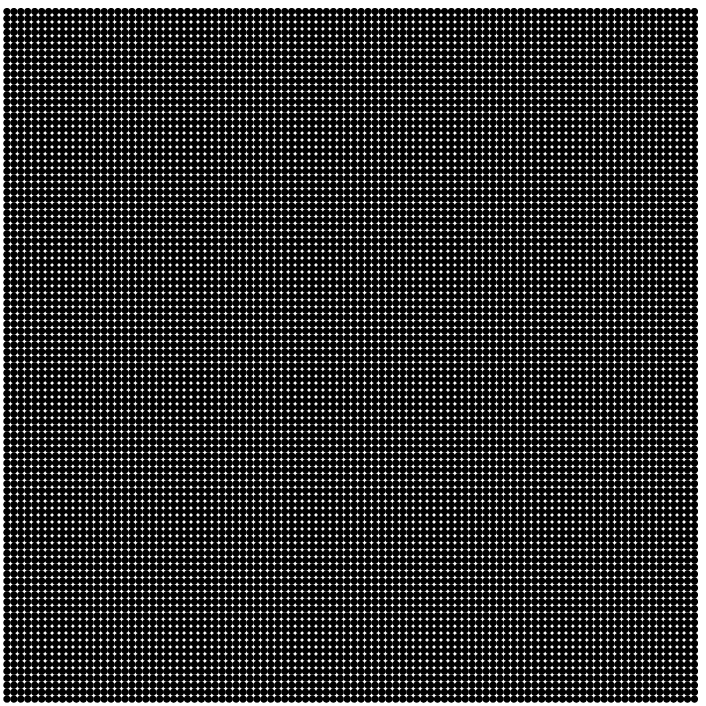}}}
}
\hspace{3cm} $t_3$ \hspace{8cm} $t_f$
\caption{Domain growth with time for two dimensional Ising spin system, after a quench to zero temperature. Here $t_1 <t_2<t_3<t_f$.
$t_f$ is the time counted after the system has reached the equilibrium(all up or all down state for $T=0$.}
\label{coarsening}
\end{figure}

It is mentioned earlier that when a system is quenched from a homogeneous high temperature disordered state to a low temperature 
state it does not order instantaneously. Broken symmetry phases compete each other to select the equilibrium state and the domains 
grow with time.  A scale-invariant morphology is developed, i.e., the network of domains is (͑statistically)͒ independent of time when 
lengths are rescaled by a single characteristic length scale that typically grows algebraically with time \cite{2bray,2gunton}.
For the zero temperature dynamics the average domain size $D$ increases in time $t$ as 
\[
 D(t)\sim t^{1/z},
\]
where $z$ is the dynamical exponent associated with the growth. A typical picture of domain coarsening following a quench to zero 
temperature is shown in figure \ref{coarsening}.

 \vskip 0.5cm

\subsection{Persistence}
\label{per}
Apart from the domain growth  phenomenon, another important dynamical behavior that has attracted considerable interest recently
is persistence. Persistence is simply the probability that the fluctuating nonequilibrium field does not change sign upto time 
$t$ \cite{satya}. The problem of persistence in spatially extended nonequilibrium systems has recently generated a lot of interest 
both theoretically \cite{der,2stauffer,krap1,watson} and experimentally \cite{marcos,tam}. 

Single spin persistence provides a natural counterpart to the survival probability in the realm of many-particle systems.
In the context of reaction processes, persistence is equivalent to the survival of immobile impurities and therefore does not 
provide information about collective properties of the bulk \cite{krap}. In Ising model, in a zero temperature quench, 
persistence is simply the probability that a spin has not flipped till time $t$ and is given by 
\[
 P(t)\sim t^{-\theta},
\]
where $\theta$ is called the persistence exponent and is unrelated to any other previously 
known static or dynamic exponents. Persistence probability is in general non-Markovian time evolution of a local fluctuating
variable, such as a spin from its initial state.

The persistence probability is hard to measure in simulations, at nonzero temperature, because one needs to distinguish spin flips
due to thermal fluctuations from those due to the motion of interfaces. The prescription for measuring the persistence probability 
for single spin flip at a finite temperature is given in \cite{2derrida}.

Apart from such $`local'$ persistence, one can also study the $`global'$ persistence behaviour by measuring the probability $P_G (t)$ 
that the order parameter does not change its sign till time $t$ \cite{global}. At the critical temperature, the probability that the 
individual spins will not be flipped till time $t$ has an exponential decay, while the global persistence shows an algebraic decay:
\[
 P_G(t)\sim t^{-\theta_G}.
\]

\section{Quenching of nearest neighbour Ising model}
\begin{figure}[hbpt]
\centerline{
 {\resizebox*{6.5cm}{!}{\includegraphics{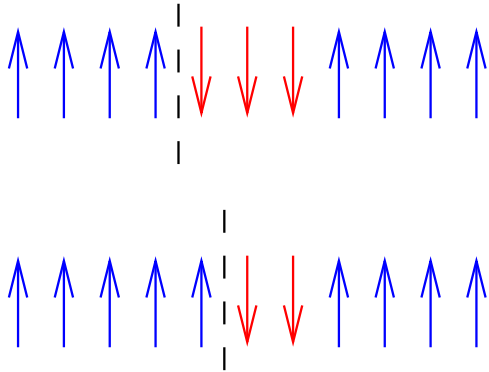}}}
}
\caption{Schematic picture of the zero temperature quenching dynamics of one dimensional nearest neighbour Ising model. The red colour 
domain is shrinking with time and the domain walls between the red and blue colour spin will annihilate each other to form all up 
state (with blue colour spins only) after few time steps.}
\label{1dIsing}
\end{figure}

The Hamiltonian of the Ising spin model we have considered here is given by equation \ref{2hamiltonian}, where $S_i = \pm 1$ and 
the sum is over all nearest-neighbor pairs of sites $\langle ij \rangle$. Now we ask the question, what is the fate of the 
Ising system after a zero temperature quench. In  one dimension, a zero temperature quench  of the  Ising model (with nearest 
neighbour interactions only) ultimately leads to the equilibrium configuration (Figure \ref{1dIsing}). Here the domain walls approach 
each other and annihilate, the system goes to its stable state (all up or all down) at very large times.

\begin{figure}[hbpt]
\centerline{
 {\resizebox*{6.5cm}{!}{\includegraphics{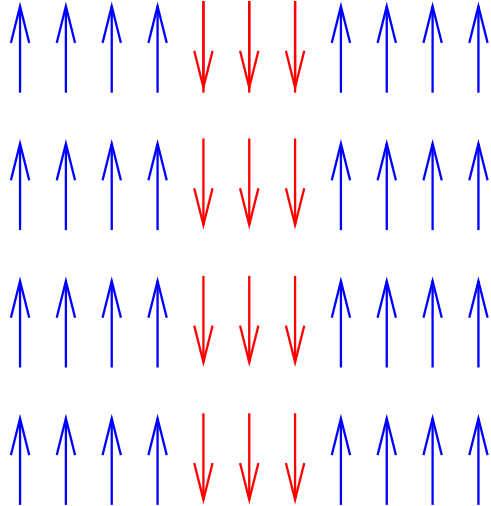}}}
}
\caption{Schematic picture to show freezing at the zero temperature Glauber dynamics of two dimensional nearest neighbour 
Ising model.}
\label{freeze}
\end{figure}

In higher dimensions, the system cannot reach the ground state for all initial configurations. In two dimensions, the system can find 
out the ground state for about $70\%$ cases \cite{freeze}. In two dimension, on the square lattice, there exist a huge number of metastable
 states that consists of alternating vertical ͑or horizontal͒ stripes whose widths are all $ \geq 2 $. These arise because in
zero-temperature Glauber dynamics, a straight boundary between up and down phases is stable (Figure\ref{freeze}).
\vskip 0.2cm
From the schematic picture \ref{freeze}, it is very clear that any spin at the boundary is supported by three neighbouring spins and 
a reversal of any spin along the boundary increases the energy. However, a stripe of width one is unstable, as there will not be any 
change of energy due to the flipping of one of the spins in the stripe.

In three dimensions the Ising spin system (with nearest neighbour interaction), never reaches the ground state by zero temperature 
single spin flip Glauber dynamics. 

\subsection{Finite size scaling}
It is not possible to reach the thermodynamic limit (means $L\rightarrow \infty$) numerically, when we are doing the simulations in 
computer. We always do our simulations in finite system size, no matter how large. So we need the finite size corrections 
for getting the results at the thermodynamic limit ($L\rightarrow \infty$). In this subsection we shall discuss the theory of finite 
size scaling of the dynamical exponent $z$ and the persistence exponent $\theta$ for the zero temperature quenching dynamics.

It is already discussed that domain size $D$ increases in time $t$ as $D(t)\sim t^{1/z}$. Now as magnetization $m\sim \sqrt{D}$, so 
magnetization grows with time as 
\[
 m(t) \sim t^{1/2z}.
\]
Let us consider a Ising spin system of $d$-dimensional geometry of linear size $L$ with nearest neighbour interaction. 
The system of spins evolve in time following the Glauber dynamics, lowering the total energy of the configuration in the process.
the persistence probability shows a power law form in time, $P(t) \sim t^{-\theta}$, as long as $t << t^*$. For $t>> t^*$ , 
the domain cannot grow any further because of the finite system size and persistence probability stops decaying, attaining a limiting
value 
\[
 P(\infty,L) \sim L^{-z \theta}.
\]
From the above behaviour of the persistence probability one can write down the dynamical scaling relation \cite{puru}
\begin{equation}
P(t,L) \sim t^{- \theta}f(L/t^{1/z}),
\label{fs}
\end{equation}
where the scaling function $f(x)  \sim x^{-\alpha}$ with $\alpha = z\theta$ for  $x <<1$.  For large $x$, $f(x)$ is a constant.
The equation \ref{fs} can also be written as 
\begin{equation}
 P(t,L) \sim L^{-\alpha}f(t/L^z)
\label{fs1}
\end{equation}
with $\alpha = z\theta$. The scaling function $f(x)  \sim x^{-\theta}$ for $x<<1$ and $f(x)$ is a constant for large $x$.
Similar finite size scaling ideas also have been used in the context of global persistence exponent for nonequilibrium critical
dynamics \cite{global}. 

The spatial correlation among the persistent sites can be quantified by the two point correlation function $C(r, t)$ defined as 
the probability that site $(x + r)$ is persistent, given that the site $x$ is persistent (averaged over $x$).
For a $d$-dimensional system, $C(r, t)$ satisfies the normalization condition 
\begin{equation}
 \int_0^L C(r, t)d^dr = L^d P (t, L).
\label{int}
\end{equation}
To calculate the integral we have to put the form of $ P(t, L)$ (equation \ref{fs1}) in equation \ref{int}. The calculation and argument 
ultimately leads to the dynamical scaling form for $ C(r, t)$ : 
\begin{equation}
  C(r, t) =r^{-\alpha}g(t/r^z),
\end{equation}
with $\alpha = z\theta$. The nature of the function $g(x)$ is same as that of the function $f(x)$. For large $t$ 
(as $t\rightarrow \infty$) 
\[
 C(r,t) \sim r^{-\alpha}.
\]
So the exponent $\alpha$ gives the spatial correlations  of the persistent spins for the dynamics. This scaling description was 
introduced in the context of 
$A + A \rightarrow \emptyset$ model \cite{puru1}. It has been shown that the exponent $\alpha $ is related to the fractal 
dimension $d_f=d-z\theta$ of the fractal formed by the persistent spins \cite{puru}. The entire picture stated above is true as 
long as 
\[
 \frac{z\theta}{d} <1
\]
For $z\theta>d$,  persistence probability will decay to zero for any lattice size $L$.

The finite size size scaling form of the correlation functions for quenches to final temperatures $T=0$ and $T=T_c$ for nearest 
neighbour Ising model in two dimensions has been proposed and discussed in \cite{scaling}.

\subsection{Known results}
Glauber found the first exact solution for the nearest neighbour kinetic Ising chain and it is proved that the dynamical exponent 
$z = 2$ for the simplest Glauber dynamics of the Ising chain \cite{glauber}. The dynamical exponent $z$ is equal to $2$ also in 
the higher dimensions \cite{priv}. 

There have been many attempts in recent years to determine the persistence exponent $\theta$ analytically for various systems and 
processes \cite{satya}. Persistence exponents belong to a new class of exponents and it cannot be derived, in general, from other 
static and dynamic exponents. Exact expression for $\theta$ is known only in one-dimensional Potts model for any Potts state 
$q$ \cite{ptsper}:
\begin{equation}
 \theta(q) = -\frac{1}{8}+\frac{2}{\pi^2}[\cos^{-1}(\frac{2-q}{\sqrt 2 q})]^2
\label{potsper}
\end{equation}
Putting $q=2$ in equation \ref{potsper} we get $\theta(2)=3/8$, the persistence exponent for one-dimensional nearest neighbour 
Ising model. Exact value of the global persistence exponent for the nearest-neighbour Ising chain is also known 
($\theta_G = 0.25$) \cite{global}.

For the higher dimensions the persistence exponent is determined numerically and by some approximation methods.
The persistence and dynamical exponents for one, two and three dimensions, for the nearest neighbour Ising model are listed below 
in the following table : 

\begin{center}
\begin{tabular}{|c|c|c|}
\hline
Ising model & $\theta$ & z \\
\hline
1-D & 3/8 & 2 \\
 \hline
2-D & 0.22 & 2\\
\hline
3-D & 0.16 & 2 \\
\hline
\end{tabular}\\
\end{center}

Further in the next chapters we shall compare these values of the exponents ($\theta$ and $z$) with the newly obtained values, 
for different types of Ising spin systems.

\chapter{Zero Temperature  Dynamics of Ising models with competating interactions}
\label{Cannni}

\section{Introduction}

In  one dimension, a zero temperature quench  of the  Ising model as mentioned earlier ultimately leads to the equilibrium
configuration, i.e., all spins point up (or down).
 The average domain size $D$ increases in time $t$ as $D(t)\sim t^{1/z}$,
where $z$ is the dynamical exponent associated with the growth \cite{3gunton,3bray}.
As the system coarsens, the magnetization also
grows in time as $m(t)\sim t^{1/2z}$ (discussed in chapter \ref{Cdyn}).
In two or higher dimensions, however, the system does not
always reach equilibrium \cite{3Krap_Redner} although these scaling relations
still hold good.

  Apart from the domain growth  phenomenon \cite{3gunton,3bray}, another important dynamical
  behavior commonly studied is persistence. In Ising model, in a zero temperature quench, persistence 
is simply the probability that a spin has not flipped till time $t$ and
is given
by $P(t)\sim t^{-\theta}$. $\theta$ is called the persistence exponent and
is unrelated to any other known static or dynamic exponents \cite{3Krap_Redner,3satya1,3derrida,3stauffer,3krap1}
 (discussed in chapter \ref{Cdyn}). 

Drastic changes in the dynamical behaviour of the Ising model
in presence of a competing next nearest neighbor interaction have been observed 
earlier \cite{3redner,3sdg_ps,3barma}. 
The one dimensional ANNNI (Axial next nearest neighbour Ising) model with $L$ spins is described by the Hamiltonian
\begin{equation}
H = -J\sum_{i=1}^L (S_iS_{i+1} - \kappa S_i S_{i+2}).
\end{equation}
Here it  was found
that for $\kappa < 1$,  under a zero temperature quench with
single spin flip Glauber dynamics,   the system does not reach  its
true ground state. (The ground state is ferromagnetic for
$\kappa < 0.5$, antiphase for $\kappa> 0.5$, and highly degenerate at $\kappa=0.5$ \cite{3selke}). On the contrary, after an initial short time,
domain walls become fixed in number but remain mobile 
at all times thereby making the persistence probability go to
zero in a stretched exponential manner. For $\kappa > 1$ on the other hand, 
although the
system reaches the ground state at long times, the dynamical
exponent and the persistence exponent are both different from 
those of the Ising model with only nearest neighbour interaction \cite{3sdg_ps}.

The above observations and  the additional fact that even in the two dimensional nearest neighbour 
Ising model, 
frozen-in striped states appear in a zero temperature quench \cite{3Krap_Redner}, 
suggest that the   two 
dimensional Ising model 
in presence 
of competing interactions could show novel dynamical behaviour. In the present work, we have introduced such
an interaction (along one direction) 
in the two dimensional Ising model, thus making it equivalent 
to the ANNNI model 
in two dimensions
precisely.  The Hamiltonian for the two dimensional ANNNI model 
on a  $L\times L$ lattice is given by
\begin{equation}
H = -J_0\sum_{i,j=1}^L S_{i,j}S_{i+1,j} - J_1\sum_{i,j=1} [S_{i,j} S_{i,j+1} -\kappa S_{i,j} S_{i,j+2}].
\label{annni2d}
\end{equation} 
Henceforth, we will assume the competing interaction to be
along the $x$ (horizontal) direction, while in the $y$ (vertical) direction,
there is only ferromagnetic interaction. 

\begin{figure}[hbpt]
\centerline{
{\resizebox*{9cm}{!}{\includegraphics{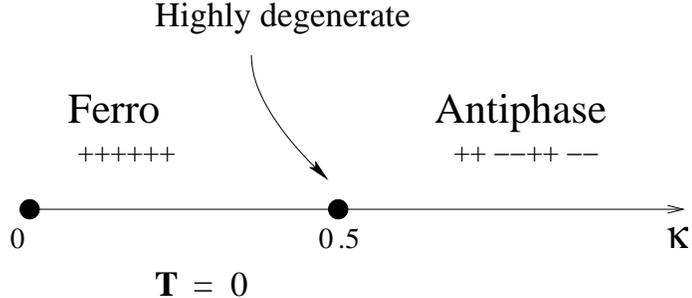}}}
}
\caption{The ground state (temperature $T=0$) 
spin configurations along the $x$ direction are shown 
for different values of  $\kappa$. In the ferromagnetic phase, there is a two fold
degeneracy and in the antiphase the degeneracy is four fold. The ground state is infinitely degenerate at the fully frustrated 
point $\kappa = 0.5$.}  
\label{phase}
\end{figure}

Although the thermal phase diagram of the two dimensional ANNNI model
is not known exactly,  the ground state is known and simple. 
If one calculates the
magnetization along the horizontal direction only, then 
for $\kappa< 0.5$, there is ferromagnetic order and antiphase order for $\kappa > 0.5$. Again, $\kappa=0.5$ is the fully
frustrated point where the ground state is 
highly degenerate. On the other hand, there is always ferromagnetic order along the
vertical direction.
In Fig. \ref{phase}, we have shown the   ground state spin configurations  along the 
$x$ direction for different values of $\kappa$.

In section 3.2, we have given a list of the quantities calculated. 
In section 3.3, we discuss the dynamic behaviour in detail.
In order to compare the results with those of a model without competition,
we have also studied the dynamical features of a two dimensional
Ising model with ferromagnetic next nearest  neighbour interaction, i.e., 
the model given by eq. (\ref{annni2d}) in which $\kappa < 0$. 
These results are also presented in section 3.3.
Discussions and concluding statements are made in the
last section of this chapter.

\section{Quantities calculated}

We have estimated the following quantities in the present work:

\begin{enumerate}
\item Persistence probability $P(t)$: As already mentioned, 
this is the probability that a spin
 does not flip till time $t$. 

In case the persistence probability shows a power law form,
$P(t) \sim t^{-\theta}$,
one can use the finite 
size scaling relation \cite{3puru}
\begin{equation}
P(t,L) \sim t^{- \theta}f(L/t^{1/z}).
\label{fss}
\end{equation}
For finite systems, the 
persistence probability saturates at a value $L^{-\alpha}$ at large times. 
Therefore,  for
 $x <<1$ , $f(x)  \sim x^{-\alpha}$ with $\alpha = z\theta$.  For large $x$,
$f(x)$ is a constant.

\begin{figure}[hbpt]
\centerline{
{\resizebox*{9cm}{!}{\includegraphics{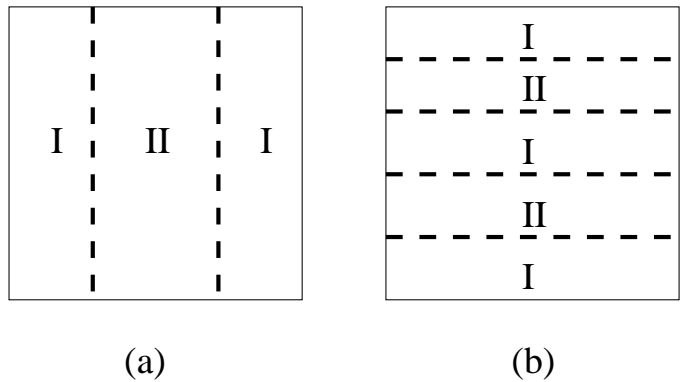}}}
}
\caption{The schematic pictures of  configurations with flat interfaces
 separating domains of type I and II  are shown:
 (a) when the interface lies parallel to $y$ axis, we have
nonzero $f_{D_x}$ ($=2/L$ in this particular case) and (b) with
interfaces parallel to the  $x$ axis we have nonzero $f_{D_y}$ ($=4/L$ here)}
\label{dxyschematic}
\end{figure}

It has been shown that the exponent $\alpha $ is related to the fractal dimension of the
fractal formed by the persistent spins
 \cite{puru}. Here we obtain an estimate 
of $\alpha$ using the above analysis.
%

\item Number of domain walls  $N_D$: Taking a single strip of $L$ spins at a time, one can calculate the number of domain walls for each strip and determine the average.
In the $L \times L$
lattice, we consider the fraction $f_D = N_D/L$ and 
study the  behaviour of $f_D$ as a function of time.
One can  take strips  along both the $x$ and $y$ directions (see Fig. \ref{dxyschematic}
 where the
calculation of $f_D$ in simple cases has been illustrated). As the system
is anisotropic, it is expected that the two measures, $f_{D_x}$ along the $x$ 
direction and $f_{D_{y}}$ along the $y$ direction,  will show different 
dynamical
behaviour in general.   
The domain size $D$  increases as $t^{1/z}$ as already mentioned and it has been observed earlier that the dynamic exponent occurring 
in 
coarsening dynamics is the same as that occurring in the finite size scaling
of $P(t)$  (eq. (\ref{fss})) \cite{3puru}. 
Although we do not calculate the 
domain sizes, the average number of domain walls per strip
is shown to follow a dynamics given by the same exponent $z$, at least
for $\kappa > 1$.

\item Distribution $P(f_D)$ (or $P(N_D)$) of the fraction (or number) 
of domain walls 
at steady state:  this  is also done for both $x$ and $y$ directions.

\item Distribution $P(m)$ of the total magnetization 
at steady state for $\kappa\leq 0 $ only. 

 We have taken  lattices of size $ L\times L$ with $L=40$,$~100$,$~200$ and $300$ 
to study the persistence behaviour and dynamics of the domain walls of the system and averaging over at least 50 configurations for each size have been made.
For estimating the distribution  $N_{D}$ we have averaged over much larger 
number of configurations (typically 4000) and restricted to system sizes 
 $40\times40$, $60\times60$, $80\times80$ and $100\times100$.
Periodic boundary condition has been used in both $x$ and $y$ directions.
 $ J_{0}= J_{1}= 1$ has been used in the numerical simulations.
\end{enumerate}

\section{Detailed dynamical behaviour} 
Before going in to the details of the dynamical behaviour let us
discuss the stability of simple configurations or structures of spins 
 which will help us in appreciating the fact that the dynamical behaviour is strongly
dependent on $\kappa$.

\subsection{Stability of simple structures}

An important question that arises in dynamics is the stability of 
spin configurations - it may happen that configurations which do not correspond to global
minimum of energy still remain stable dynamically.
This has been termed  ``dynamic frustration'' \cite{3pratap} earlier.
A known example is of course a striped state occurring in the
two or higher dimensional Ising models which is stable but not a configuration 
which has minimum energy.

In ANNNI model, the stability of the configurations depend 
very much on the value of $\kappa$. It has been previously analysed
for the one dimensional ANNNI model that $\kappa=1$ is a special point above and below which the dynamical behaviour changes completely
because of the stability of certain  structures in the system.

Let us consider the simple configuration of a single up spin
in a sea of down spins. Obviously,  it will be unstable as 
long as $\kappa < 2$. For $\kappa > 2$, although this spin is
stable, all the neighbouring spins are unstable.
However, for $\kappa < 2$, only the up spin is unstable 
and the dynamics will stop once it flips. 
When $\kappa =2$ the spin may or may not flip,
 i.e.,  the dynamics is stochastic.

Next we consider a domain of two up spins in a sea of down spin.
These two may be oriented either along horizontal or vertical direction.
These spins will be stable for $ \kappa > 1$
 only 
while all the neighouring spins are  unstable.
For $\kappa < 1$, all spins except the up spins are stable. When $\kappa = 1$,
 the dynamics is again stochastic.

\begin{figure}[hbpt]
\centerline{
{\resizebox*{9cm}{!}{\includegraphics{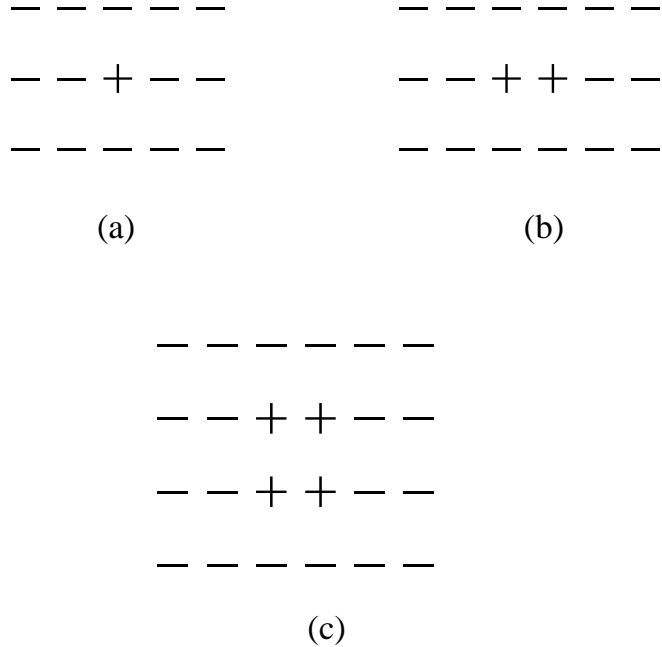}}}
}
\caption{Analysis of stability of simple structures: (a) single up spin
in sea of down spins; here for $\kappa < 2$ all the spins except the
up spin is stable (b) two up spins in a sea of down spins, all spins except the two
up spins are stable for $\kappa < 1$ (c) a two by two structure of up spins - here
all the spins are stable for $\kappa < 1$ while neighbouring spins are not (see text for details).}
\label{stability}
\end{figure}

A two by two structure of up spins in a sea of down spins on the other
hand will be stable for any value of $\kappa > 0$. But the neighbouring spins along the vertical 
direction will be unstable for $\kappa \geq 1$. This
shows that for $\kappa < 1$, one can expect that the 
dynamics will affect the minimum number of spin and therefore
the dynamics will be slowest here. A picture of the structures described above
are shown in Fig \ref{stability}.

One can take more complicated structures but the analysis of these simple ones
is sufficient  to expect that there will be different dynamical
behaviour in the regions $\kappa < 1, \kappa = 1, \kappa > 1,
\kappa =2$ and  $\kappa > 2$. However, we find that 
as far as persistence behaviour is concerned, there are only three regions
with different behaviour:
$\kappa < 1, \kappa = 1$ and $ \kappa > 1$. On the other hand, 
 when the distribution 
of the number  of domain walls in the steady state is considered,
 the three regions 
 $ 1< \kappa < 2, \kappa = 2$ and $ \kappa > 2$ have clearly distinct behaviour.

\subsection{$ 0 < \kappa < 1$}

\begin{figure}[hbpt]
\centering
\resizebox*{9cm}{!}{\includegraphics{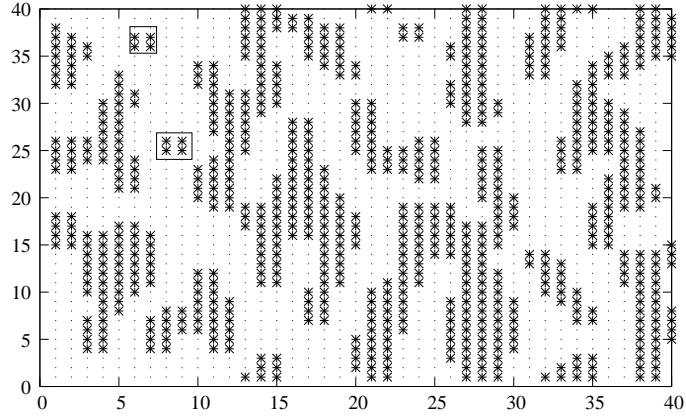}}
 \caption{
Snap shot of a $40 \times 40$ system at time $t= 10$ for $\kappa<1$
A few  simplest quasi frozen structures are highlighted.}
\label{snap1}
\end{figure}

\begin{figure}[hbpt]
\centering
\resizebox*{9cm}{!}{\includegraphics{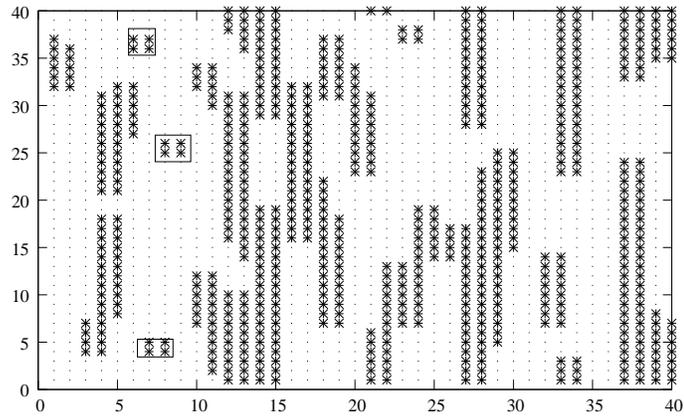}}
\caption{
Same as  Fig. \ref{snap1}  with  $t= 100$.}
\label{snap2}
\end{figure}

\begin{figure}[hbpt]
\centering
\resizebox*{9cm}{!}{\includegraphics{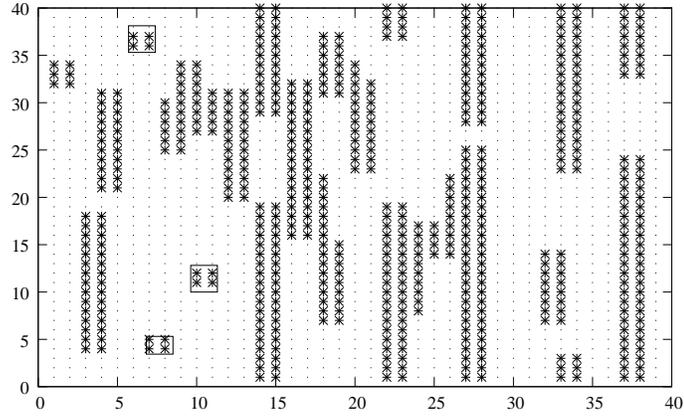}}
\caption{ Same as Fig. \ref{snap1}  with  $t= 500$.
 One of the two by two structures has melted while
another one has formed.
The ladder like structures which have formed are perfectly stable.
}
\label{snap3}
\end{figure}

\begin{figure}[hbpt]
\centering
\resizebox*{9cm}{!}{\includegraphics{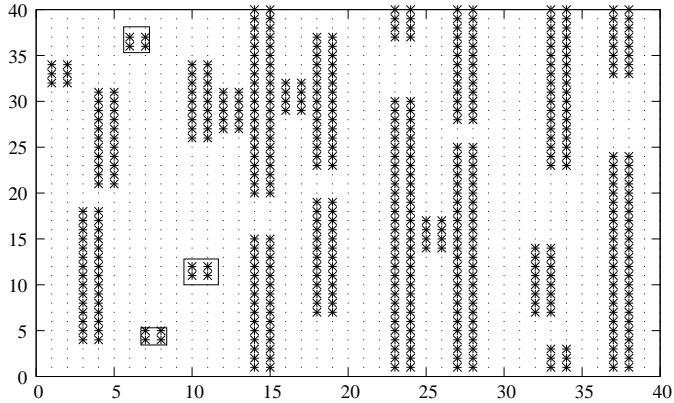}}
\caption{ Same as Fig.  \ref{snap1}  with  $t= 75000$.
This snapshot is taken after a very  long time to show that the 
system has undergone nominal changes compared to the  length of the time interval. The 
whole configuration now consists of ladders and the dynamics 
stops once the system reaches such a state.
}
\label{snap4}
\end{figure}

We find that as in \cite{3sdg_ps}, in the region $0 < \kappa < 1$, the system has identical 
dynamical behaviour for all $\kappa$. Also, like the one dimensional
case, here  the
system does not go to its equilibrium ground state. However, the dynamics
continues for a long time, albeit very slowly for reasons mentioned above. 
In Figs. \ref{snap1} - \ref{snap4}, we show the snapshots of the system at different
times for a typical quench to zero temperature.
As already mentioned, here domains of size one and two will vanish very fast
and certain
structures,  the  smallest of which is a two by two domain 
of up/down
spins in a sea of oppositely oriented spins can survive till 
very long times. These structures we call quasi-frozen as the spins inside 
these structures  (together with
the neighbourhood spins)   are locally
stable;  they can be disturbed only  when the effect of 
a spin flip occurring at a distance propagates to its vicinity which 
usually takes a long time.

The pictures at the later stages also
show that the  system tends to attain a configuration in which the domains have
straight  vertical edges, it can be easily checked that structures with kinks are not stable.
 We find a tendency to form strips of width two (``ladders'')
along the
vertical direction - this is due to the second neighbour interaction -
however,
these strips  do not span the entire lattice in general.
The domain structure
is  obviously not symmetric, e.g., ladders along the horizontal direction
will not form stable structures.
The dynamics stops once the entire lattice is spanned by only
ladders of height ${\cal{N}} \leq L$.

The persistence probability for $\kappa <1$ shows a very slow decay 
with time which can be approximated by $\frac {1}{\log(t)}$ for 
an appreciable range of time. At later times, it approaches
a saturation value in an even slower manner.
The slow dynamics of the system accounts for this slow decay.

\begin{figure}[hbpt]
\centering
\rotatebox{270}{\resizebox*{8cm}{!}{\includegraphics{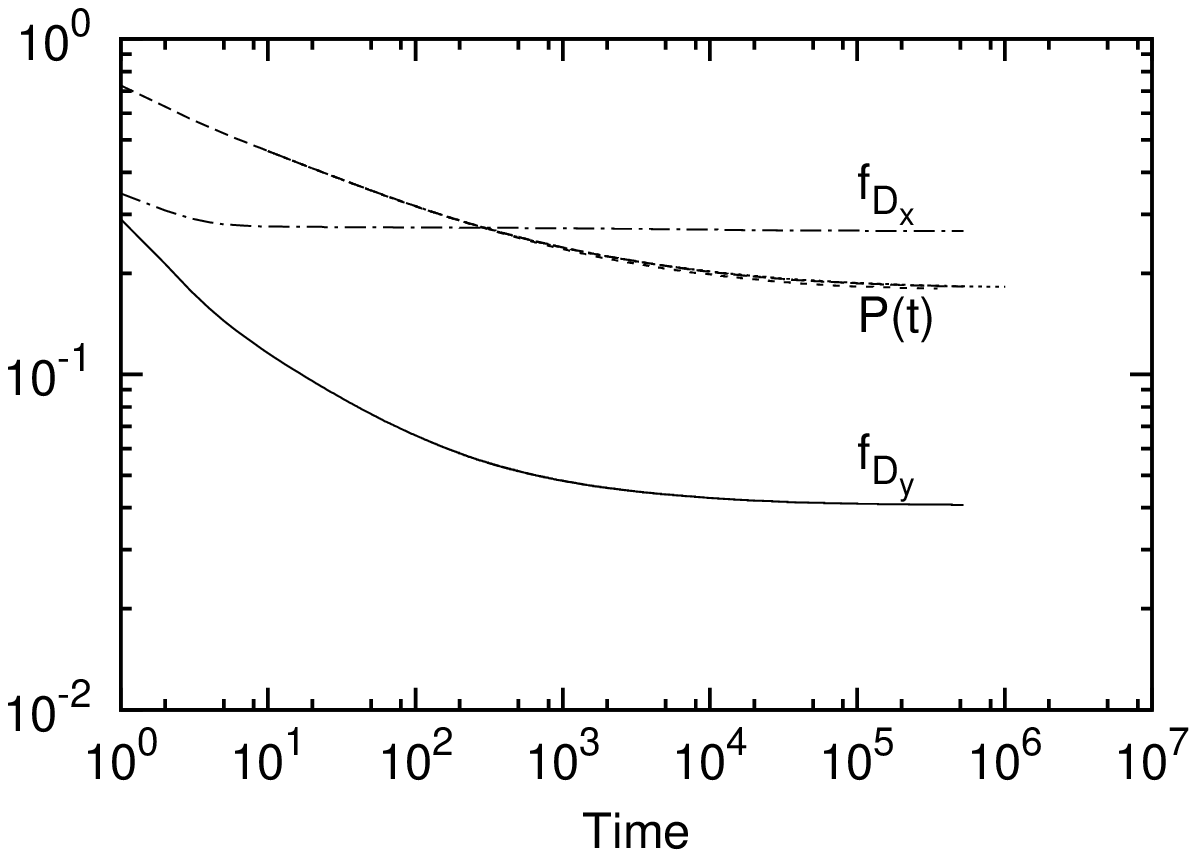}}}
\caption{
Persistence $P(t)$ and average number of domain walls per site, $f_D$ are shown for  $\kappa<1$.
}
\label{dynklt1}
\end{figure}

 The fraction of   domain walls $f_{D_x}$ and $f_{D_y}$ 
along the $x$ direction and $y$ directions
show remarkable difference as  functions of time. While that in the $x$ direction saturates quite fast, in the $y$ direction, 
it shows a gradual 
decay till very long times (see Fig. \ref{dynklt1}). This indicates 
that the dynamics essentially keeps the number of domains unchanged  along
$x$ direction while that in the other direction changes slowly
in time. The behaviour of $f_{D_x}$  is similar to what 
happens in one dimension. In fact,
the average number of domain walls $N_{D_x}$ at large times is also 
very close to that obtained for the ANNNI chain, it is about $0.27L$. 
However, in contrast to the one dimensional case where the domain walls 
remain mobile, here the mobility of the domain walls are impeded  by
the presence of the ferromagnetic interaction along the vertical direction
causing a kind of pinning of the domain walls.

\begin{figure}[hbpt]
\centering
\rotatebox{270}{\resizebox*{8cm}{!}{\includegraphics{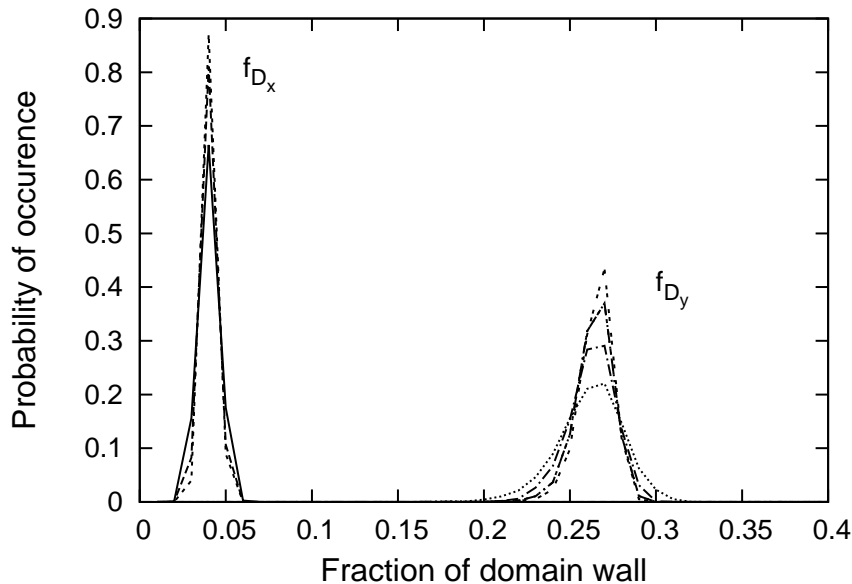}}}
\caption{
Steady state distributions of fraction  of domain walls   at $\kappa<1$ for 
different system sizes. 
The distributions become narrower as the 
system size is increased.
}
\label{distklt1}
\end{figure}

The distribution of the fraction of domain walls in the steady state shown in 
Fig. \ref{distklt1} also reveals some important features. The distribution for $f_{D_x}$ and
$f_{D_y}$ are both quite narrow with the most probable values being
$f_{D_x} \simeq 0.27$ and $f_{D_y} \simeq 0.04$  (these values are very close to the 
average values). 
With the increase in system size, the distributions tend to become 
narrower, indicating that they approach a delta function like
behaviour in the thermodynamic limit. 

\subsection{$\kappa > 1$}

It was already observed that $\kappa=1$ is the value at which the dynamical behaviour
of the ANNNI model changes drastically in one dimension. In two dimensions,
this is also true, however, we find that the additional ferromagnetic 
interaction along the vertical direction is able to affect the dynamics 
to a large extent.
Again, similar to the one dimensional case, we have different dynamical behaviour for 
$\kappa=1$ and $\kappa > 1$. In this subsection we discuss the behaviour for  $\kappa > 1$
while the $\kappa=1$ case is discussed in the next subsection.

\begin{figure}[hbpt]
\centering
{\resizebox*{10cm}{!}{\includegraphics{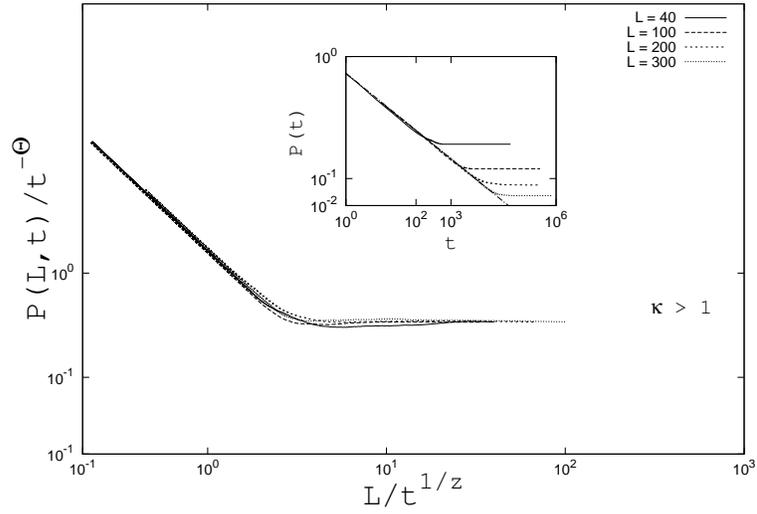}}}
\caption{The collapse of scaled persistence data versus scaled time using $\theta=0.235$ and $z=2.08$ is shown for
different system sizes for $\kappa > 1$.  Inset shows 
the unscaled data.
}
\label{collkgt1}
\end{figure}

The persistence 
probability follows a power law decay with $\theta =0.235\pm0.001$ for all $\kappa> 1$, 
while the finite size scaling analysis made according to (\ref{fss}) 
suggests a $z$ value $2.08 \pm 0.01$.
This is checked for different values of $\kappa$ ($\kappa = 1.3,1.5,2.0,20, 
100$) and 
the values of $\theta$ and $z$ have negligible variations 
with $\kappa$ which do not show any systematics.
Hence  we conclude that the exponents are independent of $\kappa$ for $\kappa > 1$.
A typical behaviour of the raw data as well as the data collapse  is shown in Fig. \ref{collkgt1}.

\begin{figure}[hbpt]
\centering
\rotatebox{270}{\resizebox*{8cm}{!}{\includegraphics{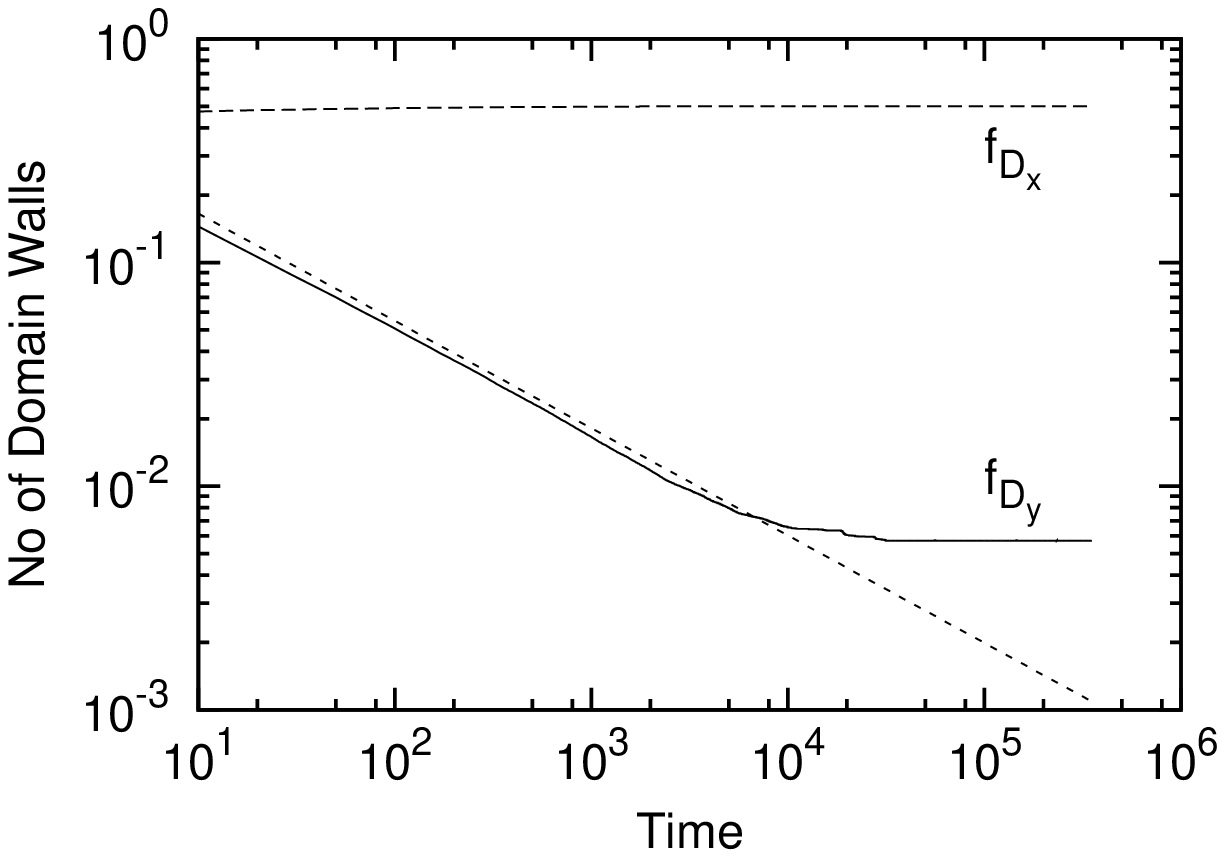}}}
\caption{
Decay of the fraction  of domain walls  with time at $\kappa>1$ are shown  along
 horizontal and vertical directions. The  dashed line has slope equal to 0.48.
}
\label{domainkgt1}
\end{figure}

The dynamics of the average fraction of domain walls along the horizontal
direction, $f_{D_x}$ again shows a fast saturation while that in the $y$ direction has a power law decay with an exponent 
$\simeq 0.48$
(Fig. \ref{domainkgt1}).
 This exponent  is also independent of $\kappa$.
As mentioned in section II, we find that there is a good agreement of the
value of this 
exponent with that of $1/z$ obtained from the finite size scaling behaviour of 
$P(t)$ implying that the average domain size $D$ is inversely proportional $f_{D_y}$. This is quite remarkable, as the fraction of domain walls 
calculated in this manner is not exactly equivalent to the inverse of
domain sizes in a two dimensional lattice; the fact that $f_{D_x}$ remains
constant may be the reason behind the good agreement (essentially the
two dimensional behaviour is getting captured along the dimension
where the number of domain walls show significant change in time). 

Although the persistence and dynamic exponents are $\kappa$ independent,
we find that the distribution of the number of domain walls
has some nontrivial $\kappa$ dependence. 

\begin{figure}[hbpt]
\centering
\rotatebox{270}{\resizebox*{7cm}{!}{\includegraphics{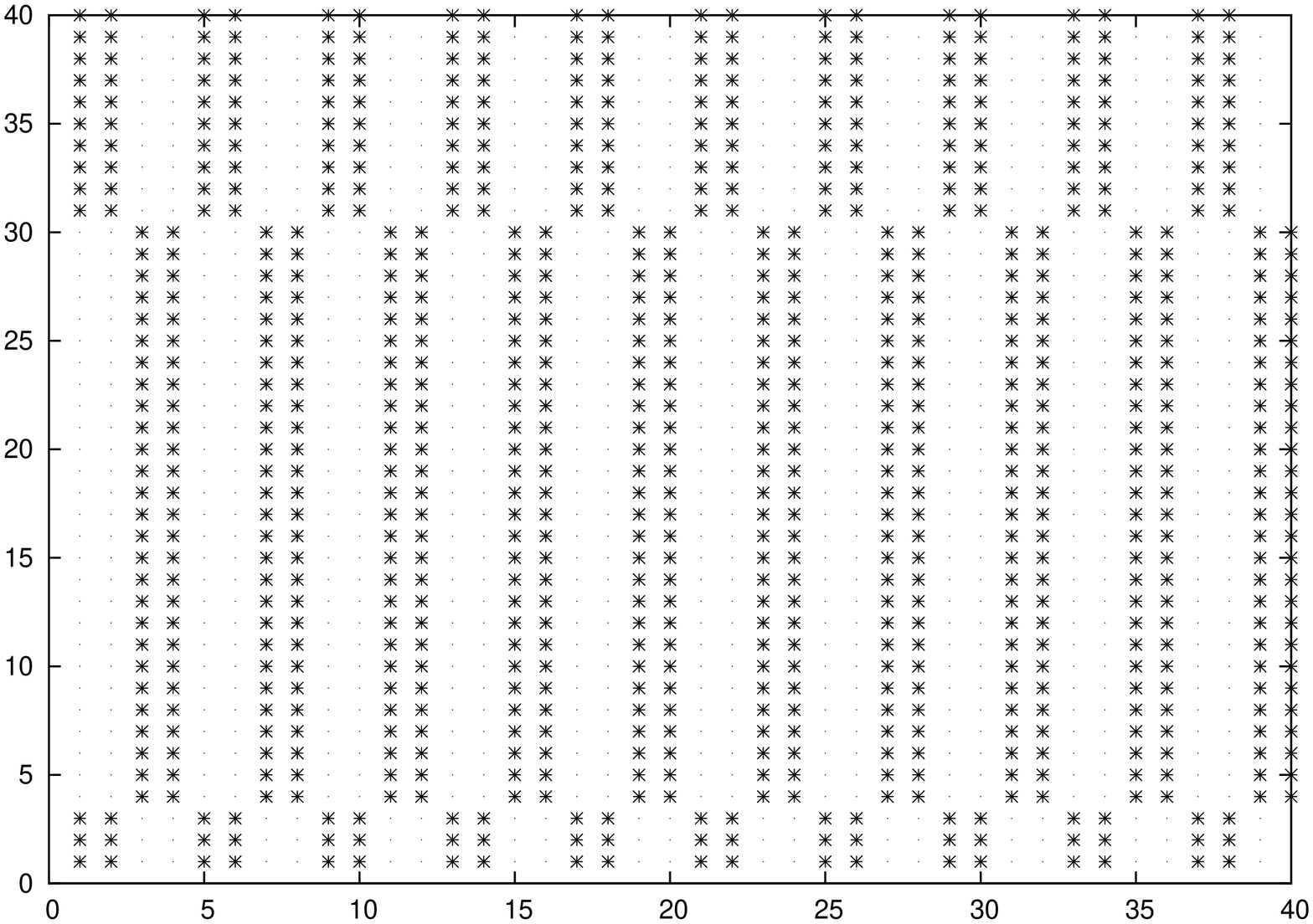}}}
\caption{A typical snapshot of a steady state configuration for $\kappa > 1$ 
with   flat horizontal interfaces 
separating two regions of antiphase ordering (see text).
}
\label{snapkgt1}
\end{figure}

Though the system, for all $\kappa > 1$, evolves to a state with
antiphase order along the horizontal direction, the ferromagnetic order along  vertical chains 
is in some cases separated by one or more  domain walls.
A typical snapshot is shown in  Fig. \ref{snapkgt1} displaying that one essentially gets a striped state
here like in the two dimensional Ising model.

Interfaces which occur parallel to the $y$ axis, separating two
regions of antiphase and 
 keeping the ferromagnetic ordering 
along the vertical  direction intact, are extremely rare, the probability 
vanishing for larger sizes. Quantitatively this means we should get $f_{D_x}=0.5$ at long times
which is confirmed by the data (Fig. \ref{domainkgt1}). Hence in the following our discussions 
on striped state will always imply flat horizontal  interfaces, i.e.,  antiphase ordering along each horizontal 
row but the ordering can be of different types (e.g., a $++--++--\cdots$
type and a $--++--++\cdots$ type, which one can call a `shifted' 
antiphase ordering with respect to the first type). 

It is of interest to investigate whether these striped states survive in the
infinite systems. To study this, we consider the distribution of the number of
domain walls rather than the fraction for different system sizes.
The probability that there are no domain walls, 
or a perfect ferromagnetic phase along the vertical direction,
turns out to be weakly dependent on the system sizes but having
different values for different ranges of values of $\kappa$. 
For
$1< \kappa < 2$, it is $\simeq 0.632$, for $\kappa = 2.0$, it is $\simeq 0.544$
while for any higher value of $\kappa $, this probability is about 0.445. 
Thus it increases for $\kappa$ although not in a continuous manner and
like the two dimensional case, we find that there is indeed a finite
probability to get a striped state. 

\begin{figure}[hbpt]
\centering
\rotatebox{270}{\resizebox*{8cm}{!}{\includegraphics{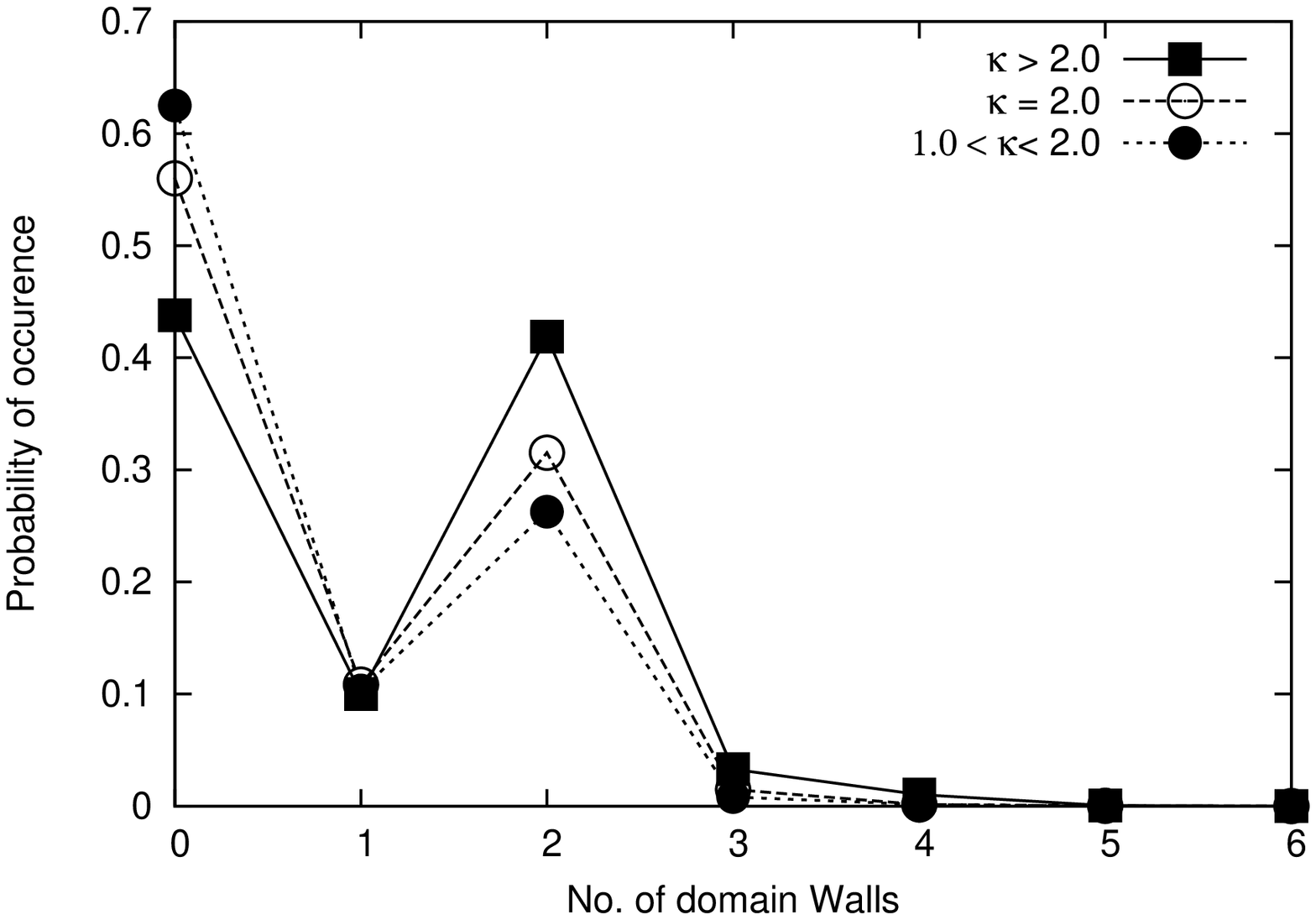}}}
\caption{
Normalized steady state distributions of number  of domain walls   for different  $\kappa>1$ show that
striped states occur with higher probability as $\kappa$ increases.
The lines are guides to the eye.
}
\label{distkgt1}
\end{figure}

While we look at the full distribution of the number of domain walls at steady state 
(Fig. \ref{distkgt1}),
 we find that there are dominant peaks at $N_{D_y}=0$ (corresponding to the unstripped state)
and at $N_{D_y}=2$ (which means there are two interfaces). 
However, we find that the distribution shows that there could be odd values 
of $N_{D_y}$ as well. This is because the antiphase has a four fold degeneracy
and the and a `shifted' ordering can occur in several ways such that 
odd values of $N_{D_y}$ are possible. In any case, the number of interfaces never exceeds 
$N_{D_y}=6$ for the system sizes considered.

\subsection{$\kappa=1$}

\begin{figure}[hbpt]
\centering
{\resizebox*{10cm}{!}{\includegraphics{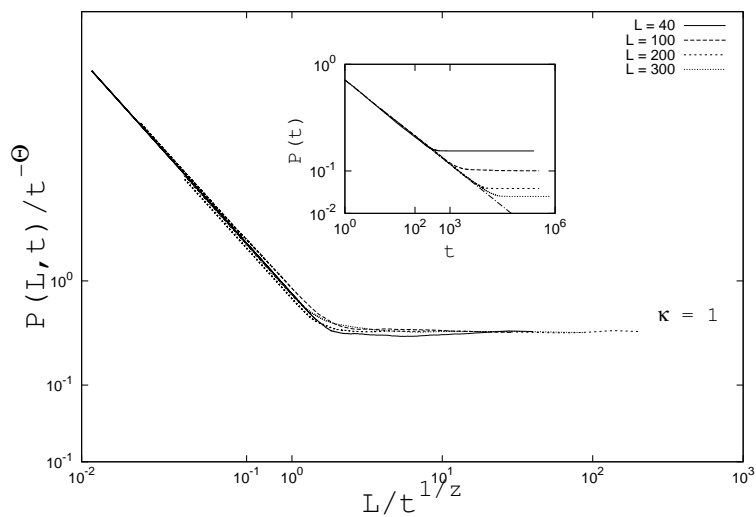}}}
\caption{The collapse of scaled persistence data versus scaled time using $\theta=0.263$ and $z=1.84$ is shown for
different system sizes at $\kappa =1$.  Inset shows
the unscaled data.
}
\label{collk1}
\end{figure}

Here  we find that the persistence 
probability follows a power law decay with $\theta =0.263\pm0.001 $. 
The finite size scaling analysis suggests a $z$ value $1.84\pm0.01$ (Fig. 
\ref{collk1}).

\begin{figure}[hbpt]
\centering
\rotatebox{270}{\resizebox*{7cm}{!}{\includegraphics{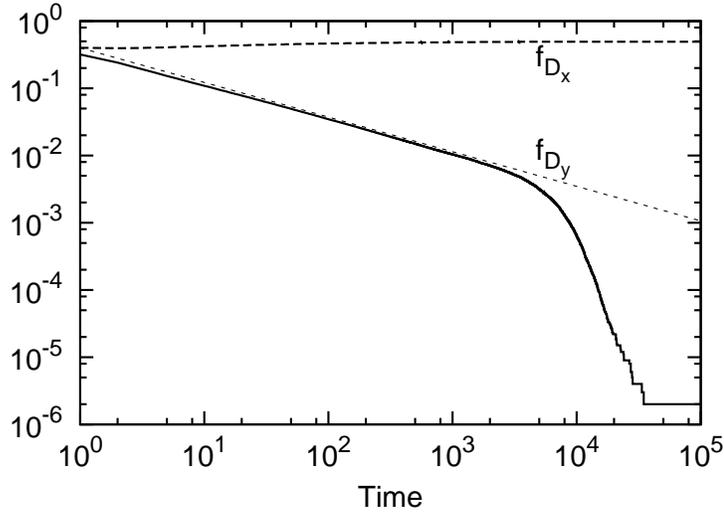}}}
\caption{Decay of the fraction  of domain walls  with time at $\kappa=1$ are shown  along
 horizontal and vertical directions. The dashed line has slope equal to 0.515.
}
\label{domaink1}
\end{figure}

We have again studied the dynamics of $f_{D_x}$ and $f_{D_y}$; the former shows a fast saturation at $0.5$ 
while the latter shows a rapid decay to  zero after an initial power law behaviour with an exponent $\approx 0.515$ 
(Fig. \ref{domaink1}). 
This value, unlike in the case $\kappa>1$, does not show very good agreement with  $1/z$ obtained from
the finite size scaling analysis. We will get back to this point in the next section.

The  results for $f_{D_x}$ and $f_{D_y}$ imply that the system reaches a perfect antiphase configuration 
as there are no interfaces left in the system  with $f_{D_x}=0.5$ and $f_{D_y} =0$ at
later times.

\subsection{$\kappa \leq 0.0$}

In order to make a comparison with the purely ferromagnetic case, we have also studied the 
Hamiltonian (\ref{annni2d}) with negative values of $\kappa$ which
essentially corresponds to the two dimensional Ising model with anisotropic next nearest 
neighbour 
ferromagnetic interaction.

\begin{figure}[hbpt]
\centering
{\resizebox*{10cm}{!}{\includegraphics{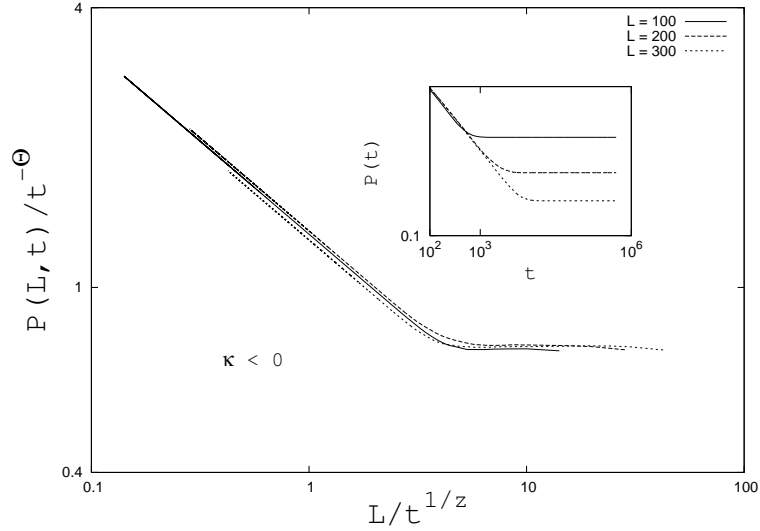}}}
\caption{The collapse of scaled persistence data versus scaled time using $\theta=0.20$ and $z=2.0$ is shown for
different system sizes for $\kappa < -1$.  Inset shows
the unscaled data.
}
\label{collklt0}
\end{figure}

$\kappa=0$ corresponds to the pure two dimensional Ising model for which
the numerically calculated value of $\theta \simeq 0.22$ is verified.
We find a new result when $\kappa$ is allowed to assume negative values, the persistence
exponent $\theta$ has a value $\simeq 0.20$ for $|\kappa| >1$ while for $0< |\kappa| \leq 1$,
the value of $\theta$ has an apparent dependence on $\kappa$, varying between 0.22 to 0.20.
However, it is difficult to numerically confirm the nature of the dependence in such a range and
we have refrained from doing it. At least for $|\kappa| >> 1$, the persistence exponent is definitely
different from that of at $\kappa = 0$.
The growth exponent $z$ however, appears to be constant and $\simeq$ 2.0 
for all values of $\kappa \leq 0$. 
A data collapse for large negative $\kappa$ is shown in Fig. \ref{collklt0} using  $\theta=0.20$ and $z=2.0$.

\begin{figure}[hbpt]
\centering
\rotatebox{270}{\resizebox*{8cm}{!}{\includegraphics{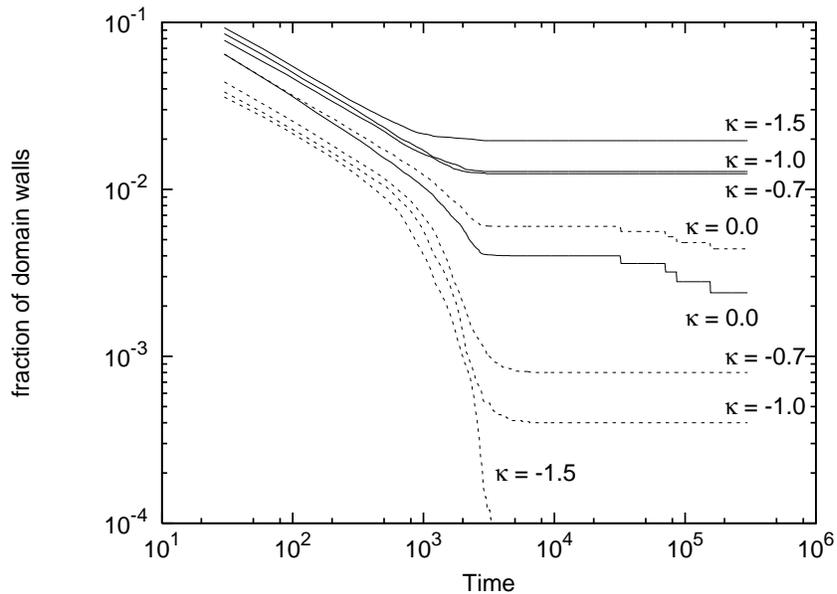}}}
\caption{
Decay of the fraction  of domain walls  with time at $\kappa \leq 0$ are shown  along
 horizontal ($f_{D_x})$, shown by dotted lines)  and vertical ($f_{D_y})$, shown by solid lines)
directions. 
}
\label{domainklt0}
\end{figure}

 The effect of the anisotropy shows up clearly 
in the behaviour of $f_{D_x}$ and $f_{D_y}$ as functions of time
(Fig. \ref{domainklt0}). For $\kappa=0$, they have identical behaviour, both
reaching a finite saturation value showing that there may be interfaces generated in either of the
directions (corresponding to the  striped states which are known to occur here). As the  absolute value of $\kappa$
is increased, $f_{D_{x}}$ shows a  fast decay to zero while  $f_{D_{y}}$ attains a constant value.
The saturation value attained by $f_{D_{y}}$ increases markedly with  $|\kappa|$ while 
 for $f_{D_{x}}$  the decay to zero becomes faster.
 One can conduct a stability analysis for striped states
to show that such states become  unstable when the interfaces are vertical
and $\kappa $ increases beyond $1$, leading to the result $f_{D_x} \to 0$.

 Extracting the $z$ value from the variations of $f_{D_{x}}$ or  $f_{D_{y}}$ is not very simple
here as the quantities do not show smooth power law behaviour over a sufficient interval of time. 

\begin{figure}[hbpt]
\centering
\rotatebox{270}{\resizebox*{8cm}{!}{\includegraphics{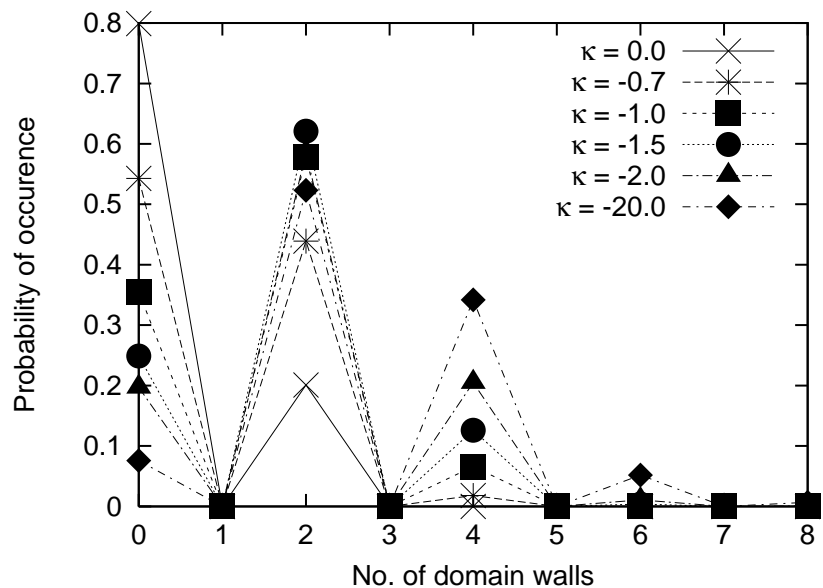}}}
\caption{
Normalized steady state distributions of number  of domain walls   for different  $\kappa \leq 0$ show that
striped states occur with higher probability as $|\kappa|$ increases.
The lines are guides to the eye.
}
\label{distklt0}
\end{figure}

The fact that $f_{D_y}$ and/or  $f_{D_x}$ reach a finite saturation value 
indicates that 
 striped states occur here as well.
The behaviour of  $f_{D_{x}}$ and  $f_{D_{y}}$ suggests that 
in contrast to the isotropic case where
interfaces can appear either horizontally or vertically, 
here
the interfaces appear dominantly along the $x$ 
direction  as $\kappa$ is increased. 
Thus the  normalized distribution of the number of domain walls along $y$ is shown in Fig. \ref{distklt0}. 
We find that as $\kappa$ is increased in magnitude, more and more interfaces
appear. However, the number of interfaces is always even consistent with the
fact that interfaces occur between ferromagnetic domains of all up and all down spins.

\begin{figure}[hbpt]
\centering
\rotatebox{270}{\resizebox*{7.5cm}{!}{\includegraphics{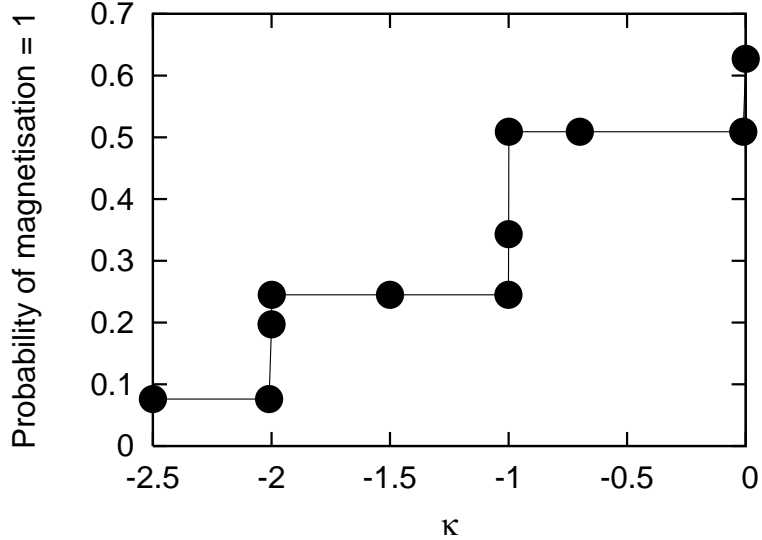}}}
\caption{
Probability that the magnetization takes a steady state value equal to unity is shown against
$\kappa$ when $\kappa \leq 0$.
}
\label{distmag}
\end{figure}

Lastly in this section, we discuss the behaviour of the magnetization
which is the order parameter in a ferromagnetic system. As striped states are formed,
the magnetization will assume values less than unity.  The probability of 
configurations with magnetization equal to unity shows a stepped behaviour, 
with values changing at $|\kappa|=1$ and $2$ and assuming constant values at
 $1< |\kappa|<2$ and above $|\kappa| = 2$ (Fig. \ref{distmag}).

\vspace{-1.2cm}
\section{Discussions and Conclusions}

We have investigated some dynamical features of the ANNNI model in two dimensions 
following a quench to zero temperature. 
We have obtained the results that the dynamics is very much dependent on the 
value of $\kappa$, the ratio  of the antiferromagnetic interaction to the 
ferromagnetic interaction along one direction. This is similar to the 
dynamics of the one dimensional model studied earlier, but here we have more 
intricate features, e.g., that of the occurrence of quasi frozen-in structures
for $\kappa < 1$ where the persistence probability shows a very slow decay
with time. 
Persistence probability is
algebraic for $\kappa\geq 1$, but exactly at $\kappa=1$, the 
exponents $\theta$  and $z$ are different from those at $\kappa > 1$. The exponents for $\kappa > 1$ 
are in fact
 very close to those of the two dimensional
Ising model with nearest neighbour ferromagnetic interaction.
(This was not at all true for the one dimensional ANNNI chain, where the
persistence exponent at $\kappa > 1$ was found to be appreciably different from
that of the one dimensional Ising chain with nearest 
neighbour ferromagnetic interaction.) 
This shows that the ferromagnetic interaction along the vertical direction
is able to negate the effect of the antiferromagnetic interaction 
to a great extent. This is apparently  a  counter intuitive phenomenon,
$\kappa =0$ and $\kappa>1$ having very similar dynamic behaviour while
in the intermediate values, the dynamics is qualitatively and quantitatively
different. At far as dynamics is concerned, the ANNNI model
in two dimensions cannot be therefore treated perturbatively. 

Although the values of $\theta$ and $z$ are individually quite close  
for $\kappa=0$ and $\kappa> 1$, the product $z\theta= \alpha $ are quite different. 
For $\kappa=0$, $\alpha \simeq 0.44$ while for 
$\kappa > 1$, it is  $0.486 \pm 0.002$. This shows that the spatial
correlations of the persistent spins are quite 
different for the two and one can safely say that the dynamical
class for $\kappa=0$ and $\kappa > 1$ are not the same.
$\kappa=1$ is the special point where the
dynamic behaviour changes radically.  
Here there appears to be some ambiguity regarding the value of $z$; estimating $\alpha$ from the finite size 
scaling analysis gives $\alpha \approx 0.484 \pm 0.005$ while using the $z$ value from the domain dynamics, the estimate is
approximately equal to 0.51. However,  the dynamics of the domain sizes
may not be very accurately reflected by the dynamics of $f_{D_y}$ in which case $\alpha \approx 0.48$
is a more reliable result. Thus we find that although the values of $\theta$ and $z$ are quite different
for $\kappa =1$ and $\kappa >1$, the $\alpha$ values are  close.

We would  like to add here  that when there is a power law
decay of a quantity related to the domain dynamics, it is highly unlikely that it will be accompanied by 
an exponent which is different from the growth exponent. Thus, even though we get
slightly different values of $z$ for $\kappa =1$ from the two analyses, it is 
 more likely that this is an artifact of the
numerical simulations.

Another feature present in the 
two
 dimensional Ising model is the finite probability with which it ends up in a striped state.
The same happens for $\kappa>1$, but here the probabilities are quite different 
and also dependent on $\kappa$. We find that there is a significant role 
of the point $\kappa=2$ here as this probability has different values 
at  $\kappa =2$,  $\kappa > 2$ and $\kappa < 2$.

Comparison of the ANNNI dynamics with that of the ferromagnetic anisotropic Ising model shows some interesting features.
In the latter, one gets a new  value of persistence exponent for $\kappa < -1$ while in the former a new value is obtained 
for  $\kappa \geq 1$.The new values    (except for $\kappa=1$) 
are in fact very close to that of the two dimensional Ising model, but simulations
done for identical system sizes averaged over the same number of initial configurations are able
to confirm the difference. 
  The 
qualitative behaviour of the domain dynamics is again strongly $\kappa$ dependent when
$\kappa$ is negative.
Another point to note 
is that  the probability that the system evolves to a pure state is $\kappa$ dependent  in both the ANNNI
model and the Ising  model. 
In both cases in fact, this probability decreases in a step like manner with increasing magnitude of $\kappa$.
We also find the interesting result that while the distribution of the number of domain walls
can have non-zero values at odd values of $N_D$  in the ANNNI model because of the four fold degeneracy of the antiphase,
for the Ising model,   odd values of $N_D$ are
not permissible as the ferromagnetic phase is two fold degenerate.

Finally we comment on the fact that although the dynamical behaviour, as far as domains  are 
concerned, reflects the inherent anisotropy of the system (in both the
ferromagnetic and antiferromagnetic models), the persistence probability is unaffected by it. In order to verify 
this, we estimated $P(t)$ along an isolated   chain of spins along $x$ and $y$ directions separately and 
found that the two estimates gave identical results  for all values of $\kappa$.

In conclusion, it is found that except for the region $0 < |\kappa| <1$, the dynamical
behaviour of the Hamiltonian (\ref{annni2d}) is remarkably similar for negative and positive $\kappa$; 
the persistence and growth exponents get only marginally affected compared to the values   of
 the two
dimensional Ising case ($\kappa=0$) and the domain distributions have similar nature.
However, the region $0 < \kappa < 1$ is extraordinary, where algebraic decay of persistence is absent. There is
dynamic frustration
as the system gets locked in a metastable state consisting of ladder-like domains and the dynamics
is very slow because of the presence of  quasi-frozen structures. 
There is in fact dynamic frustration at other $\kappa$ values also in the sense that except for $\kappa=1$,
the system has a tendency to get locked in a ``striped state''. However, even in that case, the algebraic
decay of the persistence probability is observed.  Thus algebraic decay of 
persistence probability seems to be valid only when the metastable state is a striped state. 
Although there is no dynamic frustration at $\kappa=1$ in the sense that it always evolves to a state with perfect
antiphase structure,
it happens to be a very special point where the persistence exponent and growth exponents are
unique and appreciably different from those of the $\kappa=0$ case. 

In this chapter, the behaviour of the two dimensional ANNNI model under a zero temperature has been discussed; the dynamics
at finite temperature can be in fact quite different. At finite temperatures,  the spin flipping probabilities are stochastic,  
and the  dynamical frustration may be overcome by the
thermal fluctuations. It has been observed earlier \cite{3pratap}  that in a  thermal annealing scheme of the one dimensional
 ANNNI model,  the $\kappa=0.5$ point becomes significant.  A similar effect can occur for the two dimensional case as
well. The definition of persistence being quite different at finite temperatures \cite{3derrida2}, it is also not easy to
guess its behaviour (for either the one or two dimensional model) simply from the results of the  zero temperature quench. Indeed,  the ANNNI model under a finite temperature 
quench  is an open  problem which could  be  addressed in the future.  



\chapter{Quenching Dynamics: Effect of the nature of randomness of complex networks}
\label{Crand}

\section{Introduction}
The dynamical behaviour of Ising models may change drastically when randomness is introduced in the
system.
Randomness can occur in many ways and its effect on dynamics 
can depend on   its precise nature. For example, randomness
in the Ising model can be incorporated  by introducing dilution in the site or bond occupancy in 
regular lattices and consequently the percolation transition plays an important role \cite{4stinch,4jain1}. 
Here the scaling behaviours are completely different from power laws.
One can also consider the interactions to be randomly distributed, 
either  all ferromagnetic type or mixed type (e.g., as in a spin glass) \cite{4stein};  
the system goes to a frozen state following a zero temperature quenching in both cases.
Another way to introduce randomness is to consider a random field 
in which case the scaling behaviour is also completely different from power laws \cite{4fisher}.

Here we consider Ising models on random graphs or networks where  
the nearest neighbour connections exist. In addition,  
the  spins have random  long  range interactions which are quenched in nature.
In general, here, the dynamics, instead of leading the system towards its equilibrium state, 
makes it freeze into a metastable state such that the dynamical quantities
attain  saturation values different from their equilibrium values. 

Moreover,  rather than showing a conventional power law decay 
or growth, the dynamical quantities exhibit completely different behaviour in time.

A point to be noted here is, when 
long range links are introduced, the domains are no longer well-defined as interacting neighbours 
could be  well separated in space.  
This results  in freezing of Ising spins on random graphs as well as on small
world networks \cite{4sven,4haggstrom}.   
The phase ordering dynamics of the Ising model on a Watts-Strogatz network \cite{4ws}, 
after a quench to zero
temperature, produces dynamically frozen configurations, disordered at large length scales \cite{4boyer,4haggstrom}.
Even on small world  networks, the dynamics can  depend on the nature of the randomness;  it was observed that 
while in a sparse network there is freezing, in a densely connected 
network freezing disappears in the thermodynamic limit \cite{4pratap_net}.

In this chapter, we shall present our investigation on the dynamical behaviour of an Ising system
on  two different networks  following a zero temperature quench. In these two networks, both of which are sparsely connected, 
the nature of randomness is subtly different and
we study whether this difference  has any effect   
on the dynamics. Both these networks are embedded in a one dimensional 
lattice and the nearest neighbour connections always exist and the nodes have degree four on an average. 
They differ as in one of the networks, the random long range interactions
have a spatial dependence. It may be mentioned here that quenching dynamics on such 
Euclidean networks has not been considered earlier to the best of our knowledge. 

It is also quite well known that many dynamical social phenomena 
can be appropriately mapped to dynamics of spin systems. 
 At the same time,  social systems have been shown to  behave like complex
networks (having small world and/or scale free features etc.).   
So the present study may be particularly interesting in the context of  
studying social phenomena described by Ising-type models.

In section 4.2.1 we have discussed some basic network properties and network models. In section \ref{4netmodel} we have 
described the two different networks  which we call  random model A (RMA) and random model B (RMB).
In Section 4.3 we have given a list of the quantities calculated.
In section 4.4 and 4.5 we have discussed the detailed dynamical behaviour of Ising spin systems on random model A and 
random model B respectively. The comparison of the results of the quenching dynamics  between the two models are 
discussed in section 4.6. In addition, a qualitative analysis of the quenching dynamics is also presented.
Summary and concluding statements are made in the last section.

\section{Description of the network models}
\label{netmodel}

In this section we shall first discuss some basic network properties and network models as well as the models we have considered 
for our investigation.
\subsection{Classification of network models}
A network is a set of vertices (nodes) connected via edges (links). Networks with directed edges are called directed networks 
and those with undirected edges are undirected networks. One may classify the networks studying three basic network properties 
which are average shortest paths, clustering coefficient and degree distribution.
The total number of connections of a vertex is called its degree (k). 
In a directed network the number of incoming edges of a vertex is called its in-degree $k_i$ and the number of outgoing edges 
is called its out-degree $k_o$. 
\[
 k = k_i +k_o
\]
 The {\it{clustering}} {\it{coefficient}} characterize the density of connections in the environment close to a vertex. 
 The clustering coefficient $C$ of a vertex is the ratio between the total number of the edges connecting its nearest neighbours and 
the total number of all possible edges between all this nearest neighbours.
The {\it{shortest}} {\it{path}} {\it{length }} is the shortest distance $S$ between any two nodes (say, A and B ) is the number of 
edges on the shortest path from A to B through connected nodes. 

Networks are the same as graphs. In graph theory nodes are termed as 
vertices and links as edges. Mathematicians had already classified graphs into 
two groups :
1. Regular graph : In which each vertex is linked to its $k$ nearest neighbours. For regular graph the shortest path distance 
$S \sim l$, where $l$ is the linear dimension of the graph/lattice and clustering coefficient $ C \sim 1$ (finite).

2. Random graph : In which any two vertices have a finite probability to get linked \cite{4ba,4newman}. So in general in a random graph/network each 
vertex get linked to $k$ arbitrary vertices. In this network or graph, both $D$, the diameter of the network (the largest of the 
shortest distances $S$) and $\langle S \rangle$ were found to vary as $\log(N)$. It may be mentioned here that in the random graph 
the degree distribution function $P(k)$ is given by : 
\begin{equation}
           P(k_i=k)=^{N-1}C_k  p^k  (1-p)^{N-k-1} ,
\end{equation}
which is basically Binomial distribution, where $N=$ total number of nodes, $k=$ degree of a node.
For large $N$ the above equation can be replaced by Poisson distribution : 
\begin{equation}
           P(k)=\frac{e^{-pN} (pN)^k}{k!}
               =\frac{e^{-{\langle{k}\rangle}} {\langle{k}\rangle}^k}{k!}
\end{equation}
where average degree of graph ${\langle{k}\rangle}=p(N-1) \simeq pN $
         The significance of ${\langle{k}\rangle}$ lies in the fact of forming cluster in random graphs.

Now we would like to introduce two types of network namely (1) small world network and (2) scale free network.

1. Small world Network : It is basically intermediate between the previous two networks. In simple terms this describe the fact that 
despite huge physical size of real networks, any two nodes are connected by relatively short paths, hence the network is named as 
$` small~~ world'$. These networks are characterized by large clustering coefficient  compared to the corresponding random graph 
and logarithmic dependence of shortest path \cite{4ba,4newman}.

 $ S_{sw}\sim \log (l)$ $\to$ similar to random graph.

$ C_{sw}\sim 1$ $\to$      similar to regular graph.

2. Scale free network : For this type of network, the probability that a node was connected to $k$ other nodes is proportional to 
$k^{-n}$. This means their degree distribution follows a power law for large $k$.
In general but not necessarily, scale free networks have small world properties. However the clustering coefficient of the SF model 
decreases with the network size following approximately a 
power-law. The decay is slower than the decay observed for random graphs (C = ${\langle{k}\rangle}$$N^{-1}$) and is also different 
from the behavior of the small-world models, where $C$ is independent of $N$ \cite{4ba,4newman}.

\subsection{Network models of our interest}
\label{4netmodel}
\begin{figure}[hbpt]
\centering
{\resizebox*{8cm}{!}{\includegraphics{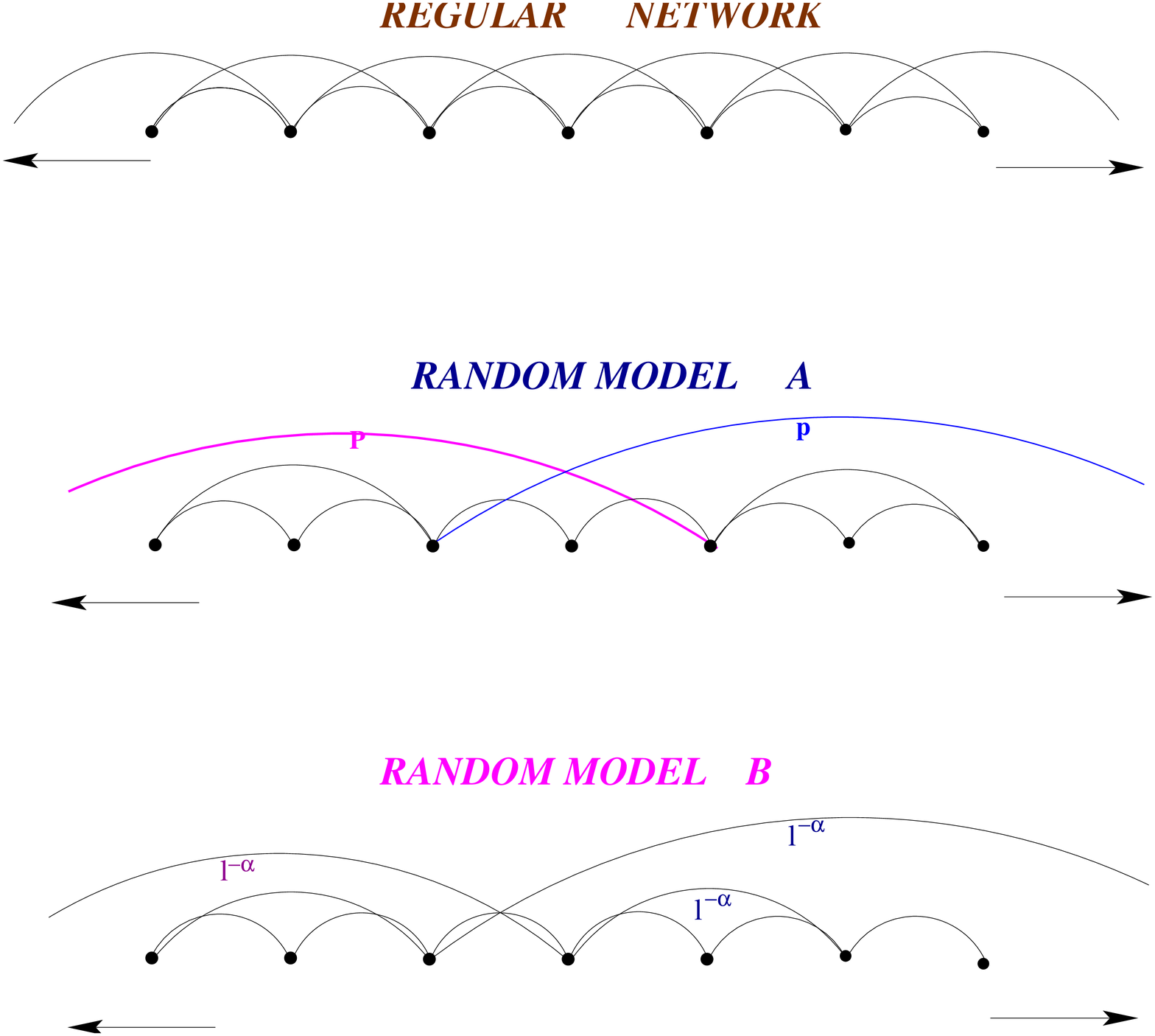}}}
\caption{ (Color online) Schematic diagram for different network models. Average degree is 2K = 4 in each network. 
In the regular network both the first and second nearest neighbours are present. 
In random model A only second neighbours are rewired with probability p.
In random model B first nearest neighbours are always linked while
other nodes are linked with the probability $l^{-\alpha}$ with $ l \geq 2 $.}  
\label{4schematic}
\end{figure}

The two network models under consideration were introduced in reference \cite{4gbs}.
The random model A  (RMA) is in fact very similar to the Watts-Strogatz network \cite{4ws}. Here  initially 
a spin is connected to its four nearest neighbours 
 and then only the second nearest neighbour links are rewired  with probability $p$ (Fig. \ref{4schematic}).
 In the RMB, each spin is connected to its  two  nearest neighbour and then
two extra bonds (on an average) are attached randomly to each spin. 
The extra bonds
are attached to spins located at a distance $l>1$ with probability $P(l) 
\propto l^{-\alpha}$ (Fig. \ref{4schematic}). 
We keep the first neighbours intact in both cases 
to ensure that the networks are connected. Average degree per node is four in both the networks.
The dynamical evolution is considered on the static networks after the process of rewiring/addition of links
is completed (for a review on spatial networks see \cite{4barth}). 

The general form of the Hamiltonian in a one dimensional Ising spin system 
for RMA and RMB can be written as
\begin{equation}
 H = -  \sum_{i<j} J_{ij}S_i S_j,
 \end{equation}
where $S_i = \pm 1$ and $J_{ij}=J$ when sites $i$ and $j$ are connected and  zero otherwise. (We take $J=1$ in this work.)
The ground state (minimum energy state at zero temperature) of the Ising spin system
 in  both RMA and RMB is a state with all spins up or all spins down.

 RMA is a variant of WS model with identical static properties.
It is regular for $p=0$, random for $p=1$, and for any $p > 0$, 
the nature of RMA is small world like \cite{4ws,4gbs}. Euclidean models of RMB type have been studied in a 
few earlier works \cite{4euclid1,4euclid2,euclid3,4gbs}.  
While it is more or less agreed that 
for $\alpha \leq 1$, the network is  random and  for $\alpha > 2$, it
behaves as a regular network, the nature of the network for intermediate values of 
$\alpha$ is not very well understood. 
According to the earlier studies \cite{4euclid1,4euclid2, euclid3,4gbs}, it may either have a small world 
characteristic or behave like a finite dimensional lattice. 
In the present work, we assume that RMB has random nature for $\alpha <1$ and for $1<\alpha<2$, it is small world like
 (at least for the system sizes considered here) following the results of \cite{4gbs}, which is  based on exact numerical evaluation of shortest distance and clustering
coefficients.
This is also because the Euclidean model considered in \cite{4gbs} is exactly
identical to RMB with average degree four, while the average degree  of the Euclidean models 
considered in the other earlier studies is not necessarily equal to four.

In case of RMA, the network is regular and random for only two extreme values $p=0$ and $p=1$ respectively, 
whereas for RMB, the random and regular behaviour of the network are observed over an  extended region.
The regular network corresponding to these two models is the one dimensional Ising spin system
 with nearest neighbour and next nearest neighbour interactions.
We have studied the zero temperature quenching dynamics for this 
 model  also, and the results for the dynamics are identical to that of the 
nearest neighbour Ising spin model.
So it will be interesting to note  how the dynamics is affected by the introduction of randomness 
in the Ising spin system and also how  the difference  in the nature of randomness
of the two models RMA and RMB shows up in the dynamics.

In the simulations, 
single spin flip Glauber dynamics is
used in both cases, the spins are oriented randomly in the initial  state.
We have taken one dimensional lattices of size $L$ with $100 \leq L \leq 1500$ to study 
the dynamics. 
The results are averaged over (a) different initial configurations and
(b) different network configurations. For each system size the number of networks considered is fifty 
and for each network the number of initial configuration is also fifty.
Periodic boundary condition has been used.

\section{Quantities Calculated}
We have estimated the following quantities in the present work.

\begin{enumerate}
\item Magnetization $m(t)$:  For a Ising spin system with regular connections and having only the ferromagnetic interaction, 
the order parameter is usually the magnetization, $m= \frac{|\sum_i S_i|}{L} $. $L $ is the size of the system. Magnetization  
can be considered as the order parameter, even when the connections are random. We have calculated the growth of 
magnetization with time and also the variation of the saturation value of the magnetization, $m_{sat}$, with $p$ and $\alpha$
for RMA and RMB respectively.

\item Persistence probability $P(t)$: As already mentioned, 
this is the probability that a spin
 does not flip till time $t$. 

\item Energy $E(t)$: In these networks, domain wall measurement is not very
significant, as domains are ill-defined. 
The presence of domain walls in regular lattices causes an energy cost \cite{4boyer}.
 So instead of the number of domain walls,
the appropriate measure for disorder is the residual energy per spin
 $\varepsilon = E - E_0 = E + 4$, where
$E_0=-4 $ is the known ground state energy per spin and $E$ is the 
energy of the dynamically evolving  state.
In fact, the magnetization is not  a good measure of the disorder either, since even
when the energy is close to the ground state, magnetization
may be very close to zero (this is also true for the models without randomness). So residual energy measurement is the best way 
to find out whether the system has reached the equilibrium ground state or 
it is stuck in a higher energy nonequilibrium steady state.
We have measured the decay of residual energy $\varepsilon$ with time and the variation of 
its saturation value, $\varepsilon_{sat}$, with $p$ and $\alpha$ for RMA and RMB respectively.

\item Freezing probability: The probability with  which any configuration freezes, i.e.,  
does not reach the ground state (the state with
magnetization $m=1$ or the state with  zero residual energy) is defined as the 
freezing probability. 

\item Saturation time : It is the time taken by the system to reach the steady state. 
It has been observed in some earlier studies \cite{4estz} that it also shows a scaling behaviour with the 
system size with the dynamical exponent $z$. This  in fact provides an alternative method to estimate $z$ 
when straight forward methods fail.

Both magnetization and energy are regarded as dimensionless quantities ($\epsilon$ and $E$ scaled by $J$) in this work.

\end{enumerate}

\section{Detailed results of quenching dynamics on  RMA} 

The results of a zero temperature quench for the Ising model on the RMA are presented in this section.
Starting from a initial random configuration following a quench to zero temperature 
the system cannot reach the ground state (the state with zero residual energy) always
for any $p\neq0$. The magnetization, energy, persistence all attain a saturation value
in time. 
The saturation values of all the quantities show  nonmonotonic behaviour as a function of $p$.

\begin{figure}[hbpt]
\centering
\rotatebox{0}{\resizebox*{11cm}{!}{\includegraphics{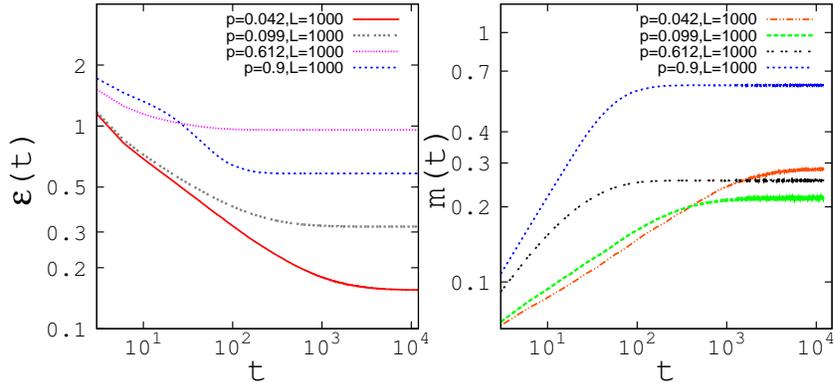}}}
\caption{(Color online) Decay of residual energy per spin and the growth of magnetization with time for RMA for different probabilities.}  
\label{4energy}
\end{figure}

Figure \ref{4energy} shows the decay of residual energy per spin and the growth of magnetization with time for different values of
the rewiring probability.

It is to be noted that the dynamic quantities do not show any obvious power law behaviour beyond a few time steps. 
For small $p$, there is apparently a power law behaviour for a larger  range of time which we believe is the effect of the
$p=0$ point where such a scaling definitely exists.

\begin{figure}[hbpt]
\centering
\rotatebox{270}{\resizebox*{7.5cm}{!}{\includegraphics{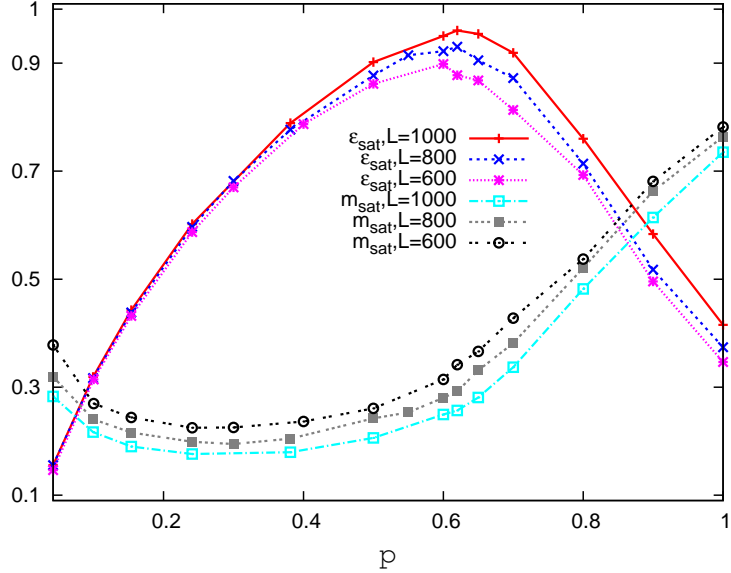}}}
\caption{(Color online) Saturation value of residual energy per spin $\varepsilon_{sat}$ and the saturation value of 
magnetization $m_{sat}$ is plotted with the probability of rewiring $p$
for Random Model A.}  
\label{4S_eng_mag}
\end{figure}

The saturation value of the residual energy per spin $\varepsilon_{sat}$ increases with the rewiring probability $p$ (for small $p$), reaches a maximum 
for an intermediate value of $p$ ($p <1 $) and then decreases again. This implies that the disorder 
of the spin system is maximum for a non trivial value of $p=p_{maxdis}$, which can be termed as the point of maximum disorder. 
The saturation value of magnetization on the other hand decreases for small $p$
and takes its minimum value for another intermediate value of $p$ ($p <1 $), and then increases again (Fig. \ref{4S_eng_mag}).

\begin{figure}[hbpt]
\centering
\rotatebox{0}{\resizebox*{8.5cm}{!}{\includegraphics{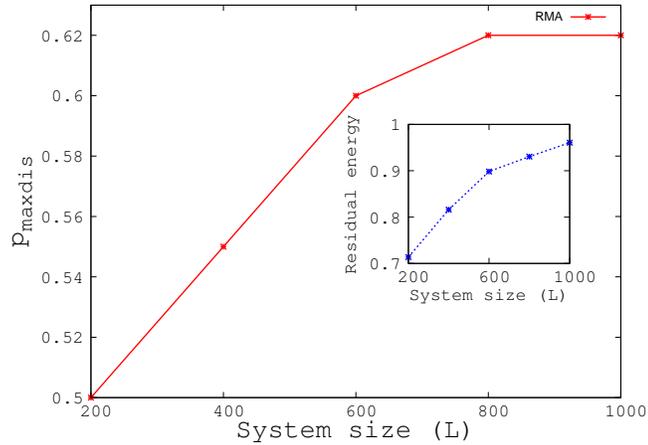}}}
\caption{(Color online) Rewiring probability at the point of maximum disorder is plotted with the system size.
Inset shows the increase of residual energy at the point of maximum disorder with the increment of the system size.}  
\label{4wsensatp}
\end{figure}

$p_{maxdis}$ increases with the system size $L$ for small $L$ 
and then appears to saturate for larger system sizes. The value of the residual energy at $p_{maxdis}$ also increases 
with the system size (Fig. \ref{4wsensatp}). This establishes  the existence of the point of maximum disorder at an 
intermediate value of $p$ ($p \simeq 0.62$) even in the thermodynamic limit.

Magnetization reaches a minimum  at a value of $p$ which is {\it less}  than $p_{maxdis}$. This implies that there exists a
region where both magnetization  and energy increase as $p$ increases. This is also apparent from Fig \ref{4S_eng_mag}.
The physical phenomena responsible for  this intriguing feature is conjectured and discussed  in detail in section 4.6.2.

\begin{figure}[hbpt]
\centering
\rotatebox{0}{\resizebox*{8.5cm}{!}{\includegraphics{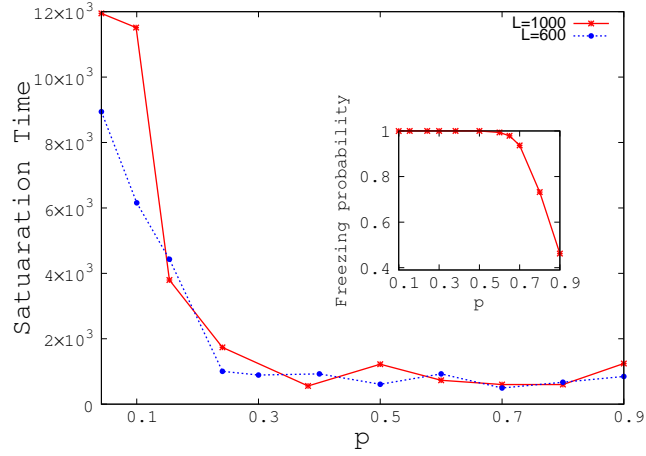}}}
\caption{(Color online) Time of saturation with the probability of rewiring is plotted for two different sizes for Random Model A.
Inset shows the variation of freezing probability with the probability of rewiring for RMA.}  
\label{4frez}
\end{figure}

The saturation time decreases very fast with the rewiring probability $p$ for small $p$ and remains almost constant 
as $p$ increases (Fig. \ref{4frez}). 
It is known that for $p=0$ the saturation time varies as $L^2$, here it appears that for any $p> 0$, there is
no  noticeable size dependence. 

For RMA, the freezing probability is almost unity for small $p$.
However, when the disorder is increased beyond $p\simeq0.5$, the  freezing probability shows a rapid decrease (Fig: \ref{4frez}, inset).
In one dimension, we checked that the freezing probability is $zero$ for the regular network ($p=0$), but here we find that 
even for very small values of $p$, the freezing probability is unity. So there is a   discontinuity in the  freezing probability at $p=0$.
This also supports the fact that any finite $p$ can make the dynamics different from a conventional coarsening process.

An interesting observation  may be made about the behaviour of the saturation value of the residual energy in the region $p < 0.5$.
If one allows $p$ to decrease from $0.5$ to $0$,  the saturation value of the residual energy also   decreases  although  the 
freezing probability is unity in the entire region. This implies that in this range of the parameter,
 although the system does not reach the real ground state in any realization of the 
network (or initial configuration), such that
$\epsilon \neq 0$ in each case, the system has a tendency to approach the    
 the actual ground state monotonically with $p$ for $p<0.5$ (Fig. \ref{4S_eng_mag}).

\begin{figure}[hbpt]
\centering
\rotatebox{0}{\resizebox*{10cm}{!}{\includegraphics{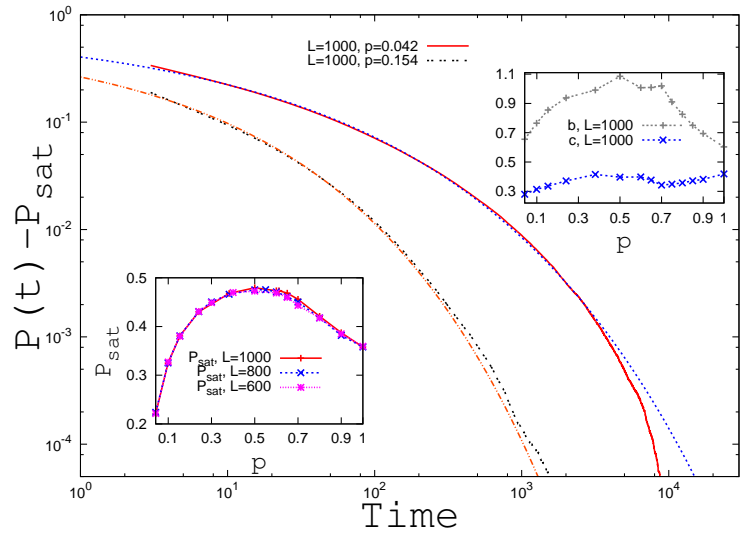}}}
\caption{(Color online) Decay of $P(t)-P_{sat}$ with time $t$ alongwith the stretched exponential function found to fit its form are 
shown. The inset in the bottom left shows the variation of the saturation value of persistence $P_{sat}$ with $p$. The other inset 
on the top right shows the variation of $b$ and $c$ with $p$.}  
\label{4per_ws}
\end{figure}

The persistence probability follows a  stretched exponential behaviour  with time for any non zero $p$,
fitting quite well to the form
\begin{equation}
 P(t) - P_{sat} \simeq a \exp(-bt^c). 
\end{equation}
The saturation value of the persistence is $P_{sat}$, and it does not depend on the system size.
 $P_{sat}$ changes with the rewiring probability $p$ and there also exists an intermediate value of $p$ 
where the value of $P_{sat}$ is maximum. 
$b$ and $c$ vary nonmonotonically with $p$ (Fig. \ref{4per_ws}).

\section{Detailed results of quenching dynamics on  RMB} 

\begin{figure}[hbpt]
\centering
\rotatebox{0}{\resizebox*{11cm}{!}{\includegraphics{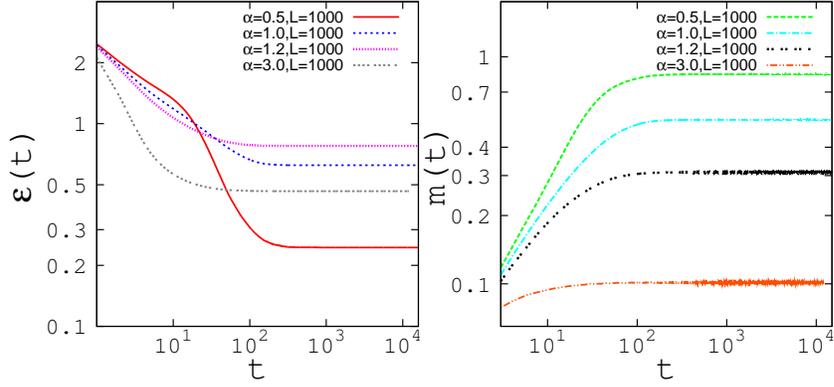}}}
\caption{(Color online) Decay of residual energy per spin and the growth of magnetization with time for RMB for different probabilities.}  
\label{4energy_eu}
\end{figure}

In this section we will present the results of the zero temperature quenching dynamics of Ising model on RMB.
Here also the system does not reach the ground state 
 always
for any finite value of $\alpha$. The magnetization, energy, persistence all attain a saturation value
in time as in RMA. Figure \ref{4energy_eu} shows the decay of residual energy per spin and the growth of magnetization with time 
for different values of $\alpha$. It is to be noted that the dynamical quantities do not show any obvious 
power law behaviour also for RMB.
 

\begin{figure}[hbpt]
\centering
\rotatebox{270}{\resizebox*{7.5cm}{!}{\includegraphics{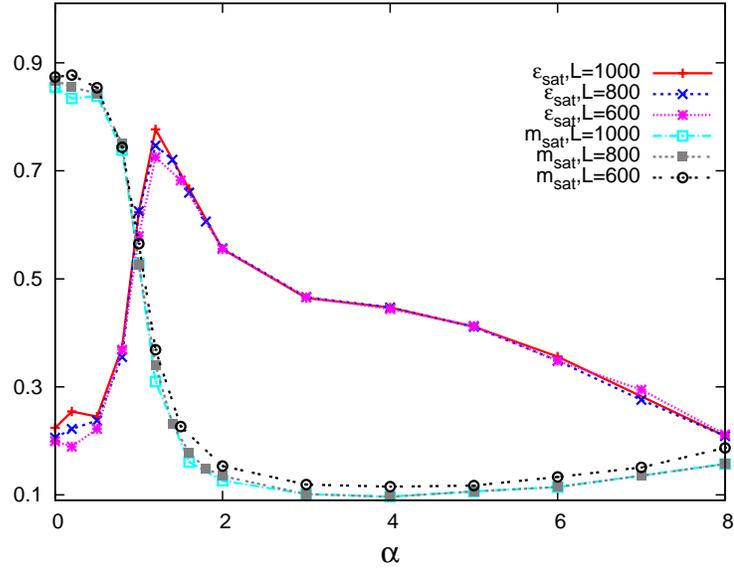}}}
\caption{(Color online) Saturation value of residual energy per spin $\varepsilon_{sat}$ and the saturation value of 
magnetization $m_{sat}$ is plotted with $\alpha$ for Random Model B.}  
\label{4S_eng_mag_eu}
\end{figure}

The saturation values of all the quantities show  nonmonotonic behaviour as a function of $\alpha$.
The saturation value of residual energy per spin $\varepsilon_{sat}$ increases with $\alpha$ for small $\alpha$, reaches a maximum 
for a finite value of $\alpha$ and then decreases again. This implies that for the RMB also, the disorder 
of the spin system is maximum for a finite value of $\alpha$, which is the point of maximum disorder here.
On the other hand, the saturation value of the magnetization  decreases for small $\alpha$
and takes its minimum value for another finite value of $\alpha$ and then slowly increases (Fig. \ref{4S_eng_mag_eu}).

\begin{figure}[hbpt]
\centering
\rotatebox{0}{\resizebox*{8.5cm}{!}{\includegraphics{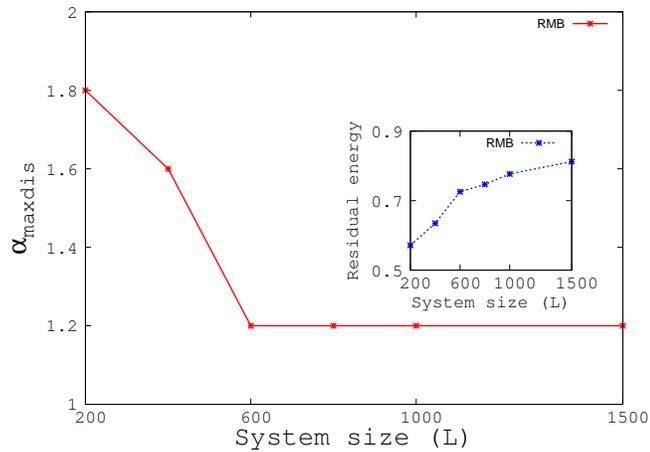}}}
\caption{(Color online) The value of $\alpha$ at the point of maximum disorder is plotted with the system size.
Inset shows the increase of residual energy at the point of maximum disorder with the increment of the system size.}  
\label{4euensatp}
\end{figure}

The value of $\alpha= \alpha_{maxdis}$, at which the  maximum disorder occurs, decreases with the system size $L$ for small $L$ 
and then saturates for larger system sizes. The value of the residual energy at  $\alpha_{maxdis}$ also increases 
with the system size (Fig. \ref{4euensatp}). This establishes  the existence of the point of maximum disorder, for the 
RMB, at a finite value of $\alpha$ ($\alpha \simeq 1.2$) even in the thermodynamic limit.
Similar to the RMA, here  is a region beyond $\alpha = 1.2$ where the energy and the magnetization both decrease,
until the magnetization starts growing again. As already mentioned, this issue is addressed in section 4.6.2.

Saturation time for RMB in the random network in the region $0\leq \alpha <1$ shows too large fluctuations to let one conclude 
whether it is a constant in this region or has a variation with $\alpha$. Beyond $\alpha =1$ and upto $\alpha =3.0$, it is 
almost independent of $\alpha$. For $\alpha > 3 $ 
the saturation time increases with $\alpha$. There is no remarkable finite size effect in the saturation time for the 
RMB for any finite value of $\alpha$. The saturation time 
varies as $L^2$ for a regular lattice corresponding to $\alpha \to \infty$, here it appears that for any finite $\alpha$, however large, there is no  remarkable size dependence. 

\begin{figure}[hbpt]
\centering
\rotatebox{0}{\resizebox*{8.6cm}{!}{\includegraphics{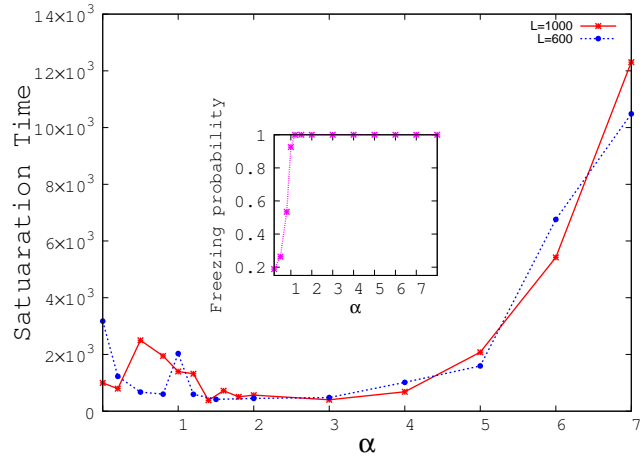}}}
\caption{(Color online) Time of saturation with the value of $\alpha$ is plotted for two different sizes for Random Model B.
Inset shows the variation of freezing probability with $\alpha$ for RMB.}  
\label{4frez_eu}
\end{figure}

The freezing probability is small for $\alpha =0$ ($\simeq 0.2$) and increases rapidly with $\alpha$ for small $\alpha$.
Freezing probability becomes almost unity beyond $\alpha \simeq 1.2$ and remains the same for large $\alpha$. It seems that 
for any finite  $\alpha > 1.2$ freezing probability remains unity and it will be zero only at  $\alpha \rightarrow \infty$ 
(Fig. \ref{4frez_eu}), as in one dimension, the freezing probability is $zero$ for the regular network. 
So for RMB there is a  discontinuity of freezing probability at $\alpha = \infty$ which corresponds to the $p=0$ point of RMA. 

Beyond $\alpha \simeq 1.2$, the energy  decreases with $\alpha$ though the freezing probability remains
unity. 
This implies that although the system definitely reaches a frozen state, it approaches  the 
real ground state monotonically as $\alpha \to \infty $  (Fig. \ref{4S_eng_mag_eu}).

The above results indicate that, though for $\alpha >2 $ the network behaves as a regular one,  dynamically the network is regular 
only at its extreme value $\alpha \rightarrow \infty$.

\begin{figure}[hbpt]
\centering
\rotatebox{0}{\resizebox*{10.2cm}{!}{\includegraphics{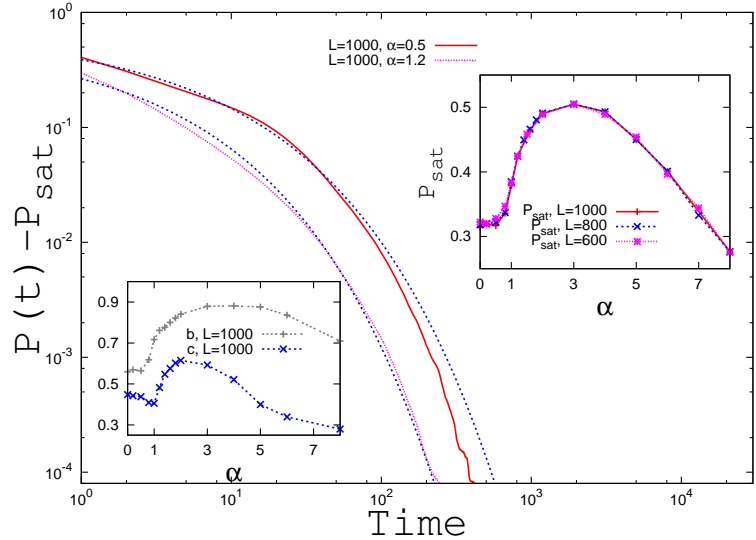}}}
\caption{(Color online) Decay of $P(t)-P_{sat}$ with time $t$ along with   the stretched exponential function found to fit it approximately  are shown. The inset in the top right 
shows the variation of the saturation value of persistence $P_{sat}$ with $\alpha$. The other inset on the bottom left shows 
the variation of $b$ and $c$ with $\alpha$.}  
\label{4per_eu}
\end{figure}

We find that the persistence probability follows roughly a stretched exponential form with time (given by equation (2)) 
for any finite $\alpha$. 
The saturation value of the persistence, $P_{sat}$, does not depend on the system size.
$P_{sat}$ changes with $\alpha$ and there exists an intermediate value of $\alpha$ 
where the value of $P_{sat}$ is maximum. 
For RMB also $b$ and $c$ vary nonmonotonically with $\alpha$ (Fig: \ref{4per_eu}). 

\section{Discussions on the results}

\subsection{Comparison of the results for RMA and RMB} 

In the last two sections the results of a quench at zero temperature for the Ising model on  RMA and RMB have been presented separately.
In this subsection we will compare the results  to understand  how the difference in the nature of randomness 
affects the dynamics of Ising spin system.

The gross features of the results are similar: in both models we have a freezing effect which makes the system
get stuck in a higher energy state compared to the static equilibrium state in which all spins are parallel.
No  power law scaling  behaviour with time is observed in the dynamic quantities in either model. There exists 
a point in the parameter space where the deviation from the static ground state is maximum. The 
behaviour of the saturation times and freezing  probability as functions of the disorder parameters are also
quite similar qualitatively.

The saturation values of magnetization
 and persistence  attain a minimum and maximum value respectively at an intermediate value of the relevant parameters in both models.
The decay of the persistence probability also follows the same functional form in the  entire parameter 
space.  The saturation values of the persistence has no size dependence for both the models.
This indicates that as a whole the dynamics is not much affected due to the change in the nature of randomness of the Ising spin system.

Let us  consider  the parameter values at which the   RMA and RMB are equivalent as a network: 
RMA and RMB  behave as random networks at $p=1$ and  $\alpha =0$ respectively. So one can expect that the saturation values of 
residual energy per spin, magnetization and the numerical value of the saturation time would be same at these values.
However, the numerical values of these quantities are quite different. 
For RMA, at $p=1$ the saturation value of the residual energy per spin $\varepsilon_{sat} \simeq 0.415$  
whereas for RMB at  $\alpha =0$ $\varepsilon_{sat} \simeq 0.224$ for $L=1000$. Similarly we found numerically that 
for RMA the value of saturation magnetization  $m_{sat} \simeq 0.735$ for RMA and $m_{sat} \simeq 0.855$ for RMB for the same system size.
This is because even though the networks are both random here, the connections have a subtle difference.  
For RMA, the number of second nearest neighbour is exactly zero at 
$p=1$ and all the other long range neighbour connections are equally probable. On the other hand, 
for RMB, second nearest neighbours can be still present in 
the network and the probability is same for this  and any other longer range connection. This difference 
in the nature of randomness affects the dynamics of the Ising spin system sufficiently to make the saturation values different. This 
means that the systems are locked at {\it different} nonequilibrium steady states.
For RMB, it is closer to the actual ground state as it is more short ranged in comparison.

The other values at which the two networks are equivalent are  $p=0$ and $\alpha > 2$ where regular network behaviour 
is found as far as the network properties are concerned. Interestingly, the behaviour of RMB even 
when $\alpha$ is finite and greater than 2, is not quite like the dynamics of a regular one dimensional lattice with  
nearest and next nearest neighbour links only. In fact, the point at which the magnetization becomes minimum is well inside 
the region $\alpha > 2$ and not within the small world region as in RMA. Actually there is an extended region of regular and 
random network behaviour for the RMB, and as a result, a few more interesting points are
possible to observe  here.  Only at the extreme point $\alpha = \infty$, the 
one dimensional Ising exponents $z = 2.0$ and $\theta = 0.375$ can be recovered as the frozen states continue to exist 
even for finite values of  $\alpha > 2$ for RMB. 
For the regular network with nearest and next nearest neighbour model,
we have checked that there is no freezing at all. So discontinuities of the freezing probabilities occur at $p=0$ 
and $\alpha = \infty$ on RMA and RMA respectively. 

Though the nature of randomness is different for RMA and RMB, for both the models there exists a point of maximum disorder where 
the saturation value of the residual energy per spin attains a maximum value. 
For RMB, maximum disorder of the Ising spin system occurs near the static phase transition point (small 
world to random phase) whereas for RMA, the point of maximum disorder is well within the small world region.

We try to explain this considering the deviation from the point $p=1$ (for RMA) and $\alpha = 0$ (for RMB). 
Two processes occur simultaneously here:
(a) Number of connections with further neighbours decreases and (b) clustering becomes more probable. 
As a result of these two processes,  freezing occurs. For RMA, the effect is $less$ as there is 
less clustering \cite{4gbs}. But for RMB, the effect is $more$ and spans the entire parameter space $\alpha >1$ and 
therefore the point of maximum disorder of Ising spin system is very close to the random - small world phase transition point
$\alpha =1$.

The question may arise whether this difference prevails when the models are made even more similar.
In RMB, the probability $p_3(\alpha)$ that $l \geq 3$ can be expressed as a function of $\alpha$:  
 \begin{equation}
 p_3(\alpha)= \frac{\sum_{l=3}^{l=L/2} l^{-\alpha}}{\sum_{l=2}^{l=L/2} l^{-\alpha}} .
 \end{equation}
 A further correspondence between the two networks can be established by imposing  
  $p=p_3(\alpha)$,  which makes the number of second neighbour links 
 in RMA and RMB also same  (but the rest of the extra links are connected differently). 

Using equation (3) we can obtain the value of $p$ corresponding to a given  value of $\alpha$ and vice versa.
But it is immediately seen that the two networks are not equivalent even after making them similar upto
the second neighbour connections. For example, for $\alpha = 2.0$, the corresponding value of $p =0.612$ in this 
scheme. But we have already seen that while the point of maximum disorder occurs close to  this value of $p$ in RMA, 
the point of maximum disorder  for RMB is considerably away from  $\alpha = 2.0$.
So the nature of randomness continues to affect the dynamics at least quantitatively.

\subsection{Analysis of some general features of the quenching  phenomena on networks}

We find several interesting features in the quenching phenomena of Ising spin systems on both the networks and  in this
subsection we attempt to provide an understanding of the same.

It is intriguing that the results indicate that the minimum amount of randomness can 
make the system freeze. What happens for small randomness? The interactions are still dominantly 
nearest neighbour type and domains in the conventional sense should grow which will be of both plus and minus signs.
The system will freeze as there will be some stable domain walls due to the few long range  interactions present.
The domains, as the system attains saturation, will be small in number and large in size irrespective of their signs.
As a result, the magnetization attains a small value while the residual energy is still small.

This effect continues for some time till something more interesting happens. Take for example the 
case of quenching on RMA. There is a distinct region  $0.4 < p < 0.6$ where the energy and 
magnetization grow simultaneously, an apparently counterintuitive result. Similar behaviour can be noted for the quenching on RMB in   a certain 
region in its 
parameter space. A problem to analyse the situation for different $p$ (or $\alpha$) values 
is that the final frozen states are not related in any way in principle. This is because the energy landscapes change as $p$
is changed and the initial configurations which undergo evolution are completely uncorrelated. 
In fact, in such a situation, even if the energy landscape is same with a number of local minima, 
different initial configurations will end up in different final nonequilibrium steady states.
 Nevertheless, one can attempt to 
explain this  counterintuitive result assuming that the final states are not largely 
different when $p$ is changed slightly in the following way. This assumption and explanation  are
supported by the actual final states obtained for small system sizes.

\begin{figure}[hbpt]
\centering
\rotatebox{270}{\resizebox*{7.5cm}{!}{\includegraphics{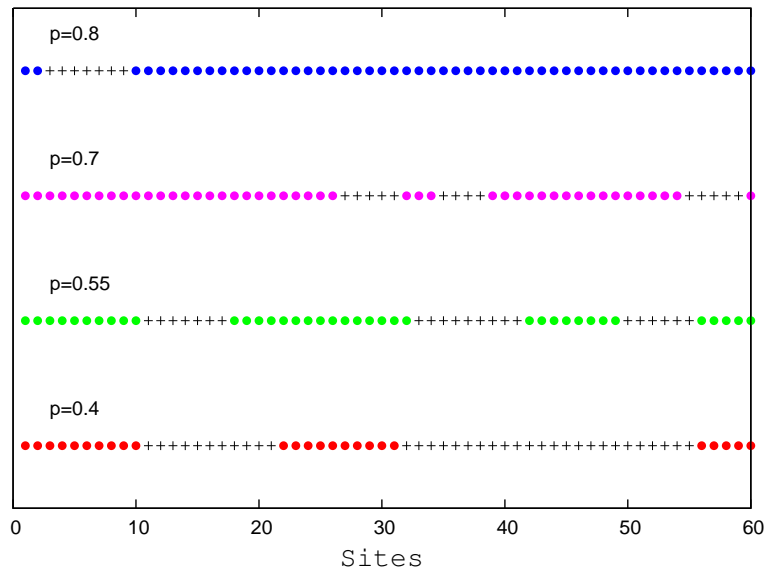}}}
\caption{(Color online) Snap shots of the final spin configurations for different values of the disorder parameter $p$ for quenching on RMA.   
The $+$ and $\bullet$ signs indicate up and down spins respectively. The domains  in the conventional sense are clearly visible.} 
\label{4snap}
\end{figure}

Let us for example consider the RMA and take two values of $p$, $p_2>p_1$, and for which  the magnetization 
and residual energy of the final state corresponding $p_2$ are both larger than those for $p_1$. 
Now this can  be possible due to the fragmentation of a larger domain into several 
domains such that the magnetization increases. This can be demonstrated with a simple example: 
let us imagine a situation where one has only two domains of size $N^+$ (of up spins) and $N^-$ 
(of down spins) for $p_1$ with magnetization equal to $ m_1 = |(N^+ - N^-)|/L$ and assume 
that for $p_2$, the domain with $N^+$ up spins remains same while the domain with $N^-$ down spins gets 
fragmented into three domains  of size $N^-{_1}, N{^+}{_1} $ and $ N^-{_2}$ in the final state. For $p_2$, 
therefore,  the magnetization is $m_2 = |(N^+ + 2N^+{_1} - N^-)|/L$  which is  larger than $m_1$. 
Here in this hypothetical case, we have assumed that $N^+ >N^-$, and $p_2$ is very close to $p_1$. 
One can also assume that the energy increases for $p_2$ as the system is still sufficiently short ranged and 
the new domain walls cause an extra energy compared to the state obtained for $p_1$.
Of course this is an oversimplified picture where we have assumed that the final states for $p_1$ and $p_2$ 
are identical except for the fragmentation of one domain. However, we find that the final configurations obtained for small systems
for different values of $p$ as shown in Fig. \ref{4snap} are consistent with our conjecture. These snapshots 
are representative of the real situation in the sense that they give a  typical picture and are not  just  
rare  cases; we have obtained a similar picture from almost all such configurations generated for small systems.

As $p$ further increases, should the domains get fragmented into even smaller pieces? Answer is no, as
the increasing number of long range interactions again help in the growth of so called domains, {\it of one particular
sign only} such that the magnetization grows and the energy decreases. However, domains of both signs still survive, 
although the sizes are no longer comparable. It can therefore be expected that the region for which both magnetization and energy increase 
as a function of $p$ or $\alpha$ would continue till the short range interactions are dominating and our results
are consistent with this expectation.

\section{Summary and concluding remarks} 

In this chapter, we addressed the question how  the quenching dynamics 
of Ising spins depend on the nature of randomness  of the underlying network by considering two networks in which the
randomness is realized differently. 
The networks are same upto the first neighbour
links and have same average degree per node. While the qualitative features are 
same, there are intricate differences occurring in the behaviour of the saturation values of the dynamical quantities.

Overall, we find some interesting features: the saturation values of the dynamical quantities do not 
have monotonic behaviour as a function of the disorder parameters. Especially, we find that increasing 
randomness does not necessarily make the system get locked in a higher energy state. The dynamics
takes the system to a steady state very fast, and the saturation times are not dependent on the system size.
No scaling behaviour is obtained from the studies either with time or with system size
for any of the dynamic quantities.
The most surprising result is perhaps the existence of a region in the parameter space where 
both the residual energy and the magnetization increase which can be explained phenomenologically.

The Euclidean model, on which the study of the  
quenching of Ising spins is done for the first time to the best of our knowledge,
shows some surprising behaviour both in the random and regular regions. We find that
 decreasing randomness makes the system end up in a higher energy state in the random region while
in the regular region, familiar behaviour of the Ising dynamics with short range interactions
are not obtained; in fact the  probability of freezing  is unity here indicating that
in none of the realization, the system could end up in the static ground state. 
The saturation time also does not show scaling with time.

As already mentioned, the present study is relevant for dynamical 
social phenomena on complex networks. For example, 
the evolution of binary  opinions on a complex network 
(where the initial states are randomly $+1$ and $-1$) 
 is analogous to the dynamical 
study reported in the present chapter. Of course, in case of the opinion dynamics,
the interactions could be more complex compared to the  the simple Ising model.  Our result indicates that the qualitative features 
of the results will  not be much different for different complex networks.

Dynamic frustration \cite{4pratap_ps} is responsible for freezing in many Ising systems where 
there is no frustration in the conventional sense. 
One interesting observation is that the nature of 
dynamic frustration in regular lattices of dimension greater than one and that in 
systems with random interaction (but no frustration) are in general quite different as in the
latter one does not encounter the familiar scaling laws.

\chapter{Opinion formation : A newly proposed dynamics}
\label{Cmodel}

\section{Introduction}

Sociophysics has emerged as one of the important areas of research during recent times. The 
concepts of statistical physics find application to many situations that occur in a social system  
with the assumption that individual free will or feelings do not take crucial role in these situations
\cite{5bkcbook,5Castellano}. One of the major issues that has attracted a lot of attention
is how  opinions evolve in a social system. Starting from random initial opinions, dynamics 
often  leads to a consensus which means a major fraction of the population support a certain 
cause, for example a motion or a candidate in an election etc.

Simulating human behaviour by models effectively implies quantifying the outcome of the behavior 
by suitable variables having continuous or discrete values. Different dynamical rules are 
proposed for the evolution of these variables, depending on how these variables change with time 
following social interactions. Thus, a social system can be treated like a physical system. For example, 
in case of opinion dynamics, if the opinions have only discrete binary values, the 
social system can be regarded as a magnetic system of Ising spins.

In this context, Schelling model \cite{5Schell}, proposed  in 1971, seems to be the very first model 
of opinion dynamics. Since then, a number of models describing the formation of
opinions in a social system have been proposed \cite{5opinion}. 
While on one hand these models attempt an understanding of how a society 
behaves and social viewpoints evolve, on the other hand, these provide rich complex
dynamical physical systems suitable for theoretical studies.  Many of these show a  close connection to familiar models of
statistical physics, e.g., the  Ising and the Potts models.

Dynamics of complex systems has become a subject of extensive research from several aspects. 
For many such systems, e.g., traffic or agent based models, one cannot define a  conventional
Hamiltonian or energy function. The only method by which one can study the 
steady state behaviour of such systems is by looking at the long time dynamics. 
Nonequilibrium dynamics involves the evolution of a system from a completely random initial 
configuration and associated with this evolution are   several phenomena of interest like domain growth
or persistence that have been studied, for example, in spin systems. Since in many sociophysics 
model, one can have variables analogous to  spin variables, these phenomena can be readily studied 
here. An important objective  is  to identify dynamical universality classes by estimating the relevant 
dynamical exponents. 
 
Another point of interest in studying dynamics is that  many systems may have identical equilibrium behaviour 
but behave differently as far as  dynamics is concerned. For example, Ising spin dynamics with or without 
conservation belong to different dynamic universality class although their equilibrium behaviour is identical.

Apart from the dynamical behaviour, different kinds of phase transitions have also been observed
in these models by introducing suitable parameters. One such phase transition can be from a 
homogeneous society where everyone has the same opinion to a heterogeneous  one with mixed 
opinions \cite{5phase-tr}.

Change in the opinion of an individual takes place in different ways in different models. For example
in the Voter model \cite{5vote}, an individual simply follows the opinion of a randomly chosen neighbour 
while in the Sznajd model \cite{5Sznajd}, the opinion of one or more individuals are changed following more 
complicated rules.

\section{Description of the proposed model}

In  a model of opinion dynamics, the key feature is
the interaction of the individuals. Usually, in all the models, it is assumed that
an individual
is influenced by its nearest neighbours.
In this chapter we propose a one dimensional 
model of binary opinion in which the dynamics 
is dependent on the {\it {size}} of the neighbouring domains as well.
Here an individual changes his/her opinion in two situations: 
first when the two  
neighbouring domains have opposite polarity,  and in this case 
the individual simply follows the opinion of
the neighbouring domain with the  larger size.
This case may arise only when the individual is at the boundary of the two 
domains.
An individual also 
changes his/her opinion when both the  neighbouring domains have an opinion 
which opposes his/her original opinion, i.e., the individual is  
sandwiched between two domains of same polarity.
It may be noted that for  the second case, 
the size of the neighbouring domains is irrelevant.
 
\begin{figure}[hbpt]
\centering
{\resizebox*{8cm}{!}{\includegraphics{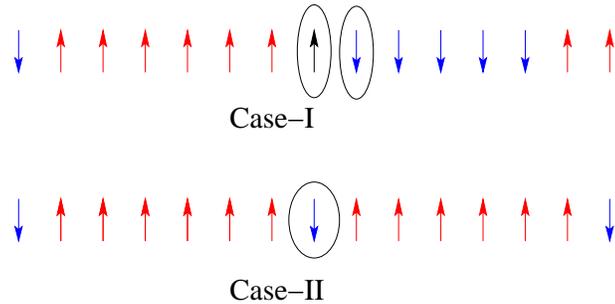}}}
\caption{ Dynamical rules for Model I: in both cases the encircled spins may change state; in case I, 
the boundary spins 
will follow the opinion of the left domain of up spins which  will grow. 
 For case II,  the 
down spin between the two up spins will flip irrespective of the size of the neigbouring domains.}
\label{5model}
\end{figure}

This model, henceforth referred to as Model I, can 
be represented by a system of Ising spins where the up and down states
correspond to the two possible opinions. 
The  two  rules followed in the dynamical evolution  
  in the equivalent spin model 
are  shown schematically 
in Fig. \ref{5model} as case I and  II. 
In the first case the spins representing  individuals  at the boundary   
between two domains 
will choose the opinion of the 
left side domain (as it is  larger in size). For the 
second case the down spin  flanked by  two neighbouring  up spins 
will flip.

 The main idea in Model I is that the size of a domain represents
 a quantity analogous to `social pressure' which is expected 
  to be proportional to the  number of  people 
supporting a cause. An individual, sitting at the domain boundary, is most exposed to
the competition between  opposing pressures and  gives in to the larger one.
This is what happens in case I shown in Fig.\ref{5model}. 
The interaction in case II on the other hand implies that it is difficult to 
stick to   one's opinion if the entire neighbourhood opposes it.  

Defining the dynamics in this way, one immediately notices that case II corresponds to what would
happen for  spins in a nearest neighbour ferromagnetic Ising model (FIM) in which the dynamics
at zero temperature is simply an energy minimisation scheme.
However,   the boundary spin in the FIM
behaves differently in case I; it may or may not flip as the energy remains same.
 In the present model, the dynamics is
deterministic even for the boundary spins
(barring the rare 
instance when the two neighbourhoods have the same size in which case 
the individual changes state with fifty percent probability).

In this
model, the important condition of changing one's opinion is the size of the
neighbouring domains which is not fixed  either in time or space.
This is the unique feature of this model, and to the best of our knowledge
such a condition has not been considered earlier. In the most familiar  
models of opinion dynamics like the Sznajd model \cite{5Sznajd} or the voter model \cite{5vote},
one takes the effect of nearest neighbours within a given radius and
even in the case of models defined on networks \cite{5v-network}, 
the influencing neighbours may be
nonlocal but always fixed in identity.

\section{Model I : Detailed dynamics}
We have done Monte Carlo simulations  to study the dynamical evolution of the 
proposed model from a given initial state.  With a system of $N$  spins representing individuals, 
at each step, 
one spin  is selected at random and 
its state updated. After $N$  such steps, one Monte Carlo time step is 
said to be completed.

If $N_{+}$ is the number of people of a particular opinion (up spin) 
and $N_{-}$ is the number of people of opposite opinion (down spin), 
the order parameter is defined as $m=\vert N_{+}-N_{-} \vert/N $.
 This is 
identical to the (absolute value of) magnetization in the Ising model.

Starting from a random initial configuration, 
the  dynamics in Model I leads to a final state with $m=1$, i.e.,   
a homogeneous state where all individuals have the same opinion. 
It is not difficult to understand this result; in absence of any fluctuation, the dominating neighbourhood (domain) 
simply grows in size ultimately spanning the entire 
system.

\begin{figure}[hbpt]
\centering
{\resizebox*{10cm}{!}{\includegraphics{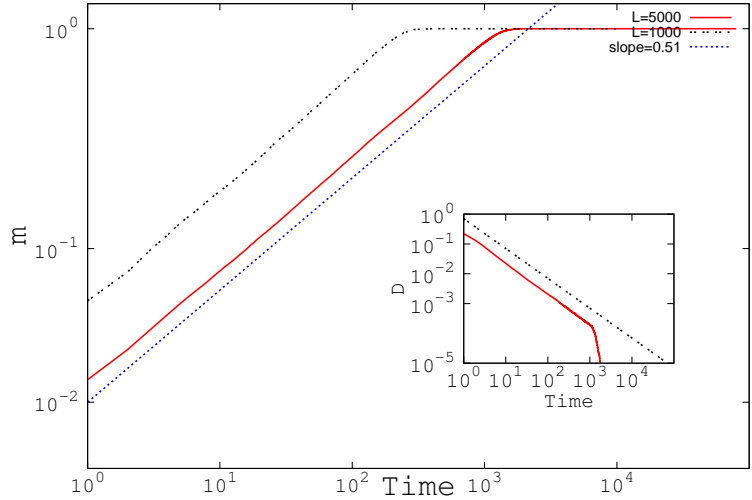}}}
\caption{ (Color online) Growth of order parameter $m$ with time for two different system sizes along with
a straight line (slope 0.51) shown in a log-log plot.
Inset shows the decay of fraction of domain wall $D$ with time.}
\label{5NP_Dw}
\end{figure}

%
We have studied the dynamical behaviour of the fraction of
domain walls $D$ and  the order parameter $m$ as the 
system evolves to the homogeneous state.
We observe that the behaviour of $D(t)$  and $m(t)$
are consistent with the usual  scaling
behaviour found in
coarsening phenomena;
  $D(t) \propto  t^{-1/z}$ with $z = 1.00 \pm 0.01 $ and  $m(t) \propto t^{1/2z}$
with  $z = 0.99 \pm 0.01 $.
These variations are shown in Fig. \ref{5NP_Dw}.

We have also calculated the persistence probability 
that a person has not 
change his/her opinion up to time $t$.
Persistence, which in general is the  probability       
that a fluctuating nonequilibrium field does not change sign upto time $t$, 
shows a power law decay behaviour 
in many physical phenomena, 
i.e., 
$P(t) \propto t^{-\theta}$, 
where $\theta$ is the  persistence exponent.
In that case, one can use the finite 
size scaling relation \cite{5puru,5biswas_sen}
\begin{equation}
P(t,L) \propto t^{- \theta}f(L/t^{1/z}).
\label{5fss}
\end{equation}
For finite systems, the 
persistence probability saturates at a value $ \propto L^{-\alpha}$ at large times. 
Therefore,  for
 $x <<1$ , $f(x)  \propto x^{-\alpha}$ with $\alpha = z\theta$.  For large $x$,
$f(x)$ is a constant (discussed in detail in chapter 2). Thus one can obtain  estimates for both  $z$ and $\theta$ using 
the above scaling form.

\begin{figure}[hbpt]
\centering
{\resizebox*{10cm}{!}{\includegraphics{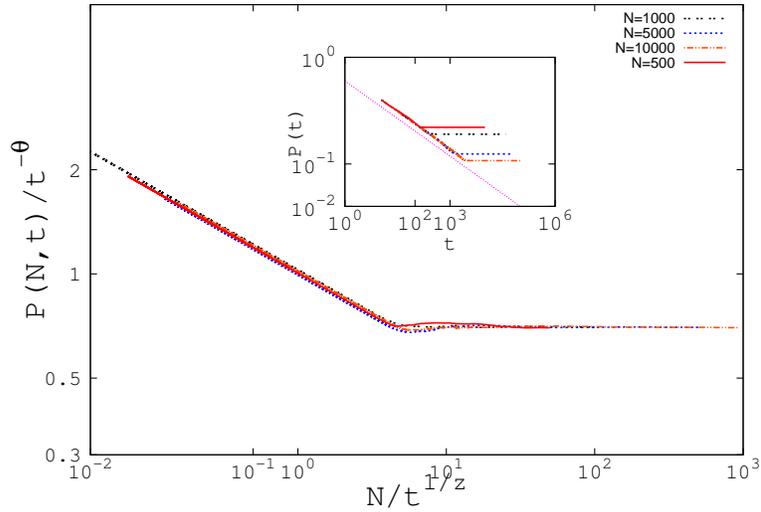}}}
\caption{ The collapse of scaled persistence probability versus scaled time using $\theta=0.235$ and $z=1.04$ is 
shown for different system sizes.  Inset shows the unscaled data.}
\label{5per_noparam}
\end{figure}

In the present model  the  persistence 
probability does show a power law decay with $\theta =0.235\pm0.003$, 
while the finite size scaling analysis made according to (\ref{5fss}) 
suggests a $z$ value $1.04 \pm 0.01$ (Fig.  \ref{5per_noparam}).
Thus we find that the values of $z$ from the three different 
calculations are consistent and conclude that the dynamic exponent 
$z = 1.02 \pm 0.02$.

It is important to note that both the exponents $z$ and $\theta$ 
are novel in the sense 
that they are   quite different from those of  
the one dimensional Ising model \cite{5Derrida} and other 
opinion/voter dynamics models \cite{5stauffer2,5sanchez,5shukla}. 
Specifically in the Ising model,  
$z=2$ and $\theta = 0.375$   
and for the   Sznajd model
the persistence exponent is  equal to that of the Ising model.
  This shows that the present model belongs to an entirely  new dynamical class.

\section{Effect of disorder : Rigidity parameter}
The Model I described so far  has no fluctuation. 
Fluctuations or disorder can be introduced in several 
ways.
We adopt a realistic outlook:   since every individual is not expected to 
succumb to social pressure, we  
 modify  Model I by introducing a parameter $\rho$ called rigidity 
coefficient which denotes the probability that people are completely rigid and never change their opinions.
Such rigid individuals had been considered earlier in \cite{5galam}.
The modified model will be called Model II in which 
there are  $\rho N$ rigid individuals  (chosen randomly at time $t=0$),  who  
retain their initial state 
throughout the time evolution. 
Thus the disorder is quenched in nature. 
The limit $\rho=1$
corresponds to the unrealistic  noninteracting case when no time evolution 
takes place; 
$\rho=1$ is in fact a trivial fixed point.
For other values of $\rho$, the system evolves to a equilibrium state.

\begin{figure}[hbpt]
\centering
{\resizebox*{10cm}{!}{\includegraphics{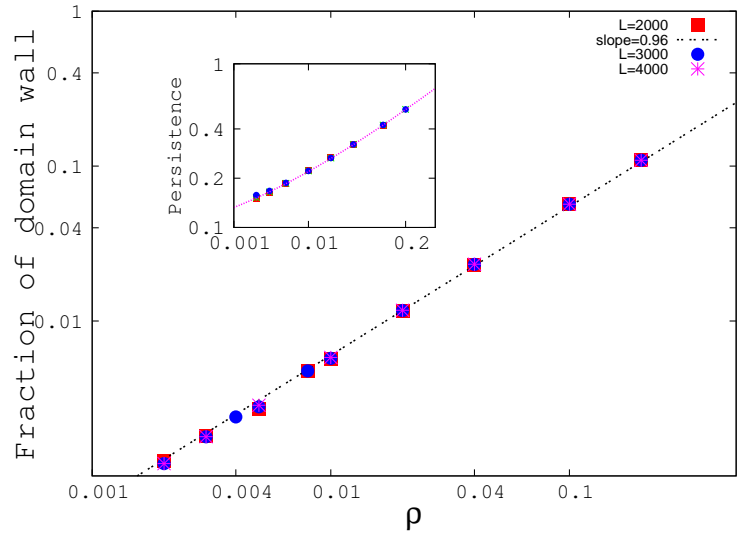}}}
\caption{ Saturation values   of fraction of domain walls  $D_s$  and  persistence probability $P_s$ (shown in inset) 
increase with rigidity coefficient $\rho$ in a power law manner. There is no system size dependence for both the quantities.}
\label{5sat_domain_per}
\end{figure}

\begin{figure}[hbpt]
\centering
{\resizebox*{10cm}{!}{\includegraphics{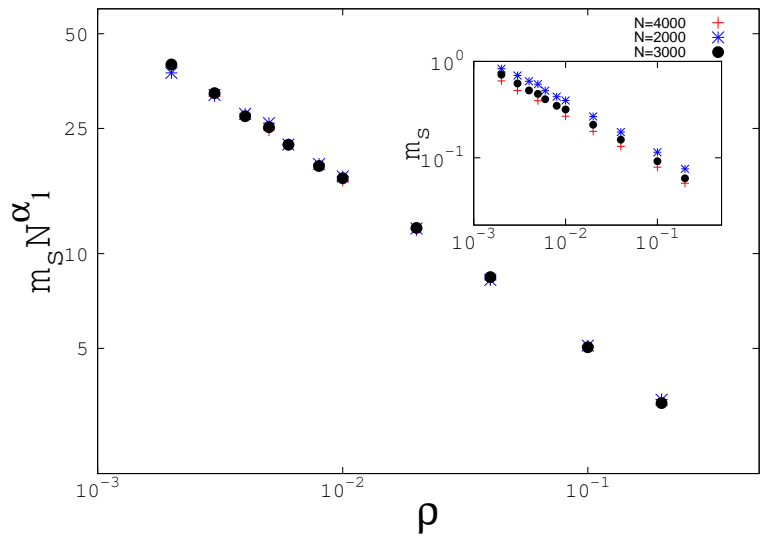}}}
\caption{ Scaled saturation value of $m_s$  decays with the rigidity coefficient $\rho$. 
 nset shows the unscaled data.
}
\label{5sat_mag}
\end{figure}

The time evolution changes drastically in nature with the 
introduction of $\rho$. All the dynamical variables like
order parameter,  fraction of domain wall and persistence
attain a saturation value at a rate which increases with 
$\rho$.  Power law variation with time can only be observed for $\rho < 0.01$ 
with the exponent values same as those for $\rho = 0$.
The saturation or equilibrium values on the other hand show
the following behaviour:
\begin{eqnarray}
&&m_s\propto N^{-\alpha_1}\rho^{-\beta_1}\nonumber\\
&&D_s \propto \rho^{-\beta_2}\nonumber\\
&&P_s  =  a + b\rho^{-\beta_3}  \label{}
\end{eqnarray}
where in the last equation $a$ is a constant  $\simeq 0.06$ independent of $\rho$.
The values of the exponents are $\alpha_1 = 0.500 \pm 0.002,$ $\beta_1 = 0.513 \pm 0.010$, $\beta_2= 0.96 \pm 0.01$ and  $\beta_3 = 0.430 \pm 0.01$.
(Figs \ref{5sat_domain_per} and \ref{5sat_mag}.)
The variation of $m_s$ with $\rho$ is strictly speaking not valid for extremely
small values of $\rho$. However, at such small values of $\rho$, it is 
difficult to obtain the exact form of the variation  numerically. 

It can be naively assumed that the $N\rho$ rigid individuals will dominantly appear at the 
domain boundaries such that  in the first order approximation (for a fixed 
population),
 $D \propto 1/\rho$. This would give $m \propto 1/\sqrt{\rho}$ 
indicating  $\beta_1 = 0.5$ and $\beta_2 =1$.   
The numerically obtained values are in fact quite close
to these estimates.

The results obtained for Model II can be explained in the following way:
with $\rho \neq 0$, 
the domains cannot grow freely  and domains with both
kinds of opinions survive making the equilibrium 
$m_s$ less than unity.
Thus  the society becomes heterogeneous for any $\rho > 0$ when people
do not follow the same opinion any longer. 
The variation of $m_s$ with $N$ shows  
that $m_s \rightarrow 0$ in the thermodynamic limit for $\rho > 0$. 
 Thus not only does the society become  heterogeneous at  the onset
of $\rho$, it goes to a completely disordered state 
analogous to the
paramagnetic state in magnetic systems.  
Thus one may conclude that a phase transition from a ordered state with $m=1$
to a disordered state ($m=0$) takes place for $\rho = 0^+$. It may be
recalled  here that  $m=0$ at the trivial fixed point $\rho=1$ 
and therefore the system flows to the $\rho=1$ fixed point for any 
nonzero value of $\rho$ indicating  that $ \rho=1$ is a stable fixed point.

That the saturation values of the fraction of domain walls do not show system size dependence for 
$ \rho=0^{+}$ further
 supports the fact  that  the phase transition occurs at $\rho=0$.

%
The effect of the parameter $\rho$ is therefore very similar to thermal
fluctuations in the Ising chain, which drives the latter 
 to a disordered state for any non-zero temperature, $\rho=1$ being
comparable to infinite temperature.
However, the role of the rigid individuals
 is  more similar to 
domain walls which are pinned rather than thermal fluctuations.
In fact, the Ising model will have dynamical evolution even at very high temperatures
while in Model II, the dynamical evolution becomes slower with $\rho$, ultimately stopping altogether 
at $\rho=1$. This is reflected in the scaling of the various thermodynamical quantities with $\rho$,
e.g., the order parameter 
shows  a power law scaling 
above the transition point. 

\begin{figure}[hbpt]
\centering
{\resizebox*{8cm}{!}{\includegraphics{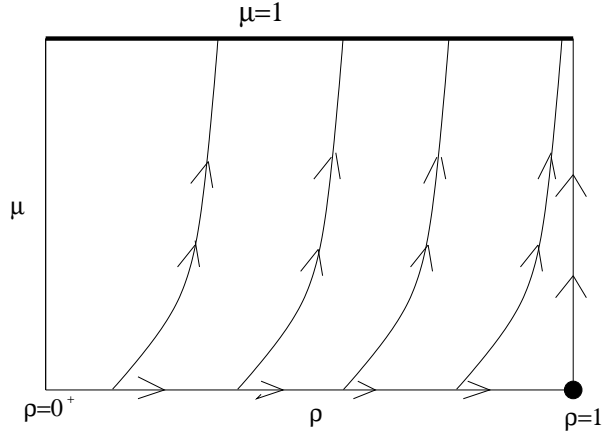}}}
\caption{
The flow lines in the $\rho-\mu$ plane:  
Any non-zero value of $\rho$ with $\mu=0$ drives
the system  to the disordered fixed point $\rho=1$. 
Any nonzero value of $\mu$ drives it to the 
ordered state ($\mu=1$, which is a line of fixed points) 
for all values of $\rho$.}
\label{5flow}
\end{figure}


Since the role of $\rho$ is similar to domain wall pinning, one can
introduce a depinning probability factor $\mu$ which in this system  
represents the probability for  rigid individuals  to become non-rigid
during each Monte Carlo step. $\mu$  relaxes the rigidity
criterion in an annealed manner in the sense 
that the identity of the individuals who become non-rigid is not fixed (in time). If  
$\mu=1$, one gets back  Model I (identical to  Model II with $\rho$=0) whatever
be the value of $\rho$, and therefore $\mu=1$ signifies a line of (Model I) 
fixed points,
where the dynamics leads the system to a homogeneous state.

With the introduction of $\mu$, one has effectively  a lesser fraction  $\rho^\prime$  of
rigid people in the society,  where
\begin{equation}
\rho^\prime = \rho(1-\mu).
\end{equation}
The difference  from Model II is, of course, that 
this effective fraction of   rigid individuals 
is not fixed in identity (over time). Thus when $\rho \neq 0, \mu \neq 0$, we have a system
in which there are both quenched and annealed disorder.
It is observed that for any  nonzero value of $\mu$, 
the system once again evolves to a homogeneous state ($m=1$) for 
all values of $\rho$. Moreover, the dynamic behaviour is also same as Model I 
with the exponent $z$ and $\theta$  having identical 
values.
This shows that the nature of randomness is crucial as 
 one cannot simply 
replace a system with parameters \{$\rho \neq 0, \mu \neq 0$\} 
by one with only quenched randomness
 \{$\rho^\prime \neq 0$, $\mu^\prime=0$\} as in the latter   case
one would end up with a heterogeneous society.
We therefore conclude that the annealed disorder
wins over the quenched disorder;
  $\mu$ effectively drives the system to
 the  $\mu=1$ fixed point  for 
any value of $\rho$. This is shown   schematically in  a flow diagram (Fig. \ref{5flow}).
It is worth remarking that it looks very similar to the flow
diagram of the one dimensional Ising model with nearest neighbour 
interactions in a longitudinal field and finite temperature. 

\section{Mapping of the opinion dynamics model to reaction diffusion  system}

The opinion dynamics model discussed so far are models where the dynamics is described 
in terms of the Ising spins that mimic the binary opinions an individual can have. In this model, 
a spin deep inside a domain does not flip. The dynamics is governed by the flipping of the spins only 
at the domain walls. The dynamics, in this respect, is reminiscent of the zero temperature Glauber 
dynamics of the kinetic Ising model. The motions of  the domain walls can be viewed as the 
motions of the particles $A$ with the reaction $A + A \rightarrow \emptyset$.  This means the 
particles are walkers and when two particles come on top of each other they are annihilated.
The annihilation reaction ensures domain coalescence and coarsening.  
Unlike that in Glauber Ising model, the walkers $A$ corresponding to Model I do not perform 
random walks. These walkers move {\it ballistically towards their nearest neighbours}. 
This bias, as we have seen before, gives rise to a new universality class than that of 
conventional reaction diffusion system \cite{5privman}.

We have also studied $A + A \rightarrow \emptyset$ model with the particles $A$ performing
random walk with a bias $\epsilon$ towards their nearest neighbors. 
We have taken $\epsilon$ as the probability that  a walker walks 
towards its nearest neighbour. Clearly, $\epsilon  = 0.5$ corresponds 
to usual reaction diffusion system with the particles performing random walk. 
On the other hand, $\epsilon=1$ is equivalent to our Model I as has been described above. 
We have studied the dynamics of the reaction diffusion system for different values of $\epsilon$ 
in the range $[0^{+},1.0]$. 

\begin{figure} [hbpt]
\begin{center}
\rotatebox{270}{\resizebox*{8 cm}{!}{\includegraphics{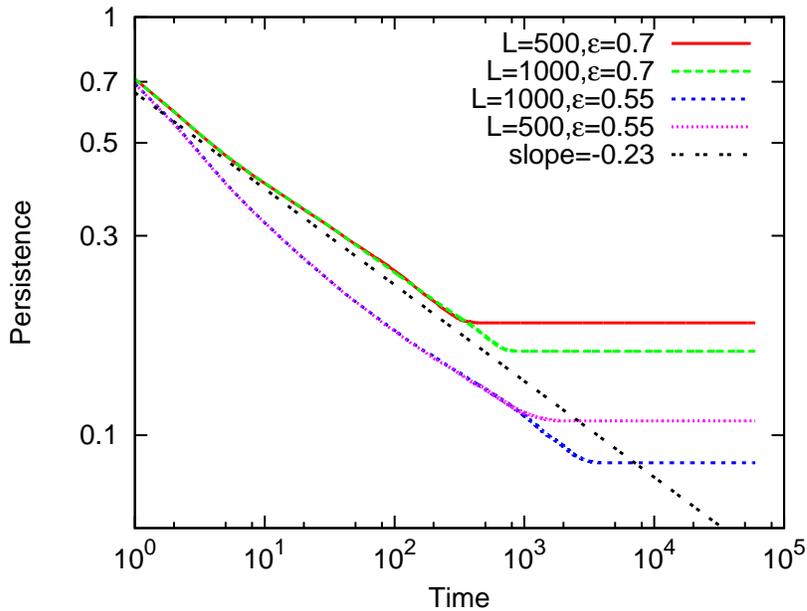}}}
\caption{Decay of persistence  with time for $\epsilon =0.7$ and $\epsilon =0.55$}
\label{5pereplt5}
\end{center}
\end{figure}

\begin{figure} [hbpt]
\begin{center}
\rotatebox{270}{\resizebox*{8cm}{!}{\includegraphics{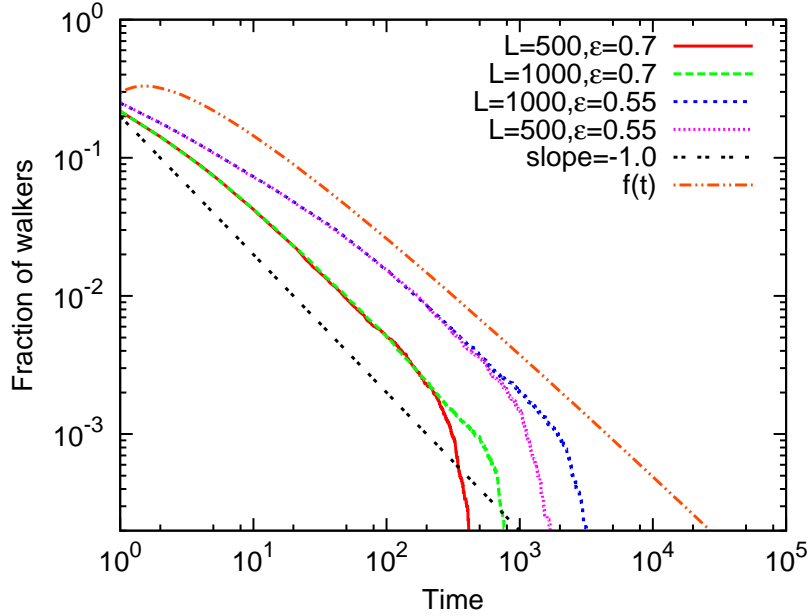}}}
\caption{Decay of number of walkers with time for $\epsilon =0.7$ and $\epsilon =0.55$. There is a logarithmic 
correction to scaling for the value of $\epsilon =0.55$. The form of $f(t)$ is $t^{-1}(1+\alpha \log(t))$ with 
$\alpha = 3.92$}
\label{5wlkeplt5}
\end{center}
\end{figure}

In reaction diffusion systems, the growth of domains is given by the 
number of surviving walkers. Persistence $P(t)$ in these systems is defined as the fraction of sites 
unvisited by any of the walkers $A$ till time $t$. Figures \ref{5pereplt5} and \ref{5wlkeplt5} show the decay of 
persistence and fraction of walkers with time for different values of $\epsilon > 0.5$. 

We find that for $\epsilon > 0.5$, the Model I behaviour is observed, namely: $z \simeq 1.0$ and 
$\theta \simeq 0.235$,  with some possible correction to the  scaling which becomes weaker as 
$\epsilon$ is increased.  For example, there is a 
logarithmic correction to scaling for the decay of the fraction of domain 
walls which takes the form $t^{-1}(1+\alpha (\epsilon) \log(t))$ where 
 $\alpha (\epsilon) \to 0$ as   $\epsilon \to 1$. 
One can compare the above model with the cases discussed in section 2.2.3, where
the introduction of stochastic dynamics also occurred with 
a bias towards the larger domain. In case of thermal disorder, the parameter 
comparable to $\epsilon$ is $\beta$. In the present case, the
exact sizes of the domains do not matter (which is important for the case with $\beta$) but the results are consistent
in the sense that any bias towards the larger domain (or nearest walker) makes the system behave 
like Model I. 

\begin{figure} [hbpt]
\begin{center}
\rotatebox{270}{\resizebox*{8cm}{!}{\includegraphics{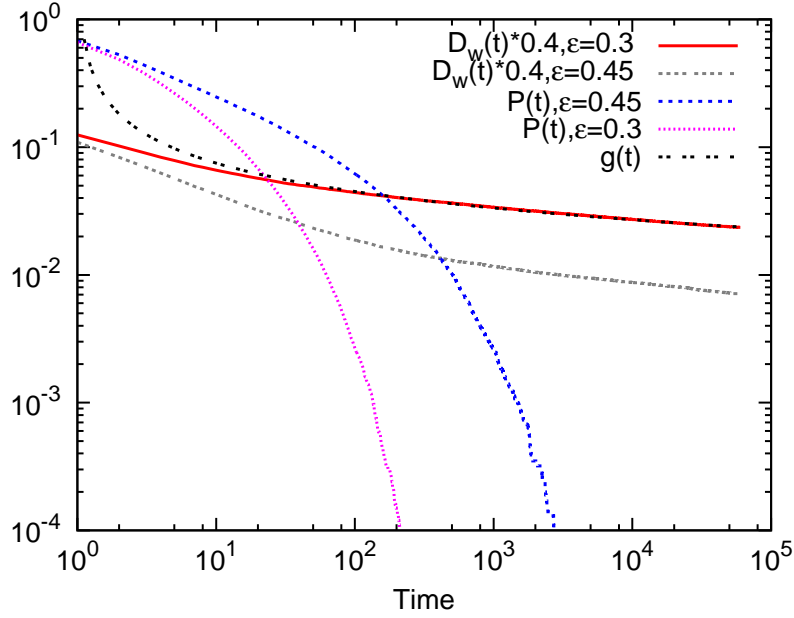}}}
\caption{Decay of persistence and number of walkers with time for $\epsilon =0.3$ and $\epsilon =0.45$. 
The form of $g(t)$ is $ a(1/ \log(t))$ where $a$ is any constant.}
\label{5eplt5}
\end{center}
\end{figure}

In this model, we have  also studied the $\epsilon < 0.5$ region 
where the opposite happens, the walker has a bias towards the further neighbour.
Obviously domain annihilations take place very slowly now, even slower than 
$1/\log(t)$ and the dynamics continues for very long times. Consequently, 
the persistence probability no longer shows a power law variation now  but falls 
exponentially to zero (Fig \ref{5eplt5}).

\section{Summary and concluding remarks} 

In summary, we have proposed a model of opinion dynamics in which
the social pressure is quantified in terms of the size of domains 
having  same opinion. In the simplest form,  the model has no disorder and
self organises to a homogeneous state in which the entire population has
the same opinion. This simple model exhibits novel coarsening exponents.
This model (Model I) in one dimension belongs to a new dynamical universality 
class with novel dynamical features not encountered in any previous models of dynamic spin system
or opinion dynamics. In the corresponding reaction diffusion system $A + A \rightarrow \emptyset$, we have 
introduced a probability $\epsilon$ of random walkers $A$ moving towards their nearest
neighbors. $\epsilon = 0.5$ corresponds to the particles $A$ performing unbiased random walks and the 
system belongs to the dynamical universality class of zero temperature Glauber Ising model. We find that 
for $\epsilon > 0.5$, the system still shows power law behavior of domain growth and persistence but with 
a universality class of that of Model I. For $\epsilon < 0.5$,  the domain grows logarithmically and the 
persistence decays exponentially in time. 

With disorder, the model undergoes a phase transition 
from a homogeneous society (with order parameter equal to one) 
to a heterogeneous one which is fully disordered
in the sense that no consensus can be reached as the order parameter
goes to zero in the thermodynamic limit.
 With both quenched and annealed 
randomness present in the system, the annealed randomness is observed
to drive the system to a homogeneous state for any amount of the quenched
randomness.

Many open questions still remain regarding  Models I and II,
the behaviour in higher dimensions being one of them.
 In fact, full understanding of the phase transition   occurring in Model II
reported here 
is   an important issue:   although   the phase transition 
has similarities with the one
dimensional Ising model, there are some distinctive features
which should be  studied in more detail. All the  models discussed in the present article can easily be extended to 
higher dimensions and its universality class determined. Phase transitions occurring at 
non-extreme values of suitably defined parameters may also be expected in higher dimensions.


\chapter{Dynamical crossover : Model with  variable range of interaction}
\label{Ccrossover}

\section{Introduction}
Dynamical  phenomena is an important topic in statistical physics. 
Physical  quantities in self organized and/or driven systems
show rich  time dependent behaviour in many cases. Some of the 
dynamical phenomena which have attracted a lot of attention are 
critical dynamics, quenching and coarsening, reaction diffusion systems, random walks etc. 

In most of these phenomena, we find there is a single timescale leading to  uniform  time 
 dependent behaviour which  in many cases 
is a power law decay or growth \cite{6bray}. However,
in some complex systems,  
it has been observed that the dynamics is governed by a distinct short
time behaviour followed by a different behavior at long times. For example, in
spin systems, at criticality, the order parameter is observed to grow 
for a macroscopically short time \cite{6early} while at longer times it decays in an  expected 
power law manner. For correlated random walks, e.g., the persistent random walk
on the other hand, one finds a ballistic (i.e., when the root mean square (rms)  displacement scales linearly
with time)
to diffusive (rms displacement varying as the square root of time) crossover in the dynamics \cite{6persrw}.
Random walks on small world networks show a completely opposite behaviour, the number of distinct sites
visited by the walker has an initial diffusive scaling followed by a ballistic variation  with time \cite{6swrw}. This is also true for a biased random walker.

In this chapter, we shall present our study on a dynamical model of Ising spins in one dimension
which is governed by a single parameter.  
The system is a generalized version of a recently 
  proposed  model
in \cite{6biswas-sen} (which we refer to as model I henceforth) 
where the state of the spins ($S = \pm 1$) may change in  two situations:
first when its two
neighbouring domains  have opposite polarity,  and in this case
the  spin  orients itself along the 
spins of the neighbouring domain with the  larger size.
This case may arise only when the spin  is at the boundary of the two
domains.
The neighbouring domain sizes are calculated excluding the spin itself, however, even if it is included, there is no change in the 
results.
A spin is also flipped when it is sandwiched between two domains of spins
with same sign.
 When  the two neighbouring domains of the spin are of the same size but have opposite polarity,
the spin will change its orientation with fifty percent probability.
 Except for this rare event the dynamics in the above model is deterministic. 
This dynamics leads to a  homogeneous state of either all spins
up or all spins down. Such evolution to absorbing homogeneous 
states are known to occur in systems belonging to
directed percolation (DP) processes, zero temperature Ising model, voter model etc. \cite{6absorb,6vote}.

Model I 
was introduced in the context 
of a social system where the binary 
opinions of individuals are 
 represented by    up and down spin states.
In opinion dynamics models, such representation of opinions by Ising or Potts
spins is quite common \cite{6opinion1}. The key feature is the interaction 
of the 
individuals which may lead to phase transitions between a homogeneous state to a heterogeneous state in many cases \cite{6opinion2}.

Model I showed the existence of 
novel  dynamical  behaviour  in a coarsening process when compared to the
dynamical behaviour of DP processes, voter model, Ising models etc. \cite{6hinrich2,6stauffer2,6sanchez,6shukla,6derrida}. The domain sizes 
were observed to grow as $t^{1/z}$ with the  exponent $z$ very close to unity. It may be noted that
the dynamics of a domain wall can be visualized as the movement of a walker and therefore the 
value $z\simeq 1$ indicated that the effective walks are ballistic.
When   stochasticity is introduced in this model, such that spin flips 
are dictated by a  so called ``temperature'' factor, it shows a 
robust behaviour in the sense that only for  the temperature going to infinity 
there is  conventional Ising model like behaviour with $z=2$, i.e., the domain wall dynamics becomes diffusive in nature \cite{6psnew}.

In this work, we have introduced the parameter $p$, which we call the cutoff factor, such that the maximum size of the neighbouring 
domains a spin can sense is given by $R = pL/2$ in a one dimensional system of $L$ spins 
with periodic boundary condition. 
It may be noted that for $p=1$, we recover the original model I where there
is effectively no restriction on the size sensitivity of the spins. $R=1$ corresponds
 to the nearest neighbour Ising model where $p\rightarrow 0$ in the thermodynamic limit.

By the introduction of the parameter $p$ we have essentially defined a
restricted neighbourhood of influence on a spin. Thus here we have a
finite neighbourhood to be considered, which is like having
a model with finite long range interaction.
In addition, here we impose the condition that within this restricted neighbourhood,
the domain structure is also important in the same way it was in 
model I.
If one considers  opinion dynamics systems (by which  model I  was originally inspired),
the domain sizes represent some kind of social pressure.
A finite cutoff (i.e., $p < 1$)
puts a restriction on the domain sizes which  may  correspond to geographical, political, cultural   boundaries etc.
The case with uniform cutoff signifies that all the individuals
have same kind of restriction;  we have also
considered the case with random cutoffs which is perhaps closer to reality.

In the next section, we describe the dynamical rule and 
quantities estimated. 
We  present the results for the case when $p$ is same for 
all spins in section 6.3 and 6.4 and in section 6.5 we consider the
case when the values of $p$ for each spin is random, lying between
zero and unity and constant over time for each spin.
In the last section, we end with concluding remarks. 

\section{Dynamical rule and quantities calculated}

As mentioned before, only the spins at the boundary of a domain wall can  change
its state. When sandwiched between two domains of same sign, it will be always flipped. On the other hand, for other boundary spins (termed the target spins henceforth), there will be
two neighbouring domains of opposite signs. For such spins, 
 we have the following dynamical 
scheme:
let $d_{up}$ and $d_{down}$ be the sizes of the two neighbouring
domains of type up and down of a target spin  (excluding itself). In model I,
the dynamical rule was like this: 
if $d_{up}$ is greater (less) than $d_{down}$, the target spin will be up (down) and if 
 $d_{up} = d_{down}$ the target spin is flipped with probability 0.5.
Now, with the introduction of $p$,  the definition of $d_{up}$ and $d_{down}$ are
modified: $d_{up} = {\rm{min}}\{R, d_{up}\}$ and 
similarly  $d_{down} = {\rm{min}}\{R, d_{down}\}$ while the same dynamical rule applies.

As far as dynamics is concerned, we investigate 
primarily the 
time dependent behaviour of the order parameter, fraction of domain walls   and
the persistence probability. The order parameter is  given by
 $m = \frac{|L_{up} - L_{down}|}{L}$
where $L_{up}~~ (L_{down})$ is the number of up (down) spins in the system
and $L = L_{up}+ L_{down}$, the total number of spins.
This is identical to the (absolute value of) magnetization in the Ising model.

The average fraction of domain walls  $D_w$, which is the average number of domain walls divided by the 
system size $L$ is identical to the 
inverse of average domain size. Hence the dynamical evolution of the order parameter and fraction of domain walls 
is  expected to be governed by the dynamical exponent $z$; $m \propto t^{1/(2z)}$ and 
$D_w \simeq t^{-1/z}$ \cite{6bray}. 

The persistence probability $P(t)$ 
of a spin is the probability that it remains in its original state 
up to time $t$ \cite{6derrida} is also estimated. $P(t)$  has been shown to have a power law decay in many systems with 
an associated exponent $\theta$.    
The persistence probability, in finite systems has been shown to obey the following scaling form \cite{6pray,6bcs}
\begin{equation}
P(t,L) \propto L^{-\alpha} f(t/L^z).
\label{6alpha}
\end{equation}
The exponent $\alpha= \theta z$ is associated with the 
saturation value of the persistence probability at $t\to \infty$ when 
$P_{sat}(L) = P(t \to \infty, L) \propto L^{-\alpha}$ \cite{6pray}. 

In the simulations, we have generated systems of size $ L \leq 6000$ with a minimum of 2000 initial 
configurations for the maximum size in general. 
Depending on the system size and time to equilibriate, maximum iteration times have been
set. Random updating process has been used to control the spin flips. In general,  the error bars in the data are less than the   
size of the data points in the figures and therefore not shown.

\section{Case with finite $R$ ($p \to 0$)}

\begin{figure} [hbpt]
\begin{center}
 
 \rotatebox{0}{\resizebox*{6cm}{!}{\includegraphics{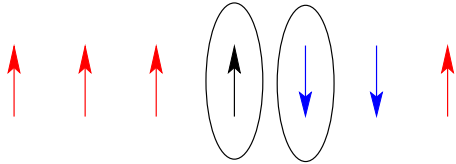}}}
\caption{ 
A schematic picture to show the dynamics in the present model for a   finite value of $R$.
Both the encircled spins will  change their state with fifty percent probability for the nearest neighbour Ising model ($R=1$). For $R=2$, the  encircled spin on the left 
will flip with probability 1/2 while  the  one on the right  will flip with probability 1. For $R=3$, the left one will not flip but the right  one will.}
\label{6schematic}
\end{center}

\end{figure}

In this section, we discuss the  case when $R$ is  finite. Effectively this 
means that $R$ does not scale with $L$ and is kept a constant for all system sizes.
Since  $R$ is kept finite,  
  expressing $R=pL/2$  implies 
$p \to 0$
in the 
  the thermodynamic limit.  
For $R=1$, the model is same as the Ising model as the dynamical rule is identical to the  zero temperature Glauber dynamics.
But it may be noted that making $R>1$ will 
make the dynamical rules different from the case of $R=1$;  as an example 
we show  in Fig. \ref{6schematic} how making $R=2$ or $3$  changes the 
dynamical rule compared to $R=1$.

\begin{figure} [hbpt]
\begin{center}
 \rotatebox{0}{\resizebox*{10cm}{!}{\includegraphics{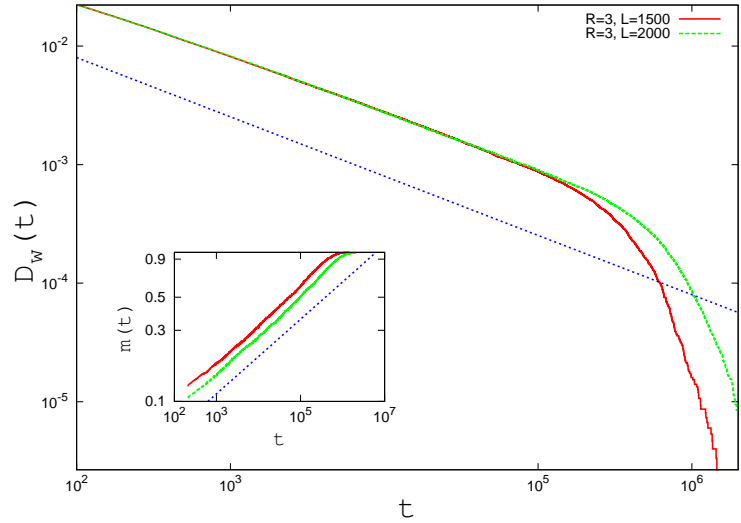}}}
\caption{
Decay of the fraction of domain walls $D_w(t)$ with time for $R=3$ and two different system sizes shown in a 
log-log plot. The dashed line has slope equal to 0.5. Inset shows growth of magnetization $m(t)$ with time for $R=3$; the  dashed line here has slope equal to 0.25.}
\label{6domainmagR3}
\end{center}
\end{figure}

\begin{figure} [hbpt]
\begin{center}
 \rotatebox{270}{\resizebox*{7cm}{!}{\includegraphics{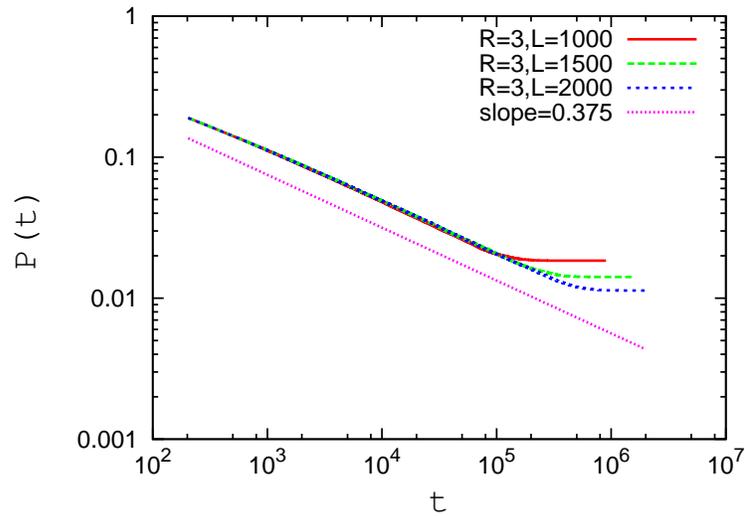}}}
\caption{
Decay of persistence probability $P(t)$ with time for three different sizes shown in a log-log plot.   The straight line has  slope 0.375.}
\label{6persR3}
\end{center}
\end{figure}

We have simulated systems with $R=2$ and $R=3$ which show
that the  dynamics leads to the equilibrium configuration of all spins up/down.
Not only that, the dynamic exponents also turn out to be identical to 
those  corresponding to the nearest neighbour Ising values (i.e.,  $\theta = 0.375$ and $z = 2$). As $R$ is increased, the finite size effects become stronger, however, it is 
indicated that the Ising exponents will prevail as the system size becomes 
larger. 
In an indirect way, we have  shown later that 
$z=2$ as $p \rightarrow 0$ using a general scaling argument. 
The behaviour of the different dynamic quantities
for $R=3$ are shown in Figs \ref{6domainmagR3} and \ref{6persR3}.

\section{Case with  $p > 0$}

In this section, we discuss the  case when $p$ is finite. We also assume that 
$p$ is uniform,  which means each spin experiences the same cutoff.

The equilibrium behavior is same for all $p$, i.e.,  starting from a random initial configuration, 
the  dynamics again leads to a final state with $m=1$, i.e., all spins up or all spins down.
For $p=1$, that is in model I, it was numerically obtained that $\theta \simeq	 0.235$ and   $z \simeq 1.0$ giving  
$\alpha \simeq 0.235$,
while in the one dimensional Ising model $\theta = 0.375$ and $z=2.0$ (exact results) giving $\alpha =0.75$.
It is  clearly indicated that  though model I and the Ising model have  identical equilibrium behaviour, they 
belong to two different dynamical classes which correspond to $p = 1$ and
$p \to 0$  limit respectively of the present model.
It is therefore of interest to investigate the dynamics in the intermediate range of $p$. 

\subsection{Results for $0<p<1$}

\begin{figure} [hbpt]
\begin{center}
 \rotatebox{0}{\resizebox*{9.5cm}{!}{\includegraphics{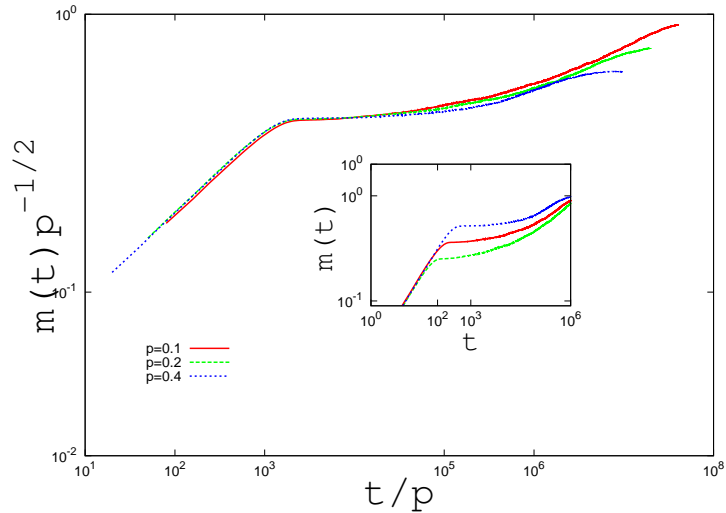}}}
\caption{ The collapse of scaled order parameter versus scaled time for different values of $p$,
shows $z=1$ for $t<t_{1}$ . Inset shows unscaled data. System size $L=3000$.}
\label{6magcolp}
\end{center}
\end{figure}

Drastic changes in the dynamics  are noted for
finite values of $ p <1$. The behaviour of all the three quantities, $m(t)$, $D_w$ and $P(t)$ shows the common feature of a power law growth or decay with time up to an initial time
$t_1$
 which 
 increases with $p$. The power law behaviour is followed by a  very slow variation 
of the quantities over a much longer interval of time, before they attain the equilibrium values. The  power law behaviour in the early time occur with
exponents consistent with model I, i.e., $z\simeq1$ and $\theta \simeq 0.235$. This early time behaviour accompanied by
model I exponents is easy to explain: it occurs while the domain sizes are less than $pL/2$ such that the size sensitivity 
does not matter and the dynamics is identical to that in model I. As the domain size increase beyond this value, the sizes of the neighbouring domains as sensed by the boundary spin   
become equal making the dynamics stochastic rather than deterministic as a result of which the dynamics becomes much slower.

\begin{figure} [hbpt]
\begin{center}
 \rotatebox{0}{\resizebox*{9.5cm}{!}{\includegraphics{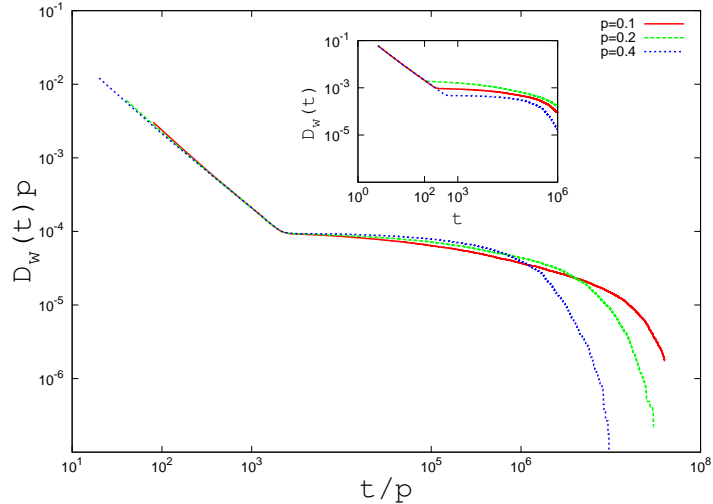}}}
\caption{The collapse of scaled fraction of domain walls versus scaled time for different values of $p$;
shows $z=1$ for $t<t_{1}$. Inset shows unscaled data. System size $L=3000$. }
\label{6domaincolp}
\end{center}
\end{figure}

\begin{figure} [hbpt]
\begin{center}
 \rotatebox{0}{\resizebox*{10cm}{!}{\includegraphics{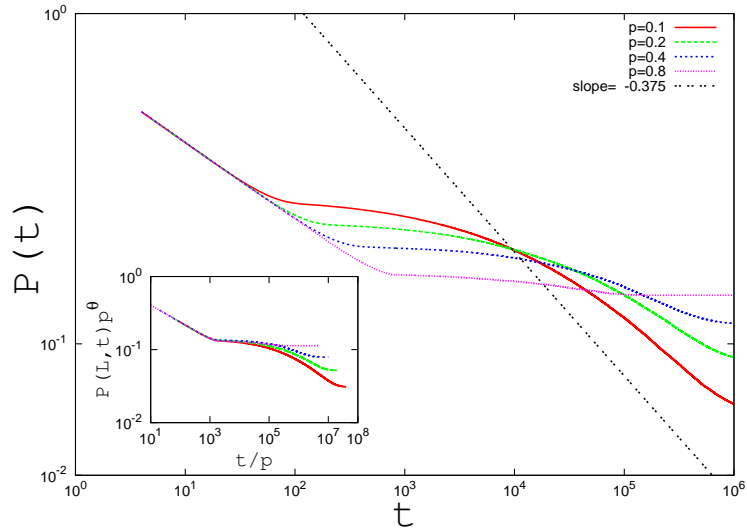}}}
\caption{ 
Persistence probability  versus time  for different values of $p$; the straight line with slope 0.375 shown for comparison.
Power law behaviour can be observed only at the initial time. 
 System size $L=3000$.
Inset shows the collapse of scaled persistence probability versus scaled time 
indicating  $z=1$ for $t<t_{1}$. 
}
\label{6percolp}
\end{center}
\end{figure}

We thus argue that since domain size $ \sim t^{1/z}$, the time up to which 
model I behaviour will be observed is $t_1= (pL/2)^{z}$. Since $z$ for model I is $1$ we expect that $t_1= pL/2$. 
For a fixed size $L$ one can then consider the  scaled time variable 
$t^\prime = t/p$, and plot the relevant scaled quantities against $t^\prime$  for different 
values of $p$ to  get a data  collapse up to $t_{1}^\prime = t_1/p$, 
independent of $p$. We indeed observe this, in Figures \ref{6magcolp}, \ref{6domaincolp} and  \ref{6percolp}, the scaling plots as well
as the raw data are shown. From the raw data, $t_1$ is clearly seen to be  different for different $p$.

Although the model I behaviour is confirmed up to $t_1$ and explained easily, beyond $t_1$, the raw data 
do not give any information about the dynamical exponents $z$ and $\theta$ as no straight forward power law fittings are possible. While an alternative method to calculate $\theta$ is not known,  one may have 
an estimate of $z$ using an indirect method. It has been shown recently that for  various dynamical Ising models,
the  time $t_{sat}$  to reach saturation varies as $L^x$ where $x$ is identical to the dynamical exponent $z$ \cite{6psnew,6pssdg}.
One may attempt to do the same here.

\begin{figure} [hbpt]
\begin{center}
 \rotatebox{270}{\resizebox*{6.8cm}{!}{\includegraphics{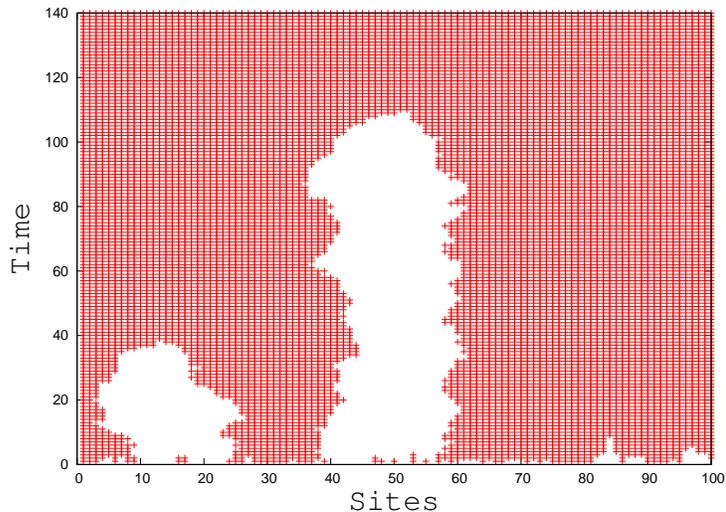}}}
\caption{ Snapshot for $p<1.0$ ($p=0.4$) for system size $L=100$.}
\label{6snap}
\end{center}
\end{figure}

Actually it  is possible to find out theoretically the form of $t_{sat}$ from the qualitative behaviour of the dynamical quantities described above
and the snapshot of the system (Fig. \ref{6snap}) at times beyond $t_1$. At $t > t_1$,  the domain sizes of the neighbours of any spin at the boundary appear equal
such that the domain walls perform random walks slowing down the annihilation process. Domain walls annihilate only after  one of the neighbouring domains
shrinks to a size $< pL/2$ again.  
In a small system, one can see that the slow process 
continues with  only two domain walls separating two domains 
remaining  in the system at later times (Fig. \ref{6snap}). 
Even in larger systems, there will be only a few  domain walls 
remaining making 
 $D_{w} ~ \propto 1/N$ at $t > t_1$ as we note from  the inset of  Fig \ref{6domaincolp}:
 $D_{w}$  remains close to $O(1/N)$ for a long time before 
going to  zero.

Thus $t_{sat}$ will have two components, $t_1$, already defined and $t_2$, the time during which there is a slow
variation of quantities over time and the last two domains remain. While $t_1 \propto pL$, one can argue that $t_2\propto (1-p)^3 L^2$.
The argument runs as follows: Let us for convenience consider the open boundary case. Here, the 
size sensitivity of the spins is $R^{open} = qL$ where $0 \leq q  \leq 1$ with the system assuming the model
I behaviour for $ q \geq 0.5$. At very late times, there
will remain only one domain boundary in the system separating two domains of size,  say, $\gamma L$ and $\beta L$,  such that
$\gamma + \beta = 1$.  With  both $\gamma , \beta > q$ the domain wall will perform random walk until either
of the domains shrinks to a size $qL$. (This picture is valid for   $ q < 0.5$ and otherwise the dynamics will be
simple model I type). Let us suppose that the domain with 
initial size $\beta L$   
shrinks to $qL$ in time $t_2^{open}$ such that the domain wall performs a random walk over a distance $s$  where
$\beta L - s = qL$. This gives
\[
t_2^{open}(\beta) \propto (\beta - q)^2L^2.
\]
Or, the average value of $t_2^{open}$ is given by
\[
t_{2}^{open} \propto \int _{q}^{1-q}  (\beta - q)^2L^2d\beta = \frac{(1-2q)^3 L^2}{3} .
\]
The result for the periodic boundary condition is  obtained by putting $q=p/2$ such that
\[
t_2 \propto (1-p)^3 L^2
\]
and therefore
\begin{equation}
t_{sat} = a pL + b (1-p)^3 L^2
\label{6tsat}
\end{equation}
The above form is also consistent with the fact that $t_{sat} \propto L^2$
for $p = 0$ and $t_{sat} \propto L$ for $p=1$.
\begin{figure} [hbpt]
\begin{center}
 \rotatebox{270}{\resizebox*{7cm}{!}{\includegraphics{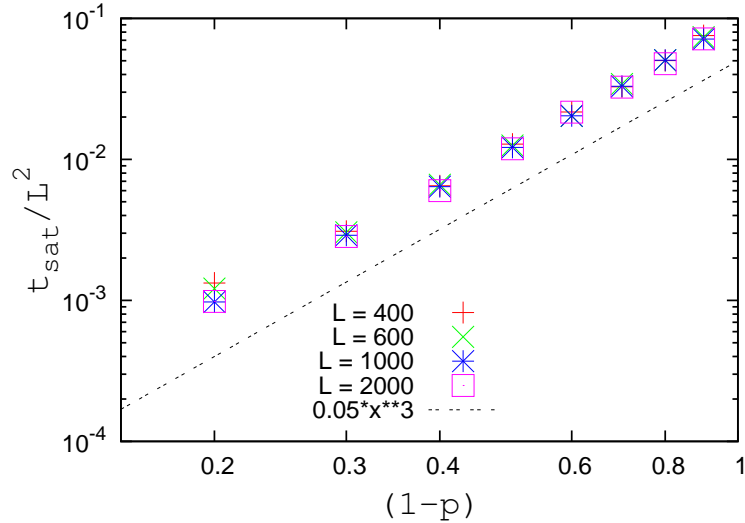}}}
\caption{ Scaled saturation time ($t_{sat}/L^2$) against $(1-p)$ for different $L$ shows collapse with $t_{sat}/L^2 \propto (1-p)^3$.}
\label{6sattime}
\end{center}
\end{figure}

For large $L$, the second term in the above equation will dominate making $t_{sat} \propto (1-p)^3 L^2$.
In order to verify this, we have numerically obtained $t_{sat}$ and plotted $t_{sat}/L^2$ against $(1-p)$
for different $L$ and found a nice collapse and a fit compatible with eq (\ref{6tsat}) (Fig. \ref{6sattime})
with  $ a ~ \sim 1$ and $ b  ~ \sim O(10^{-2})$. We conclude therefore that in the thermodynamic limit at later times,
for any $p \neq 1$, $z=2$, i.e., the dynamics is diffusive.
This argument, in fact holds for $p \rightarrow 0$ as well showing that 
for $R$ finite, $z=2$, as discussed in the preceding section.
 
\begin{figure} [hbpt]
\begin{center}
 \rotatebox{0}{\resizebox*{9cm}{!}{\includegraphics{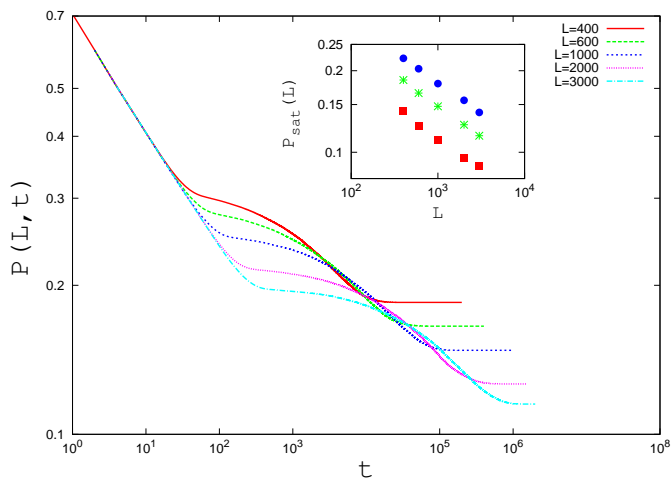}}}
\caption{ Persistence probability as a function of time for $p=0.4$ for different sizes. Inset shows
that the saturation values of the persistence probability shows a
variation $L^{-\alpha}$ for  values of $p = 0.8,0.4,0.2$ (from top to bottom) with
$\alpha \simeq 0.230$.}
\label{6per}
\end{center}
\end{figure}

We have discussed so far the time dependent behaviour and exponents only. But another exponent $\alpha$
which appears at $ t \rightarrow \infty $ for the persistence probability can also be extracted here. 
The persistence probabilities show the conventional saturation at large times, with the saturation values
depending on $L$. The log-log plot of $P(L, t \to \infty)$ against $L$ shows that power law behaviour is
obeyed here with the exponent $\alpha$ once again coinciding with the model I value, $\sim 0.23$ for any value of $p \neq 0 $
(Fig. \ref{6per}).

\begin{figure} [hbpt]
\begin{center}
\rotatebox{0}{\resizebox*{14.6cm}{!}{\includegraphics{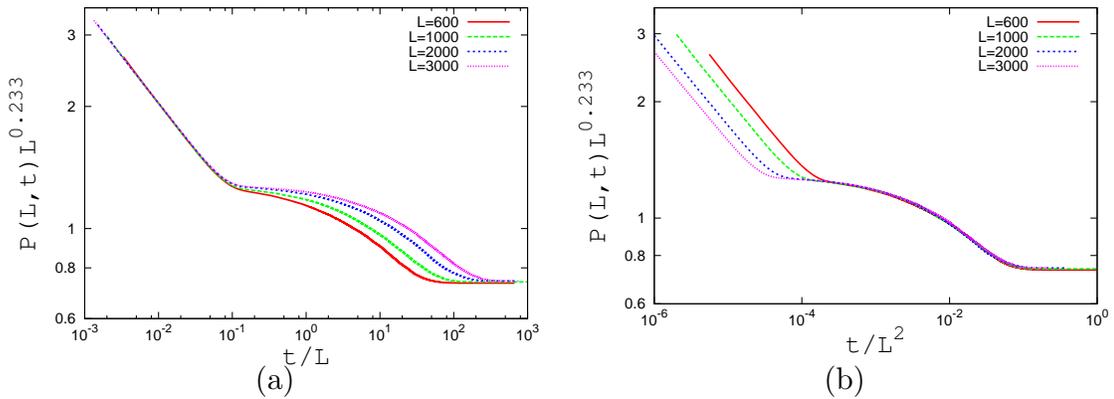}}}
\caption{ $PL^\alpha$ versus $t/L^z$ for $p=0.4$, shows a nice collapse for 
initial times up to $t_1$ using $z=1$ and $\alpha = 0.233$ (a) while using $z=2$ and the same value of $\alpha$, the data collapses over later times (b).}
\label{6colpintlt}
\end{center}
\end{figure}

Having obtained $\alpha$, we  use eq (\ref{6alpha}) with trial values of $z$ to obtain a collapse of
the data $PL^\alpha$ versus $t/L^z$ for any value of nonzero $p < 1$. As expected, an unique value of $z$ does not exist for
which the data will collapse over all $t/L^z$. However, we find that using $z=1$, one has a nice collapse for 
initial times up to $t_1$ while with $z=2$, the data collapses over later times (Fig. \ref{6colpintlt}). The significance of the result is, an unique value of $\alpha$ is good for collapse for both time regimes. However, it is not possible
to extract any value of $\theta$ for later times as $\theta$ is extracted 
from eq (\ref{6alpha}) in the limit  $t/L^z < 1$ only. 

\subsection{Discussions on the results}

At this juncture, several comments and discussions are necessary. We have obtained a crossover behaviour
in this model where an initial ballistic behaviour for macroscopic time scales is followed by a diffusive 
late time behaviour. 
However, the diffusive behaviour at later times is not apparent in the
simple log-log plots of the variables and can be extracted only from the
study of the total time to equilibriate. This is due to the fact that the 
initial ballistic dynamics leaves the system
into a non-typical configuration which is evidently far from those on 
diffusion paths. In fact in the  diffusive regime, the coarsening 
process hardly continues in terms of domain growth as only few domain walls 
remain at $t> t_1$.

A consequence of this is evident in the behaviour of the persistence at later times.  One may expect that 
the persistence exponent 3/8 may be obtained at very late times as here one has
independent random walkers, few in number, which annihilate each other 
as they meet much like in a reaction diffusion process. However, such an exponent 
is not observed from the data (Fig. \ref{6percolp}).  
 Although with $z=2$ we can obtain a collapse at later times, 
it is not possible to obtain a value of $\theta$. 
Since persistence is a non-Markovian phenomena and it depends on the history, the exponent may not be apparent even 
if the phenomena is reaction diffusion like. Therefore
to analyse the dynamical 
scenario further, we study the persistence in a different way.
In order to study the persistence dynamics beyond $t=t_1$, 
we reset the zero of time at $t=t_1$. 
 In case the number of domain walls left in the system at $t_1$ is of the order of the system size ($O(L)$), 
the behaviour of persistence should be as in the case of Ising model, i.e., a power
law decay with exponent 3/8.
On the other hand, if the number of independent random walkers is {\it{finite}} (i.e., vanishes in the $L \to \infty$ limit) which
can not annihilate each other, the persistence probability is
approximately
\begin{equation}
P_{rand}(t,L) = 1-ct^{1/2}/L ,
\label{6single}
\end{equation}
where we have assumed that number of distinct sites visited by the walker is proportional to the distance traveled, which is   
 $O(t^{1/2})$.

\begin{figure} [hbpt]
\begin{center}
 \rotatebox{0}{\resizebox*{10cm}{!}{\includegraphics{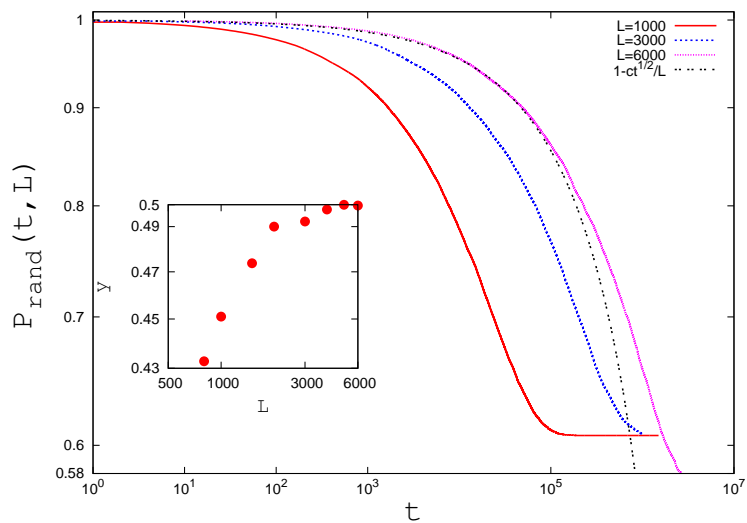}}}
\caption{ Persistence probability shows a decay as a function of time when $t_1$ is set as the initial time. 
The $L = 6000$ curve is fitted to the form $P_{rand}(t,L) = 1- ct^y/L$ with $y = 0.5$ (shown with the broken line). 
Inset shows the variation of $y$ with system size. $p=0.4$ here.}
\label{6perlt}
\end{center}
\end{figure}

 We find that in the present case, resetting the zero of time at $t_1$, the persistence probability
shows a decay before attaining a constant value.
The decay for a large initial time interval can be fitted to a form $\tilde P(t) = 1- c t^y$
where the exponent $y$ increases with $L$ and 
clearly tends to saturate at 0.5 as the system size is increased. This shows that the persistence 
probability is identical to (\ref{6single}) in form (Fig. \ref{6perlt}). 
This  signifies that 
at $t> t_1$, the dynamics only involves the motions of random walkers which
do {\it{not}} meet and annihilate each other for a long time and explains the fact that 
domain walls remain a constant over this interval. Only at very large times close to 
equilibriation the domain walls meet and the persistence probability
starts deviating from the behaviour given by (\ref{6single}). 
Actually once one of the neighbouring domains becomes less than $pL/2$ in size, the 
random walk will cease to take place and will become ballistic, which finally 
leads to  annihilation within a very short time. Therefore although we have at later times independent walkers 
performing random walk, the power law behaviour with exponent 3/8 will never be observed (even when the origin of the time is shifted) as the annihilation 
here is not taking place as in a usual reaction diffusion system but determined 
by the model I like dynamics. It may also be noted that beyond $t=t_1$,  annihilations 
occur only when the system is very close to equilibriation unlike in 
a reaction diffusion system where annihilations occur 
over all time scales.

The reason why a single value of $\alpha$ is valid
for both $t>t_1$ and $ t<t_1$ is also clear from the above study. We expect that at $t=t_1$, the number of persistent sites $\propto L^{- \alpha}$ with 
the value of $\alpha \simeq 0.235$ as in model I. The additional number of sites which become non-persistent beyond $t_1$ is proportional to $(t-t_1)^{y}/L$ and therefore at $t=t_{sat}$ expected number of persistent site is 
\[
 c_1 L^{- \alpha} - c_2 (t_{sat}-t_1)^{y}/L = c_1 L^{- \alpha} - c_2 t_2^{y}/L~,
\]
where $c_1,c_2$ are  proportionality constants. Since in the thermodynamic limit $y\rightarrow 1/2$ and $t_2 \propto L^2$, the number of persistence sites remains $\propto L^{-\alpha}$. 
Here we have assumed $c_2$ to be independent of $L$, the assumption is justified by the result.
 
\section{The case with quenched randomness}

In this section, we briefly report the behaviour of the system
when each spin is assigned a value of $p~(0<p \leq 1)$ randomly from a uniform
distribution. The randomness is quenched as the value of
$p$ assumed by a spin is fixed for all times. 

Here we note that the equilibrium behaviour, all spins up or down is
once again achieved in the system. However the time to reach equilibrium values are  larger
than the $p=1$ case. 

\begin{figure} [hbpt]
\begin{center}
 \rotatebox{0}{\resizebox*{10cm}{!}{\includegraphics{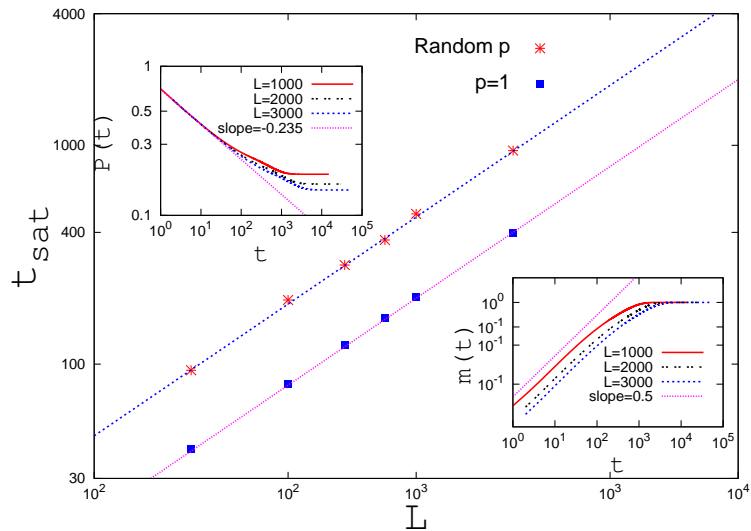}}}
\caption{ Saturation time $(t_{sat})$ against system size $L$ shows $z=1$. Inset on the top left shows the  persistence probability $P(t)$ with time which follows a power law decay with exponent $\sim 0.235$ initially.
The other inset on the bottom right shows the growth of magnetization $m(t)$ with time where the initial variation is like $m(t) \sim t^{1/2}$.}
\label{6inset3}
\end{center}
\end{figure}

The entire dynamics of the system, once again, can be
regarded as  walks performed  by the domain walls.
For $p=1$ for all sites, the walks are ballistic with the tendency
of a domain wall being to move towards its nearest one.
For $0< p \leq 1$ but same for all sites, as discussed in the previous section,
the walk is either ballistic (at initial times) or diffusive (at later times) 
but identical for all the walkers. 
 When $p$ is different for each site, one expects that 
when a site with a relatively large $p$ is hit, the corresponding  
domain wall will move towards its nearest domain wall  while when a site with relatively small $p$ 
is hit, the dynamics of the domain wall will be  diffusive.

It has been previously noted  that
model I with noise (of a different kind) 
which induces similar mixture of diffusive and ballistic
motions shows an overall ballistic behaviour (for finite noise) with the value of the 
dynamic exponent equal to  unity \cite{6psnew}.  
In the present model with quenched randomness also,
we find, by analyzing the saturation times that $z=1$. However, the variation of the magnetization, domain walls and persistence show 
power law scalings with exponents  corresponding to  model I only for an initial range of time (Fig \ref{6inset3}).

\section{Summary and concluding remarks}

In summary, we have  proposed a model in which a cutoff is introduced in
the size of the neighbouring domains as  sensed  by the spins. The cutoff $R$ is expressed in terms of a parameter $p$.
 At $p \rightarrow 0$ (finite $R$) and $p=1$ 
it shows pure diffusive and ballistic behavior respectively. In the uniform case where $p$ is same for all spins, a ballistic to 
diffusive crossover occurs in time for any nonzero $ p \neq 1$. 
Usually in a crossover phenomenon, where a power law behaviour occurs with two 
different exponents, the crossover is evident from a  simple log-log plot.
In this case, however, 
the crossover phenomena  is not apparent 
as a change in exponents in simple log-log plots does not appear.
The crossover occurs between two different types of phenomena, the first
is pure coarsening in which domain walls prefer to move towards their nearest
neighbours as in model I and one gets the expected power law behaviour. At $t_1$, 
as mentioned before, some special configurations are generated and therefore the second phenomena involves pure diffusion of a few  domain walls (density of domain walls 
going to zero in the thermodynamic limit) which remain non-interacting
up to  large times. Naturally, the only dynamic exponent in the diffusive 
regime 
is the diffusion exponent $z=2$ which  is {\it {distinct}} from the growth 
exponent $z=1$. So the two dynamic exponents not only differ
in magnitude, they are connected to distinct phenomena as well.  
This
crossover behaviour is therefore a striking feature for the model.
For $R$ finite ($p \to 0$), there is no crossover effect, as the time 
$t_1$ is too small to generate these special configurations and usual 
reaction diffusion type of behaviour prevails.

Persistence probability, in whichever
way one sets the zero of time, does not show  any power law behaviour in the 
second time regime.
 At the same time, a single value of $\alpha$ is required for the
collapse in the two regimes.
 
Another point of interest is that while $z=2$ is expected for nonzero $p\neq 1$ values at later times, 
the behaviour of the total time to equilibriate as a function of $p$ is not obvious. Our
calculation shows that it is proportional to $(1-p)^3$, which is another important result 
of the present work. 


We also found that making $p$ a quenched random variable taken from an uniform
distribution,  one gets back
model I like behaviour to a large extent. 
However, choosing a different distribution might lead to different results. 
The fact that the model has different 
behaviour with uniform  $p$ and with quenched random
value of $p$ is reminiscent of the different behaviour observed in agent based models 
with savings in econophysics \cite{6ccm}.


\end{document}